\documentstyle[12pt]{article}
\begin{document}
\newtheorem{DD}{Definition}[subsection]
\newtheorem{TT}{Theorem}[subsection]
\newtheorem{PP}{Proposition}[subsection]
\newtheorem{CC}{Corollary}[subsection]
\newtheorem{LL}{Lemma}[subsection]
\def\CC{{\cal C}}
\def\cn{{\cal C}^{\infty}(U)\otimes \!\!
          \makebox(12,9)[t]{ 
          \footnotesize $\bigwedge$} ({I\!\!R}^n)'}
\def\cnn{{\cal C}^{\infty}(U')\otimes \!\!
          \makebox(12,9)[t]{ 
          \footnotesize $\bigwedge$} ({I\!\!R}^{n})'}
\def\Z{Z\!\!\!Z}
\def\ZZ{Z\!\!\!Z_2}
\def\R{I\!\!R}
\def\Rn{I\!\!R^n}
\def\Rm{I\!\!R^m}
\def\Rnn{I\!\!R^{n'}}
\def\Rmm{I\!\!R^{m'}}
\def\ra{\rightarrow}
\def\lra{\longrightarrow}
\def\ve{\varepsilon}
\def\vp{\varphi}
\def\D{\Delta}
\def\a{\alpha}
\def\b{\beta}
\def\g{\gamma}
\begin{flushright}
                ITPUWr 917/97\\
\end{flushright}
\vspace{1cm}
\begin{center}
{\huge \bf Smooth coalgebras}
\end{center}
\bigskip
\begin{center}
{\bf  Zbigniew Jask\'{o}lski}
\end{center}
\begin{center}
Institute of Theoretical Physics\\
University of Wroc{\l}aw\\
pl. Maxa Borna 9\\
PL - 50 - 206 Wroc{\l}aw
\end{center}
\bigskip
\begin{abstract}
A complete mathematical framework for coalgebraic formulation
of supergeometry and its infinite-dimensional extension
is proposed. Within this approach a supermanifold is defined 
as a graded coalgebra endowed with a smooth structure.
The category of such  coalgebras 
is constructed and analysed. It is shown that it contains as its full
subcategories both the category of
smooth Fr\'{e}chet manifolds and the category of finite-dimensional
Berezin-Leites-Kostant supermanifolds.
\end{abstract}

\section{Introduction}

There are basically two different approaches to supergeometry:
the algebraic approach introduced
by Berezin and Leites \cite{berlei} and further developed by
Kostant \cite{Kostant} and Leites \cite{Leites},
cf.\cite{Gawedzki,Schmitt,Berezin,Manin}, and the geometrical approach
proposed by Rogers \cite {Rogers} and DeWitt \cite {DeWitt},
cf.\cite{Pilch,Boyer,Rabin,Bruzzo,BBH}.
The  theoretical framework incorporating both approaches
was   first proposed  by Rothstein in the form of axiomatic
definition of a supermanifold \cite{Rothstein}
and further analysed and improved by Bruzzo at al \cite{BBHP}.
The Berezin-Leites-Kostant (BLK) theory provides the simplest
realisation of this axiomatic definition  perfectly sufficient to derive
all nontrivial results of finite-dimensional supergeometry
including theories of:
Lie supergroups \cite{Kostant},
complex supermanifolds \cite{Green,Penkov,Rot2},
supersymplectic supermanifolds \cite{Giachetti,Rot3}, or
moduli of super Riemann surfaces
\cite{BMFS,Bryant,LeBrun,Giddings}.

All these results supported by methods  of algebraic and
analytic geometry along with  conceptual simplicity
of the BLK approach (it does not contain any spurious
Grassman algebra of constants) make it a
perfect mathematical language  for all physical applications
in which a finite-dimensional geometry is involved.
In supersymmetric classical and quantum field theories however
one has to deal with infinite-dimensional superspaces
of supergeometric structures. A typical problem one
encounters in this type of applications is to analyse the global
structure  of supermoduli intuitively constructed
as a quotient of an infinite-dimensional
supermanifold of field configurations by an infinite-dimensional
supergroup of gauge transformations. This construction
well known for "bosonic" models was never
made rigorous in the super case. The only method available is
based on finite-dimensional
techniques of deformation theory \cite{BMFS,LeBrun,Giddings}.
Although it usually provides quite a lot of information about
global geometry of the supermoduli, analysing induced
structures seems to require constructing the quotient. 
The lack of rigorous and efficient
methods of infinite-dimensional supergeometry
is also responsible for the informal heuristic
way one treats anticommuting classical fields in physical models.
As a  consequence the understanding of global geometry of
supermanifolds of field configurations as well as
actions of gauge supergroups on these supermanifolds
is in sharp contrast with  sophisticated methods
of standard global functional analysis \cite{Palais,Hamilton}
and detailed knowledge about similar problems in
"bosonic" models.

The aim of the present paper is to
construct an infinite-dimensional extension of
the BLK supergeometry.
Before discussing a possible solution to this problem,
let us briefly consider what kind of  
examples of infinite-dimensional
supermanifolds one should expect in physical models.
In the standard smooth geometry
the most important and interesting class of objects
studied via methods of functional non-linear analysis
are manifolds of maps with possibly additional properties
like that carried by sections of bundles.
In particular manifolds of various geometrical structures
belong to this class which is actually
essential for all physical and most of
mathematical applications of
infinite-dimensional geometry \cite{Palais,Hamilton}.
One can expect that also in supergeometry,
supermanifolds of maps are fundamental
for a geometric formulation
of supersymmetric models.
For an excellent heuristic discussion of the
notion of map between supermanifolds
in  the context of physical applications
we refer to the paper by Nelson \cite{Nelson}.
A special case of maps from the supermanifold
$S^{1,1}$ to a manifold was also analysed by Lott \cite{Lott1,Lott2}.
The main point is that
supergeometry requires {\it a notion of map
essentially wider than the notion of morphism}.
 This is in contrast to the standard smooth
geometry where both notions coincide.

In the BLK category morphisms are defined as even
$\ZZ$-graded algebra morphisms. For instance for
any pair ${\cal A}, {\cal B}$ of supermanifolds
all BLK morphisms from ${\cal A}$ to ${\cal B}$
form an ordinary (not graded)
infinite-dimensional manifold ${\rm Mor}({\cal A},{\cal B})$.
One would rather expect a supermanifold of maps
${\rm Map}({\cal A},{\cal B})$ with ${\rm Mor}({\cal A},{\cal B})$
playing the role of its underlying manifold.
In particular one would like to interpret the
$\ZZ$-graded space of real valued superfunctions
on a finite-dimensional supermanifold ${\cal A}$ as
a model space of a linear infinite-dimensional
supermanifold of maps from ${\cal A}$ to $\R$.
Certainly morphisms from ${\cal A}$ to $\R$ form only
the even part of this superspace.

The fact that morphisms are not enough to capture
the intuitive notion of "odd" maps one needs
in physical application is sometimes referred to as
the main shortcoming of the BLK theory (see for example
the discussion in \cite{Smolin}).
In its simplest form the argument says
that coefficients of a superfunction are
ordinary real-valued functions which do not
anticommute and therefore cannot provide a working model
for anticommuting classical fields one needs in physics.

This apparent drawback can be simply overcome by
regarding anticommuting classical fields as
odd coordinates on an   infinite-dimensional
supermanifold of fields configurations \cite{Schmitt0}.
Within the BLK approach the odd
variables $\{\theta_\alpha \}_{\alpha = 1}^n$ can be seen as
 a basis in the odd part
of a  $\ZZ$-graded model space $\R^{m,n}=\R^m \oplus \R^n$.
As such they are
genuine "commuting" objects of standard linear algebra.
The "anticommuting" nature shows up when elements of
the dual basis  $\{\theta^\alpha \}_{\alpha = 1}^n$ are interpreted
as generators of the exterior algebra
$ \makebox(12,9)[t]{
  \footnotesize $\bigwedge$} {{I\!\!R}^n}'$
over ${\R^n}'$. Following this line of thinking one can
regard classical fermion fields as elements
of the ordinary linear space ${\cal F}_1$ of sections of an
appropriate bundle, with ${\cal F}_1$  being the odd part
of an infinite-dimensional $\ZZ$-graded
model space ${\cal F}={\cal F}_0\oplus {\cal F}_1$.
Elements of ${\cal F}_1$
anticommute as arguments of
functionals from the exterior algebra
$ \makebox(12,9)[t]{
  \footnotesize $\bigwedge$} {\cal F}_1'$
and play essentially the same role as
$\theta$-variables in the finite-dimensional case.

It should be stressed that a reasonable
extension of an ordinary manifold
${\rm Mor}({\cal A},{\cal B})$
to supermanifold ${\rm Map}({\cal A},{\cal B})$
requires a global construction.
Indeed according to the basic idea of the
BKL approach one can think of "odd maps" as
odd  coordinates of an infinite dimensional supermanifolds of
maps ${\rm Map}({\cal A},{\cal B})$ rather then elements of some set.
This means in particular that also
the notion of composition cannot be defined
point by point but rather as a morphism of supermanifolds
$$
\circ: {\rm Map}({\cal A},{\cal B})
\times {\rm Map}({\cal B},{\cal C})
\longrightarrow {\rm Map}({\cal A},{\cal C})
$$
where $\times$ stands for the direct product in the category
of infinite-dimensional supermanifolds.
The obvious requirement for composition $\circ$ is that its
underlying map coincides with the standard composition of morphisms
of finite-dimensional supermanifolds.

Another problem of constructing an infinite-dimensional supergeometry
is to choose an appropriate  class of model spaces.
Since the composition of morphisms in
the BLK category involves differentiation of their
coefficient functions with respect to even coordinates,
smoothness is the minimal possible requirement
for morphisms and functions.
In consequence supermanifolds of supergeometrical
structures which are expected to be most interesting
objects  of infinite-dimensional supergeometry
should be  modelled on Fr\'{e}chet spaces.

The first systematic formulation
of infinite-dimensional supergeometry
was given by Molotkov \cite{Molotkov}. In his  approach
Banach supermanifolds are defined as functors from
the category of finite-dimensional real Grassman
superalgebras
$          \makebox(12,9)[t]{
          \footnotesize $\bigwedge$} {I\!\!R}^n
\; ( n=1,2,...)$ to the category of
smooth Banach
manifolds
$$
{\cal M}:           \makebox(12,9)[t]{
          \footnotesize $\bigwedge$} {I\!\!R}^n \longrightarrow
{\cal M}(          \makebox(12,9)[t]{
          \footnotesize $\bigwedge$} {I\!\!R}^n)
$$
For each Grassman algebra
$\makebox(12,9)[t]{
 \footnotesize $\bigwedge$} {I\!\!R}^n$,
${\cal M}(          \makebox(12,9)[t]{
          \footnotesize $\bigwedge$} {I\!\!R}^n)
$
can be identified
with smooth manifold of morphisms
${\rm Mor}({\cal P}_n, {\cal M})$ where
${\cal P}_n$ denotes finite dimensional
supermanifold with zero even dimension and
the odd dimension $n$ ( $n$-dimensional
superpoint).

The idea to regard supermanifolds as
point functors was first introduced by Schwarz in
his attempt to reconcile the standard sheaf
formulation of BLK finite-dimensional geometry
with the intuitive informal language used by physicists
\cite{Schwarz}. The equivalence
of Schwarz's  approach with the BLK theory
was shown by Voronov in \cite{Voronov}.
Molotkov's formulation can
be seen as a proper generalisation of the Schwarz description
to infinite-dimensions (i.e.
the BLK category of finite-dimensional
supermanifolds is a full subcategory of the category
of smooth Banach supermanifolds).

For any two supermanifolds ${\cal A}, {\cal B}$
(not necessarily finite-dimensional)
the supermanifold of maps can be defined by
the functor
$$
{\rm Map}({\cal A}, {\cal B}): \makebox(12,9)[t]{
          \footnotesize $\bigwedge$} {I\!\!R}^n
          \longrightarrow
{\rm Mor}({\cal P}_n  \times {\cal A}, {\cal B})
$$
The formalism also allows for a construction of a
composition with the required properties and
applies as well to
smooth supermanifolds modelled on
locally convex or tame Fr\'{e}chet superspaces.
In principle Molotkov's formulation satisfies
all requirements a mathematically rigorous
infinite-dimensional supergeometry should
satisfy. It has been in  fact implicitly used in several
papers when a rigorous treatment of
elements of infinite-dimensional
supergeometry was unavoidable \cite{BMFS,LeBrun,Lott1,Lott2}.
However, technical and conceptual difficulties
of this approach make its wider application in
physics highly problematic. According to the
basic idea of the functorial approach 
an object ${\cal A}$ in the category is fully described
by morphisms ${\rm Mor}({\cal P}_n,{\cal A})$ from a 
sufficiently large family
$\left\{ {\cal P}_n \right\}_{n\in I}$ of
other objects. Such description  is in sharp
contrast with the intuitive physical
understanding of space or superspace.
Also technicalities involved are in
contrast with relatively simple heuristic formalism
used by physicists.

Another approach to infinite-dimensional
supergeometry aimed to avoid
the functorial definition of supermanifolds
was developed by Schmitt \cite{Schmitt1,Schmitt2}.
The basic idea is to define an infinite-dimensional
supermanifold as a ringed space.
Although not functorial, this approach is technically
even more complicated. The main reason is that
in the infinite-dimensional supergeometry
the language and methods of algebraic geometry are
essentially less efficient and less powerful than in the
finite-dimensional case. In the standard BLK
approach one has a very simple
algebraic description of morphisms between supermanifolds,
either as morphisms of sheaves of graded algebras
or as morphisms of graded algebras of functions.
Also vector fields on a supermanifold can be described in
a purely algebraic way as graded derivations of the
graded algebra of superfunctions.
Proceeding to infinite-dimensional geometry one can
still consider sheaves of smooth functionals but the
simple algebraic descriptions of morphisms and vector
fields are no longer available. Additional
conditions involving topology as well as differential
calculus on the infinite-dimensional model spaces
are required in both cases.
In fact technical difficulties involved were
overcome only in the case of real-analytic and complex-analytic
supermanifolds \cite{Schmitt1,Schmitt2} which essentially restricts
possible physical applications of the theory.
It is also not clear how to construct supermanifolds
of maps and composition within this approach.

The formalisms of both approaches seem
to be technically too difficult when compared
with relatively simple heuristic rules
used by physicists even in most complicated geometrical
supersymmetric models. This suggests that there might be
a simpler theory incorporating all desired features
of the hitherto formulations but better suited for
constructing and analysing
examples arising in physical applications.

The aim of the present paper is to construct an alternative
coalgebraic formulation of infinite-dimensional
supergeometry which
allows to avoid at least some of the technicalities of
the functorial and the sheaf descriptions.
The idea of such approach was first proposed in the
excellent paper by Batchelor \cite{Batchelor} where
a candidate for
the dual coalgebra of the supermanifold of maps
between finite-dimensional supermanifolds was
constructed and analysed.

For any associative algebra with unit $(A,m,u)$
let us denote by $A^{\circ}$ the largest  subspace
of the full algebraic dual $A'$
such that $m'(A^{\circ})\subset A^{\circ}
\otimes A^{\circ}$,
where  $m':A'\rightarrow (A\otimes A)'$ is the map
dual to the multiplication $m: A\otimes A \rightarrow A$.
$A^{\circ}$ with comultiplication given by $m'$ and
counit given by $u'$ is called the dual coalgebra of $A$.
In the case of the algebra $C^{\infty}(M)$ of smooth
functions on a finite-dimensional manifold $M$,
$C^{\infty}(M)^{\circ}$ is called the dual coalgebra
of $M$ and consists of all finite linear combinations of
Dirac delta functions and their partial derivatives.
In the context of finite-dimensional supergeometry
dual ($\ZZ$-graded) coalgebras were first analysed and
extensively used by Kostant in his theory of
Lie supergroups \cite{Kostant}.

The idea of Batchelor's approach is to consider
the dual algebra of a supermanifold as a fundamental
object.  The crucial notion introduced in \cite{Batchelor} is
that of mapping coalgebra $P({\cal A},{\cal B})$ defined
for any two finite-dimensional
supermanifolds ${\cal A}, {\cal B}$
in terms of universal
coalgebra measuring the algebra of superfuncions on ${\cal B}$
to the algebra of superfunctions on ${\cal A}$.
Although the full structure theorem for
$P({\cal A},{\cal B})$ has not been proven,
the mapping coalgebra has many expected properties
of the dual coalgebra of the
supermanifolds of maps ${\rm Map}({\cal A},{\cal B})$.
In particular the space of group-like elements
of $P({\cal A},{\cal B})$ coincides with
${\rm Mor}({\cal A},{\cal B})$. Moreover there exists
a simple definition of composition which leads to the expected
Hopf algebra structure in the case of superdiffeomorphisms.
Batchelor's construction can be also extended to
coalgebras corresponding to supermanifolds of sections.

In the original paper \cite{Batchelor} only the algebraic structure
of mapping coalgebra has been analysed. This is
certainly not enough to define
supermanifold in terms  of its dual coalgebra.
A pure coalgebra structure
has to be supplemented by analytic data encoding
a smooth structure on a supermanifold.
This additional data is also necessary to
select those coalgebra morphisms which correspond
to smooth morphisms of supermanifolds.
Extra conditions in the definition of morphisms
may seem to be a shortcoming of the coalgebraic approach
in comparison to the algebraic one where smooth morphisms
can be defined in a purely algebraic way. Let us however
recall that the simple algebraic definition does not work 
in the case of infinite-dimensional model spaces.
Moreover the detailed discussion of morphisms within Schmitt's sheaf
formulation of infinite-dimensional supergeometry shows that
the coalgebraic structure is essential
for an appropriate definition of smoothness or analyticity
\cite{Schmitt1}.
On the other hand (as we shall see in the following)
the extra conditions one has to impose on
coalgebraic maps are essentially identical to
the differentiability
condition one imposes on maps of sets in the traditional
definition of smooth morphisms between manifolds.

In the present paper we propose an intrinsic
way to handle the additional analytic data
necessary to describe smooth structures.
The main result is the construction of the category
of smooth coalgebras which contains as its full
subcategories both the BLK category of finite-dimensional
supermanifolds and the category of smooth Fr\'{e}chet
manifolds. This provides a complete theoretical framework
of the coalgebraic formulation of supergeometry.
Although Batchelor's results \cite{Batchelor} were
our main motivation, we leave the construction
of smooth structure on the mapping coalgebra 
$P({\cal A},{\cal B})$
for future publications.
This involves in particular a construction of smooth
atlas on the manifold ${\rm Mor}({\cal A},{\cal B})$,
which goes far beyond the scope of this paper.

It should be stressed that the coalgebraic
description of supermanifolds has its advantages even
in the finite dimensions. First of all the general structure
of the theory is similar to that of the  standard smooth geometry:
supermanifolds are defined as sets
with extra structure and morphisms as maps of sets 
(with arrows in the "right" direction)
preserving these structures.
Secondly the direct p roduct in the category is just
the algebraic tensor product of coalgebras which
makes many of the standard geometric constructions
much simpler and more intuitive than in the sheaf
or the functorial approaches.
Finally the coalgebraic techniques proved to
be very useful in Kostant's theory of Lie supergroups
\cite{Kostant}.
In fact this theory gets much simpler when
smooth coalgebra morphisms are defined
in the intrinsic coalgebraic language without
referring to the algebraic formulation.
\bigskip

The content of the paper is as follows. Section 2 contains
preliminary material necessary for further constructions.
In Subsect.2.1 the basic facts about symmetric tensor algebra
$S(X)$ of
$\ZZ$-graded vector space $X=X_0\oplus X_1$ are presented.
In particular Hopf algebra structure on $S(X)$ is
described and a less known universal property of $S(X)$
with respect to its coalgebraic structure is proven. This
property is crucial for our description of smooth coalgebraic maps.
In Subsections 2.2 and 2.3 we recall some properties of
the model category ${\bf fm}$ of Fr\'{e}chet manifolds and the model
category ${\bf sm}$ of BLK supermanifolds,
respectively. In Subsection 2.4 the category of BLK finite-dimensional
supermanifolds is briefly presented.
This well known material is included for notational purposes
as well as for providing some motivation
for further constructions.

In Section 3  model category
${\bf sc}$ of smooth coalgebras is defined.
In Subsection 3.1. we introduce open coalgebras as
objects of the model category.
In Subsection 3.2. we present a crucial (for
all coalgebraic formulation) notion of smooth coalgebra
morphism and prove that it satisfies all the required
properties. In particular the component description
of morphisms is introduced and the formula for the composition
is derived. In Subsection 3.3 the construction of model category
${\bf sc}$ is completed and the direct product in ${\bf sc}$
is analysed. Finally in Subsection 3.4 we prove
that the model category ${\bf fm}$ of Fr\'{e}chet manifolds
can be identified with the full subcategory ${\bf sc}_0$
of even open coalgebras,
and the model category ${\bf sm}$ of BLK supermanifolds
can be identified with the full subcategory ${\bf sc}^{<}$
of finite-dimensional open coalgebras. This shows
that ${\bf sc}$ is an appropriate extension of
${\bf fm}$ and ${\bf sm}$.
In Subsection 3.5. the notion of superfunction
on an open coalgebra is introduced and analysed.

In Section 4 we describe construction and main properties of
the  category ${\rm \bf SC}$ of smooth coalgebras.
In Subsection 4.1 the smooth coalgebra is defined as a collection
of objects from the model category ${\bf sc}$ glued together by
by a collection of compatible morphisms from ${\bf sc}$.
Smooth morphisms of smooth coalgebras are defined along standard lines
by requiring that their local expressions are morphisms from
${\bf sc}$. The notion of superfunction on a smooth coalgebra
is defined and the functor from the category of
smooth coalgebras ${\rm \bf SC}$ to the category of sheaves of
$\ZZ$-graded algebras is constructed.
In Subsection 4.2. the direct product in the category
${\rm \bf SC}$ is analysed. In Subsection 4.3. we prove
that the full subcategory ${\rm \bf SC}_0$ of even
smooth coalgebras is isomorphic with the category of
Fr\'{e}chet manifolds. In Subsection 4.4. the corresponding
result for full subcategory ${\rm \bf SC}^{<}$ of
finite-dimensional smooth coalgebras and the category
of BLK supermanifolds is derived.

Appendix contains
definitions and basic facts about $\ZZ$-graded
spaces (A.1),
algebras (A.2), coalgebras (A.3), and bialgebras (A.4). Also some
elementary material on dual coalgebras of finite-dimensional
supermanifolds is briefly presented (A.5).
\bigskip

\noindent{\bf Acknowledgements}
This work has been written during
last few years with many long breaks
and at many places. 
I would like to thank to
Marjorie Batchelor, Ugo Bruzzo,
Alice Rogers,  Sergio Scarlatti, and
Alberto Verjovsky for enlightening discussions
on supergeometry.
I am very grateful to Adolfo S\'{a}nchez Valenzuela
for many hours of interesting  discussions on 
supergeometry and its coalgebraic formulation.
I owe special thanks to Sylvie Paycha for
for her enthusiasm and criticism
on  all stages of this project, for many discussions
and comments,
and for her careful critical reading
of the first version of this paper. 

The hospitality of the  International Centre of Theoretical
Physics, Trieste; Centro de Investigaci\'{o}n en Matem\'{a}ticas,
Guanajuato;  Institute de Recherche sur la Math\'{e}matique
Avanc\'{e}e,  Universit\'{e} Louis Pasteur et CRNS, Strasbourg;
Laboratoire de Math\'{e}matiques Appliquees, Universit\'{e}
Blaise Pascal, Clermont-Ferrand;
 and  Physikalisches Institut, Universit\"{a}t Bonn where various
parts of this work were completed is gratefully acknowledged.

\section{Preliminaries}

\subsection{Symmetric algebra of  graded vector space}

\begin{DD}
Let $X = X_0 \oplus X_1$ be a $\ZZ$-graded space. A  symmetric
 algebra of $X$  is a pair $(S(X),\theta)$ where $S(X)$ is a 
$\ZZ$-graded
 commutative algebra and $\theta:X\rightarrow S(X)$ a morphism of
 $\ZZ$-graded spaces such that the following universal property is 
satisfied.

For every $\ZZ$-graded commutative algebra $A$ and every 
morphism $F_0:X \rightarrow A$ of $\ZZ$-graded spaces there 
exists a unique $\ZZ$-graded algebra
morphism $F:S(X)\rightarrow A$ making the diagram

\begin{center}
\begin{picture}(120,75)(0,10)
\put(5,65){\makebox(35,10){$
%
   X
$}}
\put(50,75){\makebox(20,10)[b]{\small $
%
 \theta
$}}
\put(80,65){\makebox(40,10){$
%
 S(X)
$}}
\put(107,35){\makebox(20,10)[l]{\small$
%
 F
$}}
\put(80,5){\makebox(40,10){$
%
 A
$}}
\put(37,28){\makebox(20,10)[tr]{\small$
%
 F_0
$}}
\put(35,70){\vector(1,0){45}}
\put(35,61){\vector(4,-3){54}}
\put(100,28){\vector(0,-1){8}}
\multiput(100,60)(0,-8){4}{\line(0,-1){4}}

\end{picture}
\end{center}

\noindent commute.
\end{DD}
The uniqueness of $S(X)$ is a standard consequence of the universal
property.
The existence can be shown by explicit construction of $S(X)$ as the
quotient
algebra
$$ \Pi:T(X) \longrightarrow T(X)/I(X) \equiv S(X)\;\;\;, $$
where $(T(X),\theta_T : X  \rightarrow T(X) )$ is the tensor algebra 
of $X$
and $I(X)$ is the ideal generated by elements
$$\hspace{50pt} a\otimes b - (-1)^{\left| a \right| \left| b \right|} 
b
\otimes a
  \; \; \; ,
$$
where $a , b \in X_0 \cup X_1$, and $|.|$ denotes the parity of
an element.
The
ideal $I(X)$ is homogeneous with respect to the canonical
$ \ZZ \oplus \Z_+$
bigrading  on $T(X)$ and 
$S(X)$ acquires the structure of bigraded commutative algebra
$$
S(X)  =  \parbox[t]{35pt}{   \scriptsize $
                                    \hspace{5pt} \bigoplus \\
                                    \;\; k \geq 0  \\
                                    \;\;i  =   0,1
                                   $  }
                 \!\!\!\!           S^k_i(X) \;\;\; ,
$$
where $S^k(X)=S^k_0(X)\oplus S^k_1(X)$ is the $k$-th
symmetric tensor power of $X$.
Since $I(X)$
is generated by elements of degree 2 , one has the identifications
\begin{eqnarray*}
 S^0(X) & = & T^0(X) \;= \;\R \;\;\;,\\
 S^1(X) & = & T^1(X) \;=\; X\;\;\;,
\end{eqnarray*}
and the canonical map
$
\theta= \Pi\circ\theta_T : X \longrightarrow S(X)
$
is injective.

  The algebra $S(X)$ is generated by the set $ \{1\} \cup
X_0 \cup X_1 $ i.e. every element $\omega \in S(X)$ can be 
represented as
a finite sum of monomials of homogeneous elements of $X$ and $1$.

\begin{PP}
Let $X$, $Y$ be $\ZZ$-graded 
spaces and $\theta_X : X \rightarrow S(X)$ ,
$\theta_Y : Y \rightarrow S(Y)$ the canonical inclusions into the
corresponding symmetric algebras. Then the universal extension
$$
\kappa : S(X \oplus Y) \longrightarrow S(X) \otimes S(Y)
$$
of the map
$$
\kappa_0 : X \oplus Y \ni (a,b) \longrightarrow
\theta_Xa \otimes 1 +  1 \otimes \theta_Yb \in S(X) \otimes S(Y)
$$
is an isomorphism of bigraded algebras.
\end{PP}

\noindent {\bf Remark 2.1.1} 
 By Prop.2.1.1 for each $\ZZ$-graded space $X=X_0 \oplus X_1$
 there is the canonical
isomorphism of bigraded algebras
$$
\overline{\kappa} : S(X_0 \oplus X_1) \longrightarrow S(X_0) \otimes
\makebox(12,9)[t]{\footnotesize $\bigwedge$}(X_1) \;\;\;,
$$
where $S(X_0)$ is the usual symmetric algebra with its canonical
$\Z_+$-grading and the
trivial $\ZZ$-grading ( $S(X_0)_1 =  \{o\}$), and
 $\makebox(12,9)[t]{\footnotesize $\bigwedge$}(Y_1)$
 is the usual exterior  algebra
of the vector space $Y_1$
with its canonical $ \ZZ \oplus \Z_+$
bigrading.

\begin{PP}
Let $X$ be a $\ZZ$-graded space and $S(X)$ its symmetric algebra.

\noindent  Let 
$\Delta : S(X) \ra S(X) \otimes S(X)$ be the universal extention of
the map
$$
d : X \ni a  \longrightarrow  a \otimes 1 +
1 \otimes a \in S(X) \otimes S(X)\;\;\;, 
$$
and $\varepsilon :S(X) \rightarrow \R$
the universal extention of the map
$
0 : X \ni a  \longrightarrow  0 \in \R$.

Then  $(S(X),\Delta,\varepsilon)$ is a commutative cocommutative 
$\ZZ$-graded bialgebra.
\end{PP}

\noindent {\bf Remark 2.1.2}
We shall introduce some notational conventions which will be
used in various contexts in the following.

A  {\it $k$-partition} of the index set $\left\{ 1,...,n\right\}$
is defined as a sequence ${\cal P} = \left\{ P_1, \ldots ,P_k \right\}$ 
of (possibly empty) disjoint subset of the index set such that
$\left\{ 1, \ldots ,n \right\} = P_1 \cup \ldots \cup P_k$. 
A $k$-partition is {\it nonempty} if $P_i \neq \emptyset$ for
all $i=1,...,k$. Note that a nonempty $n$-partition of
the index set $\left\{ 1,...,n\right\}$ is a permutation of
$\left\{ 1,...,n\right\}$.

 Let ${\cal X} = \left\{ a_i \right\}_{i=1}^n$ be a
sequence of
nonvanishing homogeneous elements of a graded space $X$. For every
nonempty subset $P$ of the index set $\left\{ 1,...,n\right\}$ we 
define
$$
a_P = a_{p_1} \cdot \ldots \cdot a_{p_l} \in S^l(X)
$$
where $\left\{ p_1 , \ldots , p_l \right\} = P$ and $p_1 \leq \ldots 
\leq
p_l$. We denote the number of elements of $P$ by $\left| P \right|$.
If $P$ is empty we set $a_P = 1$ and $\left| P \right| = 0$.

For every
$k$-partition ${\cal P} = \left\{ P_1, \ldots ,P_k \right\}$ 
of the  index set 
$\left\{ 1, \ldots ,n \right\}$ 
we define the number $\sigma({\cal X},{\cal P})=\pm 1$ 
uniquely determined by the relation
$$
a_1 \cdot \ldots \cdot a_k = \sigma({\cal X},{\cal P}) a_{P_1} \cdot 
\ldots
\cdot a_{P_k} \;\;\;.
$$

\begin{PP}
Let $(S(X),\Delta,\ve)$ be the coalgebra of {\rm Prop.2.1.2}, and
 ${\cal X} = \left\{ a_i \right\}_{i=1}^n$ be a sequence of 
nonvanishing
homogeneous elements of a graded space $X$. Then:
\begin{enumerate}
\item $\Delta (1 )= 1 \otimes 1$, and $\varepsilon (1) = 1$.
\item For every $k \geq 1$
\begin{equation}
\label{delta}
\Delta^k(a_1 \cdot \ldots \cdot a_n) = \sum_{{\cal P} =
                                       \left\{ P_1, \ldots ,P_{k + 1}
\right\} }
          \sigma({\cal X},{\cal P}) a_{P_1} \otimes \ldots \otimes
a_{P_{k+1}}
\end{equation}
where the sum runs over all $(k+1)$-partitions of the index set
$\left\{ 1, \ldots , n \right\}$.
\item $\varepsilon (a_1 \cdot \ldots \cdot a_n) = 0$.
\end{enumerate}
\end{PP}

\begin{PP}
The symmetric algebra $S(X)$ of a $\ZZ$-graded space $X$ with
the coalgebraic structure of {\rm Prop.2.1.2} is a strictly
bigraded cocommutative coalgebra.
\end{PP}

\noindent{\bf Remark 2.1.3}
It follows from Prop.2.1.4 that  $S(X)$ is a pointed 
irreducible coalgebra. The relation between the $\Z_+$-grading
$
S(X) = \parbox{30pt}{\scriptsize $ \hspace{5pt} \infty \\
                               \makebox[2pt]{ } \bigoplus  \\
                                i = 0                           $}
\!\!\!\!\! S^i(X)
$
 and the coradical filtration 
$
S(X) =  \parbox{35pt}{\scriptsize $\hspace{6pt} \bigcup \\
                                    \;\; k \geq 0$}
\!\!\!\!\!\! S^{(k)}(X)
$
is given by
$$
S^{(k)}(X) = \parbox{30pt}{\scriptsize $ \hspace{7pt} k \\
                               \makebox[2pt]{ } \bigoplus  \\
                                i = 0                           $}
\!\!\!\!\! S^i(X)\;\;\;.
$$
\noindent The coalgebraic structure of $S(X)$ introduced 
in Prop.2.1.2 is universal in the following sense

\begin{TT}
Let $S(X)$ be the  symmetric algebra  of a $\ZZ$-graded
vector space $X$. 
There exists a unique extension of the bigraded
commutative algebra structure on $S(X)$ to  a strictly bigraded
commutative cocommutative Hopf algebra structure on $S(X)$.
\end{TT}

\noindent{\bf Remark 2.1.4}
The  antipode $s:S(X)\ra S(X)$ 
is given by the universal extension of the
map
$$
- : X \ni a  \longrightarrow  -a \in S(X)_{\rm op}\;\;\;,
$$
where $S(X)_{\rm op}$ is the bigraded space $S(X)$ with the
"opposite" algebra structure given by $
M_{\rm op}(a\otimes b) = (-1)^{|a| |b|}M(b\otimes a)$,
$u_{\rm op} = u$. One can easily show that for 
arbitrary homogeneous elements $a_1,...,a_n\in X$,
$
s(a_1 \cdot \ldots \cdot a_n) = (-1)^na_n \cdot \ldots \cdot a_1$.
\medskip 

\noindent
{\bf Definition 2.1.2}
Let $({\cal C},\Delta,\varepsilon)$ be a $\ZZ$-graded coalgebra,
$({\cal A},M,u)$ a $\ZZ$-graded algebra, and  ${\rm Hom}({\cal C},{\cal A})$ the
space of linear maps from ${\cal C}$ to ${\cal A}$.
For any  $f,g\in{\rm Hom}({\cal C},{\cal A})$  we define
$$
f \ast g \equiv M\circ (f\otimes g) \circ \Delta\;\;\;.
$$
By definition $f \ast g\in {\rm Hom}({\cal C},{\cal A})$ and $|f \ast g| =
|f| + |g|$ (with respect to the standard $\ZZ$-grading in 
${\rm Hom}({\cal C},{\cal A})$). One easily verifies that ${\rm Hom}({\cal C},{\cal A})$
with the multiplication $\ast$ and the identity $1_{\ast} \equiv
u\circ \ve$ is a $\ZZ$-graded algebra. The multiplication
$\ast$ is called {\it convolution}. \medskip

\noindent
The next theorem describes the universal property of $S(X)$
with respect to its coalgebra structure. This result is essential
for our description of smooth coalgebra morphisms given in the
subsequent  section.

\begin{TT}
Let $S(X)$ be the symmetric algebra of a $\ZZ$-graded space $X$ and
$\pi^P : S(X) \rightarrow S^1(X) = X$ the projection with respect to
the $\Z_+$-grading in $S(X)$. Let ${\cal C}$ be a pointed
irreducible $\ZZ$-graded cocommutative coalgebra and ${\cal C}^+$ the kernel
of the counit $\varepsilon_C$ in ${\cal C}$. Denote by $\rho^+ : {\cal C}^+
\rightarrow
{\cal C}$ the inclusion and by $\pi^+ : {\cal C} \rightarrow {\cal C}^+$ the projection
with
respect to the direct sum decomposition ${\cal C} = \R\{p\} \oplus {\cal C}^+$,
where $p$ is the unique group-like element of ${\cal C}$. Then:
\begin{enumerate}
\item
For every morphism $\Phi^+ : {\cal C}^+ \rightarrow X$ of $\ZZ$-graded
spaces there
exists a unique morphism $\Phi : {\cal C} \rightarrow S(X)$ of $\ZZ$-graded
coalgebras
such
that the diagram
\begin{center}
\begin{picture}(120,75)(0,10)
\put(0,65){\makebox(40,10){$
%
   X
$}}
\put(50,75){\makebox(20,10)[b]{\small $
%
 \pi^P
$}}
\put(80,65){\makebox(40,10){$
%
 S(X)
$}}
\put(107,35){\makebox(20,10)[l]{\small$
%
 \Phi
$}}
\put(80,5){\makebox(40,10){$
%
 {\cal C}
$}}
\put(50,15){\makebox(20,10)[b]{\small $
%
  \rho^+
$}}
\put(0,5){\makebox(40,10)[c]{$
%
 {\cal C}^+
$}}
\put(-7,35){\makebox(20,10)[r]{\small$
%
\Phi^+
$}}

\put(80,70){\vector(-1,0){45}}
\put(35,10){\vector(1,0){50}}
\put(20,20){\vector(0,1){40}}
\put(100,52){\vector(0,1){8}}
\multiput(100,20)(0,8){4}{\line(0,1){4}}

\end{picture}
\end{center}
is commutative.
\item
The universal extension $\Phi$ is given by
$$
\Phi = \ast\exp \Phi^+ \equiv \sum_{k \geq 0} \frac{1}{k!} \Phi^{+k} 
\;\;,
$$
where
$$
\begin{array}{lclcl}
\Phi^{+0} & \equiv & u \circ \varepsilon_C & , &   \\
\Phi^{+k} & \equiv &
\underbrace{\Phi^+ \circ \pi^+ \ast \ldots \ast \Phi^+ \circ 
\pi^+}_{k}
& , & \;\; k \geq 1 \;\;,
\end{array}
$$
and $\ast$ is the convolution in ${\rm Hom}({\cal C},S(X))$.
\end{enumerate}
\end{TT}

\noindent
Note that in the purely even case $X = X_0 \oplus \{0\}$, $S(X)$ with
respect to its $\Z_+$-graded coalgebra structure is 
isomorphic with
the universal pointed irreducible cocommutative coalgebra considered 
in
\cite{sweedler}. Part 1) of the theorem above is a $\ZZ$-graded
version of Th.12.2.5 in \cite{sweedler}. In the special case $C = S(Y)$
where
$Y$ is
a $\ZZ$-graded space the explicit formula for 
the universal extension has 
been
derived in \cite{Schmitt1}.  The proof given here is a
generalization of   Schmitt's method.


\begin{LL}
With the notation of Th.2.1.2, for every $k \geq 0$ the following
relation holds
$$
\Delta \circ \Phi^{+k} = \sum_{i=0}^k
                           \left(
                           \parbox{14pt}{\scriptsize $\makebox[4.5pt]{} k \\
                           \makebox[5.5pt]{} i    $ }
                           \right)
(\Phi^{+i} \otimes \Phi^{+k-i}) \circ \Delta_C \;\;\;.
$$
\end{LL}

\noindent {\bf Proof.}
The case $k=0$ is straightforward. For $k=1$ one has
\begin{eqnarray*}
\Delta \circ \Phi^{+1}(c) & = &
\Phi^{+1}(c) \otimes 1 + 1 \otimes \Phi^{+1}(c) \\
& = & \sum_{(c)} \left(
      \Phi^{+1}(c_{(1)}) \otimes u \circ \varepsilon_C(c_{(2)}) +
      u \circ \varepsilon_C(c_{(1)}) \otimes \Phi^{+1}(c_{(2)}) \right) \\
& = & \sum_{i=1}^1 \left( \parbox{14pt}{\scriptsize $\makebox[5.5pt]{} 1 \\
                                     \makebox[5.5pt]{} i    $ }
  \right) (\Phi^{+i} \otimes \Phi^{+1-i}) \circ \Delta_C(c) \;\;\;.
\end{eqnarray*}
By definition of $\Phi^{+k}$
\begin{equation}    \label{eq:pA}
\Phi^{+k} = M \circ (\Phi^{+k} \otimes \Phi^{+0}) \circ \Delta_C =
M \circ (\Phi^{+0} \otimes \Phi^{+k}) \circ \Delta_C \;\;\;,
\end{equation}
and
\begin{equation}   \label{eq:pB}
\Phi^{+k+1} = M \circ (\Phi^{+k} \otimes \Phi^{+1}) \circ \Delta_C =
M \circ (\Phi^{+1} \otimes \Phi^{+k}) \circ \Delta_C \;\;\;,
\end{equation}
for all $k \geq 1$. Then by the induction hypothesis and (\ref{eq:pB})
\begin{eqnarray*}
\Delta \circ \Phi^{+k+1} & = &
             \Delta \circ M \circ (\Phi^{+k} \otimes \Phi^{+1}) \circ \Delta_C \\
& = & (M \otimes M) \circ (id \otimes T \otimes id) \circ
        \left( ( \Delta \circ \Phi^{+k} ) \otimes ( \Delta \circ \Phi^{+1} )
        \right) \circ
        \Delta_C \\
& = & (M \otimes M) \circ (id \otimes T \otimes id) \circ                    \\
& &    \sum_{i=0}^k  \left( \parbox{14pt}{\scriptsize $\makebox[4.5pt]{} n \\
                                     \makebox[5.5pt]{} i    $ }
                                  \right)
                          \left( \Phi^{+i} \otimes \Phi^{+k-i}   \otimes
                               ( \Phi^{+1} \otimes \Phi^{+0}  +
                                 \Phi^{+0} \otimes \Phi^{+1}) \right) \circ \\
& & (\Delta_C \otimes \Delta_C) \circ \Delta_C  \\
& = & (M \otimes M) \circ   \\
& &        \sum_{i=0}^k
                      \left( \parbox{14pt}{\scriptsize $\makebox[4.5pt]{} n \\
                                     \makebox[5.5pt]{} i    $ }
                      \right)
                      \left(
         \Phi^{+i} \otimes \Phi^{+1} \otimes \Phi^{+k-i} \otimes \Phi^{+0} +
         \Phi^{+i} \otimes \Phi^{+0} \otimes \Phi^{+k-i} \otimes \Phi^{+1}
                     \right) \circ \\
& &       (id \otimes T \otimes id) \circ
           (\Delta_C \otimes \Delta_C) \circ \Delta_C\;\;\;.
\end{eqnarray*}
 As a consequence of the coassociativity and
cocommutativity of $\Delta_C$ one has
$$
(id \otimes T \otimes id) \circ (\Delta_C \otimes \Delta_C) \circ \Delta_C =
(\Delta_C \otimes \Delta_C) \circ \Delta_C \;\;\;.
$$
Using this relation  and  formulae~(\ref{eq:pA}),
(\ref{eq:pB}) one finally gets
\begin{eqnarray*}
\Delta \circ \Phi^{+k+1} & = &
           \sum_{i=0}^k
                      \left( \parbox{14pt}{\scriptsize $\makebox[4.5pt]{} n \\
                                     \makebox[5.5pt]{} i    $ }
                      \right)
                      \left(
         \Phi^{+i+1} \otimes \Phi^{+k-1}  +
         \Phi^{+i} \otimes \Phi^{+k+1-i}
                     \right)  \circ \Delta_C    \\
& = & \sum_{i=0}^{k+1}
                      \left( \parbox{20pt}{\scriptsize $\makebox[0.5pt]{} k+1 \\
                                     \makebox[8.5pt]{} i    $ }
                      \right)
         ( \Phi^{+i} \otimes \Phi^{+k+1-i} )
                       \circ \Delta_C \;\;\;.  \;\;\;\;\;\;\; \Box
\end{eqnarray*}
\bigskip


\noindent {\bf Proof of Th.2.1.2.}
{\it Existence.}
Let $\left\{ {\cal C}^{(k)} \right\}_{k \geq 0}$ be the coradical filtration
on ${\cal C}$. For any given $k \geq 0$ all terms $\frac{1}{l!} \Phi^{+l}$ with
$l \geq k$ vanish on ${\cal C}^{(k)}$. By the structure theorem
Th.A.3.2 (see Appendix)
for every $c \in {\cal C}$
there exists $k$ such that $c \in {\cal C}^{(k)}$, hence
$\ast \exp \Phi^+$ is well defined on ${\cal C}$. By construction
$\Phi = \ast \exp \Phi^+$ is a morphism of graded spaces
such that $\pi^P \circ \Phi \circ \rho^+ = \Phi^+$.
 Using Lemma 1.1 one has
\begin{eqnarray*}
\Delta \circ \Phi & = & \sum_{k \geq 0} \frac{1}{k!}
                      \left( \parbox{14pt}{\scriptsize $\makebox[4.5pt]{} k \\
                                     \makebox[5.5pt]{} i    $ }
                      \right)
         ( \Phi^{+i} \otimes \Phi^{+k-i} ) \circ \Delta_C \\
& = & \sum_{n \geq o} \sum_{m \geq 0}
                                       \frac{1}{n!} \frac{1}{m!}
        ( \Phi^{+n} \otimes \Phi^{+m} ) \circ \Delta_C \\
& = & ( \Phi \otimes \Phi ) \circ \Delta_C \;\;\;\;.
\end{eqnarray*}
For $k \geq 1$, $\Phi^{+k} \subset S(X)^+ = \ker \varepsilon $ and
$$
\varepsilon \circ \Phi = \varepsilon \circ u \circ \varepsilon_C +
\sum_{k \geq 0} \frac{1}{k!} \varepsilon \circ \Phi^{+k} =
\varepsilon_C \;\;\;.
$$
It follows that $\Phi$ is a morphism of graded algebras which completes
the proof of existence.

{\it Uniqueness.}
For any other extension $\Phi'$ we have $\pi^P \circ \Phi' =
\pi^P \circ \Phi$ and therefore ${\rm Im}(\Phi' - \Phi) \cap
P(S(X)) = \{ o\}$. Since $S(X)$ is pointed irreducible,
$\Phi' = \Phi$ by Proposition A.3.2 (see Appendix). $\;\;\; \Box$
\bigskip

\subsection{Model Category of Fr\'{e}chet manifolds}

Let $U\subset X, V \in Y$ be open subsets of the Fr\'{e}chet spaces
$X$ and $Y$, respectively, and $\psi: U\rightarrow V$ a map.
\begin{DD}
The derivative of $\psi: U\rightarrow V$ at a point $u\in U$
in the direction $x\in X$ is defined by
$$
D^1\psi(u;x) = \lim_{\epsilon \rightarrow 0}
{\psi(u + \epsilon x) - \psi(u) \over \epsilon}\;\;\;.
$$
whenever the corresponding limit exists.
One says that $\psi$ is continuously differentiable
or $C^1$ on $U$ if the limit exists for all $u\in U$ and
$x\in X$ and if the map
$$
D^1\psi : U\times X \longrightarrow Y
$$
is jointly continuous (as a map on a subset of the product).
\end{DD}

\begin{DD}
The higher order  derivatives ($k\geq2$)
are inductively defined by 
$$
D^{k}\psi(u;x_1,...,x_{k}) 
= \lim_{\epsilon \rightarrow 0}
{D^{k-1}\psi(u + \epsilon x_{k}; x_1,...,x_{k-1}) - 
D^{k-1}\psi(u;x_1,...,x_{k-1}) \over \epsilon}\;\;\;,
$$
whenever the corresponding limit exists.

One says $\psi$ is $C^k$ if $D^{k}\psi(u;x_1,...,x_{k}) $ exists
for all $u\in U$ and
$x_1,...,x_{k} \in X$ and is jointly continuous as a map
$$
D^k\psi : U\times \underbrace{X\times...\times X}_{k}
 \longrightarrow Y\;\;\;.
$$
A map $\psi$ is smooth ($C^{\infty}$) if it is $C^k$ for
all $k>0$.
\end{DD}

\noindent
If $\psi$ is $C^k$ than 
$D^{k}\psi(u;x_1,...,x_{k})$  is totally
symmetric and linear separately in $x_1,...,x_{k}$ [Hamilton].
It can be therefore extended in the second variable to the
map
$$
D^k\psi : U\times S^k(X)
\ni (u;x_1\cdot\ldots\cdot x_{k}) \longrightarrow 
D^{k}\psi(u;x_1,...,x_{k}) \in Y\;\;\;,
$$
where $S^k(X)$ is the $k$-th symmetric tensor power of $X$.
In the following the same symbol $D^k\psi$ will be used for 
the derivatives and for their extensions defined above.
\bigskip

\noindent 
Using the chain rule and the Leibnitz rule for the first derivative 
\cite{Hamilton}
as well as an induction on $k$ one gets
\begin{PP}
Let $U,V,W$ be  open subsets of  Fr\'{e}chet space $X,Y,Z$, respectively,
and
$\phi: U \rightarrow V, \psi: V \rightarrow W$, $C^k$ maps. Then
the composition $\psi\circ\phi$ is a $C^k$ map and for all
$1\leq l \leq k$, $u\in U$, and $x_1,\ldots,x_k \in X$
one has
\begin{eqnarray}
\label{chr}
D^l(\psi\!\!\!\!&\circ & \!\!\!\! \phi)(u;x_1,\ldots,x_l) = \\
 &=&\sum\limits_{i=1}^l {1\over i!} \!
\sum\limits_{
               \parbox[t]{50pt}{ \scriptsize $
                                     \{ P_1,\ldots,P_i\} \\
                                    \makebox[10pt]{}| P_i| > 0 
                                   $  }}
\!\!\! D^i\psi(\phi(u);
D^{|P_1|}\phi(u;x_{P_1}),\ldots,D^{|P_i|}\phi(u;x_{P_i}))
\;, \nonumber
\end{eqnarray}
where the sum runs over all ordered nonempty $i$-partitions of the 
index set $\{1,...,l\}$.
\end{PP}

\begin{PP}
 For all $C^k$ functions  $f, g :U \rightarrow I\!\!R $,
$1\leq l \leq k$ and 
$x_1,\ldots,x_l \in X$
one has
\begin{equation}
\label{lr}
D^l(f\cdot g)(u;x_1,\ldots,x_k) =
\sum\limits_{\{P_1,P_2\}} 
D^{|P_1|}f(u;x_{P_1})\cdot D^{|P_2|}g(u;x_{P_2})
\;\;\;,
\end{equation}
where the sum runs over all ordered $2$-partitions of the index set
$\{1,\ldots,l\}$ and the convention $D^0f(u;x_{\emptyset}) \equiv 
f(u)$ is used.
\end{PP}

\begin{DD}
The objects of the model category ${\bf fm}$
of smooth Fr\'{e}chet manifolds are  open subsets
$U \subset X$ where $X$ runs over the category of Fr\'{e}chet spaces.

For any two objects $U,V \in {\rm O}{\bf fm}$
the space of morphisms ${\rm M}{\bf fm}(U,V)$ consists of all
smooth maps $\psi: U \rightarrow V$. The composition of morphisms is defined as
the composition of maps. An isomorphism in the category
 ${\bf fm}$ is called a diffeomorphism.
\end{DD}
We denote by ${\bf fm}^<$ the subcategory of ${\bf fm}$ 
consisting of all
open subsets of finite-dimensional Fr\'{e}chet spaces
and all ${\bf fm}$-morphism between them. \bigskip

\noindent
Let  ${\cal C}^{\infty}_U
= \left( U, {\cal C}^{\infty}(.) \right)$ be the sheaf of smooth functions
on $U$. A smooth map $\psi: U \rightarrow V$ 
 induces a morphism of sheaves of commutative algebras
$$
\overline{\psi} = ( \psi, \psi^*_{.}) : 
{\cal C}^{\infty}_U \longrightarrow {\cal C}^{\infty}_V
$$
where
for each open $V'\subset V$ the algebra map $\psi^*_{V'}$
is given by
$$
\psi^*_{V'}: {\cal C}^{\infty}(V') \ni f
\longrightarrow f\circ \psi \in {\cal C}^{\infty}(\psi^{-1}(V'))
\;\;\;.
$$
A morphism $
F=(F^0,F_.) :
  {\cal C}^{\infty}_U \rightarrow {\cal C}^{\infty}_V$
of sheaves of commutative algebras
 is not in general  of this form. However for the 
finite-dimensional Fr\'{e}chet spaces one has

\begin{PP}Let $U\in {I\!\!R}^m,V\in {I\!\!R}^{m'}$ be open subsets.

\begin{enumerate}
\item
For any morphism of sheaves of  algebras
$$
F=(F^0,F_.) :
  {\cal C}^{\infty}_U \longrightarrow {\cal C}^{\infty}_V\;\;\;,
$$
$F^0:U\rightarrow V$ is smooth and 
$F = \overline{F^0}$. 
\item For any algebra morphism
$A: {\cal C}^{\infty}(V) \rightarrow {\cal C}^{\infty}(U) $
there exists a unique smooth map $\psi:U\rightarrow V$
such that $\psi^*_U = A$.
\end{enumerate}
\end{PP}

\noindent
It follows  that for $U,V \in {\rm O}{\bf fm}^<$
one has the 1-1 correspondence
$$
{\rm M}{\bf fm}(U,V) =
{\rm M}{\bf fm}^<(U,V) \ni \psi \longrightarrow
\psi^*_V \in {\rm Alg}({\cal C}^{\infty}(V),{\cal C}^{\infty}(U))
\;\;\;.
$$

\noindent
This simple algebraic description of morphisms either
as morphisms of sheaves of algebras or as morphisms
of algebras of function is not longer valid 
for infinite dimensional Fr\'{e}chet spaces. In this case
the space of morphisms
of sheaves of algebras is essentially bigger than the space of
smooth maps of open sets. In order to characterize
the sheaf morphisms corresponding to smooth maps one 
needs some additional not algebraical conditions.
This makes the idea of ringed spaces  in  infinite-dimensional
geometry rather awkward and difficult to deal with
\cite{Schmitt1,Schmitt2}.
This is also the main difficulty in developing a working
infinite-dimensional extension of the Berezin-Leites-Kostant
theory of supermanifolds which was originally developed as a theory of
ringed spaces \cite{berlei,Kostant}.

\subsection{Model category of BLK supermanifolds}

\begin{DD}
The objects of the model category ${\bf sm}$
of supermanifolds are sheaves of ${ Z\!\!\!Z}_2$-graded algebras
$$
{\cal S}^{m,n}_U =\left( U, \cn \right)
\;\;\;,
$$
where $m, n$ are arbitrary nonnegative integers and 
$U$ runs over all open sets of ${I\!\!R}^m$.

For any two objects ${\cal S}^{m,n}_U, 
{\cal S}^{m',n'}_V \in {\rm O}{\bf sm}$
the space of morphisms ${\rm M{\bf sm}}
({\cal S}^{m,n}_U, {\cal S}^{m',n'}_V)$
consists of all morphisms of sheaves of  ${ Z\!\!\!Z}_2$-graded 
algebras. The composition of morphisms is defined as the composition
of  morphisms of sheaves. An isomorphism in ${\bf sm}$ is called
a superdiffeomorphism.
\end{DD}

\noindent
$(\Rn)'$ in the definition above denotes the space dual to 
$\Rn$.
An object  ${\cal S}^{m,n}_U \in {\rm O}{\bf sm}$
is called a {\it superdomain} of  the superspace
$\Rm\! \oplus \!\Rn$. 
 For each open subset
$U'$ of  the {\it underlying set} $U$
the elements 
of the ${ Z\!\!\!Z}_2$-graded  algebra ${\cal S}^{m,n}(U')
= \cnn$ are 
{\it superfunctions on $U'$}. According to the ${ Z\!\!\!Z}_2$-grading
$$
{\cal S}^{m,n}(U) = {\cal S}^{m,n}(U)_0 \oplus
{\cal S}^{m,n}(U)_1 \;\;\;,
$$  
each superfunctions can be uniquely represented as the
sum $f=f_0 + f_1$ of 
the {\it even} $f_0$ and the {\it odd} $f_1$ {\it parts}. 
A superfunction $f$ is called {\it even} ({\it odd}) if
$f_1=0$ ($f_0=0$).

For each 
${\bf sm}$-morphism  $F =(F^0,F_.):
{\cal S}^{m,n}_U \rightarrow {\cal S}^{m',n'}_V$
the  map $F^0: U\rightarrow V$ is called the {\it underlying part
of $F$}. In our notation $F_.$ denotes the family of
$\ZZ$-graded algebra morphisms
$F_{V'} :{\cal S}^{m',n'}(V') \rightarrow
{\cal S}^{m,n}_V({F^0}^{-1}(V')$,
where $V'$ runs over all open subsets of $V$.

\noindent
As a "super" counterpart of Prop.2.2.3 one has [Berezin, Leites,
Schmitt]

\begin{PP}
Let ${\cal S}^{m,n}_U, {\cal S}^{m',n'}_V$ be
superdomains.
\begin{enumerate}
\item
For any ${\bf sm}$-morphism  $F =(F^0,F_.)
:{\cal S}^{m,n}_U \rightarrow {\cal S}^{m',n'}_V$ 
the underlying map $F^0 : U \rightarrow V$ is smooth.
\item For any morphism $A : 
{\cal S}^{m',n'}(V) \rightarrow {\cal S}^{m,n}(U)$
of  ${ Z\!\!\!Z}_2$-graded algebras there exists a unique
${\bf sm}$-morphism  $F = (F^0,F_.):
{\cal S}^{m,n}_U \rightarrow {\cal S}^{m',n'}_V$  
 such that
$A=F_V$.
\end{enumerate}
\end{PP}

\noindent {\bf Remark 2.3.1}
It follows from Prop.2.2.3 and Prop.2.3.1 that the 
covariant functor 
\begin{eqnarray*}
{\rm O}{\bf sm} \ni {\cal S}^{m,n}_U
 &\longrightarrow & U\in {\rm O}{\bf fm}^{<} \;\;\;\\
{\rm M}{\bf sm} \ni F = (F^0, F_.)
&\longrightarrow & F^0 \in{\rm M}{\bf fm}^{<}\;\;\;
\end{eqnarray*}
has the  right inverse 
\begin{eqnarray}
\label{uf}
{\rm O}{\bf fm}^{<} \ni U &\longrightarrow &
{\cal C}^{\infty}_U = {\cal S}^{m,0}_U
\in {\rm O}{\bf sm}\;\;\;\\
{\rm M}{\bf fm}^{<}\ni \psi &\longrightarrow &
\overline{\psi} = (\psi, \psi^*_.) \in {\rm M}{\bf sm}
\nonumber\;\;\;.
\end{eqnarray}
The image of the functor above coincides with the subcategory
 ${\bf sm}_0$  of ${\bf sm}$
consisting of all objects of the form ${\cal S}^{m,0}_U$ and
all ${\bf sm}$-morphisms between them. It follows that 
the model category of
finite dimensional manifolds 
${\bf fm}^{<}$ can be regarded as the full
subcategory ${\bf sm}_0$ of the model category of BLK supermanifolds.
\bigskip

\noindent
Let $\left\{ \overline{u}_{\mu} \right\}_{\mu=1}^{m}$ be a standard
basis in ${I\!\!R}^m$. 
The functions $u^{\mu}:{I\!\!R}^m\supset U\ni u \rightarrow u^{\mu}
 \in {I\!\!R}$ uniquely defined by 
 $ u = \sum_{\mu =1}^{m} u^{\mu}\overline{u}_{\mu}$ 
are called  the {\it standard coordinates}
on $U\subset {I\!\!R}^m$. Let  $\left\{ \theta^{\alpha}
 \right\}_{\alpha=1}^{n}$ be the standard basis in $({I\!\!R}^n)'$.
The collection $\left\{ u^1,...,u^m,
\theta^1,...,\theta^n \right\}$ regarded as a subset of 
${\cal S}^{m,n}(U) = \cn $
is called 
the  {\it standard coordinate system  on} ${\cal S}^{m,n}_U$.
In the standard coordinates  every superfunction
$f\in {\cal S}^{m,n}(U)$ has a unique representation
\begin{equation}
\label{def}
f = f(u,\theta)
= f^0(u) + f^{\wedge}_0(u,\theta)
+ f^{\wedge}_1(u,\theta)\;\;\;,
\end{equation}
where
\begin{eqnarray*}
f^{\wedge}_0(u,\theta)&=& 
\sum\limits_{\parbox{30pt}{\scriptsize $k=2$ \\
even}}^n {1\over k!}
\sum\limits_{\alpha_1,...,\alpha_k =1}^n 
f^{\wedge}_{\alpha_1...\alpha_k}(u)\theta^{\alpha_1} 
\wedge...\wedge\theta^{\alpha_k}\;\;\;,\\
f^{\wedge}_1(u,\theta)&=&
\sum\limits_{\parbox{30pt}{\scriptsize $k=1$ \\
odd}}^n {1\over k!}
\sum\limits_{\alpha_1,...,\alpha_k =1}^n 
f^{\wedge}_{\alpha_1...\alpha_k}(u)\theta^{\alpha_1} 
\wedge...\wedge\theta^{\alpha_k}\;\;\;,
\end{eqnarray*}
and $f^0(u), f^{\wedge}_{\alpha_1...\alpha_k}(u)
\in {\cal C}^{\infty}(U)$. The coefficient functions
$f^{\wedge}_{\alpha_1...\alpha_k}(u)$ are assumed to be totally antisymmetric
in their indices.
 In the decomposition (\ref{def}) $f^0(u)$ is called the
{\it underlying} and $f^{\wedge}(u,\theta) =f^{\wedge}_0(u,\theta)
+ f^{\wedge}_1(u,\theta)$ the
{\it exterior part} of  the superfunction
$f$. 

Note that for an
arbitrary ${\bf sm}$-morphism  $F =(F^0,F_.):
{\cal S}^{m,n}_U \rightarrow {\cal S}^{m',n'}_V$ and any
superfunction $g\in {\cal S}^{m',n'}(V)$ one has
\begin{eqnarray*}
(F_Vg)^0 &=& 
F_V(g^0) \;= \;g^0\circ F^0\;\;\;,\\
(F_Vg)^{\wedge}_0 &=&F_V(g^{\wedge}_0)\;\;\;,
\end{eqnarray*}
but in general
$
(F_Vg)^{\wedge}_1  \neq F_V(g^{\wedge}_1)
$.

\begin{DD}
Let $\left\{ u^1,...,u^m,
\theta^1,...,\theta^n \right\}$ be the standard coordinate
system on  $ {\cal S}^{m,n}_U$. For each open
subset $U'\subset U$ the partial derivatives of
a superfunction  $f \in {\cal S}^{m,n}(U')$
are defined by
\begin{eqnarray*}
{\partial \over \partial u^{\mu}}f(u,\theta) &=& 
{\partial \over \partial u^{\mu}}f^0(u) + 
\sum\limits_{k=1}^n {1\over k!}
\sum\limits_{\alpha_1,...,\alpha_k =1}^n 
{\partial \over \partial u^{\mu}}
f^{\wedge}_{\alpha_1...\alpha_k}(u)\,\theta^{\alpha_1}\! 
\wedge\!...\!\wedge\!\theta^{\alpha_k}\;\;\;,\\
{\partial \over \partial {\theta}^{\alpha}}f(u,\theta) &=&  
\sum\limits_{k=1}^n {1\over k!}
\sum\limits_{\alpha_1,...,\alpha_k =1}^n 
\sum\limits_{i=1}^{k} (-1)^{i+1}
 \delta_{\alpha \alpha_i}
f_{\alpha_1\alpha_2...\alpha_k}(u)\,\theta^{\alpha_1}\!
\wedge\!...\!\wedge\!\widehat{\theta^{\alpha_i}}\!
\wedge\!...\!\wedge\!\theta^{\alpha_k}\;\;\;,
\end{eqnarray*}
where $\widehat{\theta^{\alpha_i}}$ means that 
$\theta^{\alpha_i}$ is omitted.
\end{DD}

\noindent
In the following we shall also use the  compact notation 
$\{ x^I \}_{I=1}^{m + n}$ for the standard coordinate
system $\left\{ u^{\mu} \right\}_{\mu=1}^{m} \cup
\left\{ {\theta}^{\alpha} \right\}_{\alpha=1}^{n}$ on 
${\cal S}^{m,n}_U$, where
$$
x^I = \left\{\parbox{200pt}{$u^{I}\;\;\;\;\;\;\; 
              {\rm for}\;\;\;I=1,...,m\\
              \theta^{I-m}\;\;\;{\rm for} \;\;\;I=m+ 1,...,m+n$}
\right.\;\;\;.
$$
Accordingly, for the partial derivatives one has
$$
\partial_I = \left\{\parbox{200pt}{
$\displaystyle {\partial\over  \partial u^{I}}\;\;\;\;\;\;\;
 {\rm for}\;\;\;I=1,...,m\\
{\partial \makebox(15,7){} \over 
 \partial \theta^{I-m}}\;\;\;{\rm for} \;\;\;I=m+1,...,m+n  $}
\right.\;\;\;.
$$
\bigskip

\noindent
Let $\{ 
\overline{{\theta}}_{\alpha} \}_{\alpha=1}^{n}$
be the basis in $\Rn$ dual to the basis 
$\left\{ {\theta}^{\alpha} \right\}_{\alpha=1}^{n}$.
Then $\{ \overline{x}_I \}_{I=1}^{m + n}
= \left\{ \overline{u}_{\mu} \right\}_{\mu=1}^{m} \cup
\{ \overline{{\theta}}_{\alpha} \}_{\alpha=1}^{n}$ is the 
standard basis in the $\ZZ$-graded space $\Rm\!\oplus\! \Rn$.

\begin{DD}
Let  $f \in {\cal S}^{m,n}(U)$ and $a= \sum
a^I \overline{x}_I \in  \Rm\!\oplus\!\Rn$. The first derivative
of the superfunction $f$ in the direction $a$ is defined by
$$
D^1f(u,\theta;a) = \sum\limits_{I =1}^{m+n}
a^I\partial_I f(u,\theta)\;\;\;.
$$
For $a_1,...,a_k \in  \Rm\!\oplus\!\Rn$ the higher order 
directional derivatives are given by
$$
D^kf(u,\theta;a_{1},...,a_{k}) = 
\sum\limits_{I_1,...,I_k =1}^{m+n} a_1^{I_1}...a_k^{I_k}
\partial_{I_1}...\partial_{I_k} f(u,\theta)\;\;\;.
$$
The $k$-th order differentiations ($k\geq 1$) are defined as the underlying
parts of the $k$-th order directional derivatives:
$$
\widetilde{D}^kf(u;a_{1},...,a_{k}) = 
\sum\limits_{I_1,...,I_k =1}^{m+n} a_1^{I_1}...a_k^{I_k}
\left(\partial_{I_1}...\partial_{I_k} f\right)^0(u)\;\;\;.
$$
\end{DD}

\noindent
The higher directional derivatives 
and differentiations are totally antisymmetric 
in the variables $a_1,...,a_k$ and therefore can be uniquely 
extended (in the second variable) to linear functions
on the $k$-th graded symmetric tensor power $S^k(\Rm\!\oplus\!\Rn)$
of the $\ZZ$-graded space $\Rm\!\oplus\!\Rn$.
With this  interpretation one can use the notation of Rem.2.1.2
for arguments of multilinear functions.
\bigskip

\noindent
Using the graded Leibnitz rule for first derivatives and induction on $k$
one gets the following multiple Leibnitz rule for superfunctions.

\begin{PP}
Let $f,g \in {\cal S}^{m,n}(U)$ be superfunctions 
and ${\cal X} = \left\{ a_i \right\}_{i=1}^k$ a sequence of homogeneous
elements of the graded space $\Rm\oplus\Rn$. Then
\begin{eqnarray}
\label{slr}
D^k(f\cdot g)(u
\!\!\!\!\!\!&,&\!\!\!\!\!\!\theta;a_1,\ldots,a_k) =\\
&=&\sum\limits_{{\cal P}=\{P_1,P_2\}} \sigma({\cal X},{\cal P}) 
(-1)^{|f||a_{P_2}|}
D^{|P_1|}f(u,\theta;a_{P_1})\cdot D^{|P_2|}g(u,\theta;a_{P_2})
\;\;\;,\nonumber
\end{eqnarray}
where the sum runs over all $2$-partitions of the index set
$\{1,\ldots,k\}$, and the notation of Rem.2.1.2 as well as
the convention $D^0f(u,\theta;x_{\emptyset})
\equiv f(u,\theta)$ are used.
\end{PP}

\noindent
The standard coordinate system on ${\cal S}^{m,n}_U$
 generates
the  subalgebra of superfunctions with polynomial coefficients.
With the topology of uniform convergence on compact subsets this
is a dense subalgebra of ${\cal S}^{m,n}(U)$. The following
proposition 
 \cite{Leites,Schmitt,Berezin,Manin} says that the standard coordinates
behave as algebraic generators with respect to
${ Z\!\!\!Z}_2$-graded  
algebra morphisms. This property is essential  for the coordinate
description of morphisms in ${\bf sm}$.

\begin{PP}
Let $\{x^I \}_{I=1}^{m+n}$ and 
$\{ y^J \}_{J=1}^{m'+n'}$ be 
the standard coordinate systems on
${\cal S}^{m,n}_U$ and ${\cal S}^{m',n'}_V$, respectively.
Let $\{ F^{J}\}_{J=1}^{m'+n'}  $
be a collection of superfunctions on  ${\cal S}^{m,n}_U$
such that $F^{J}$ is even for $J=1,...,m'$,
$F^J$ is odd for $J=m' + 1,..., m'+n'$, and the map
$$
F^0: U \ni u \longrightarrow 
( F^{10}(u),...,F^{m'0}(u)) \in {I\!\!R}^{m'}
$$
has its image in $V \subset  {I\!\!R}^{m'}$.

 Then there exists a unique ${\bf sm}$-morphism 
$F =(F^0,F_.) :
{\cal S}^{m,n}_U \rightarrow {\cal S}^{m',n'}_V$
such that 
\begin{equation}
\label{sch}
F_V(y^J) = F^J(u,\theta)\;\;\;,
\;\;\;J= 1,...,m' + n'\;\;\;.
\end{equation}
\end{PP}
The superfunctions $ F^J(u,\theta)=F_V(y^J)$ are called
the {\it coordinate representation} of the ${\bf sm}$-morphism
$F =(F^0,F_.)$. Formula (\ref{sch}) is
frequently written in the following somewhat
 incorrect but more intuitive  form
\begin{eqnarray*}
v^{\nu} &=& F^{\nu 0}(u) +
\sum\limits_{\parbox{30pt}{\scriptsize $k=2$ \\
even}}^n {1\over k!}
\sum\limits_{\alpha_1,...,\alpha_k =1}^n 
F^{ \nu\wedge}_{\alpha_1...\alpha_k}(u)\;\theta^{\alpha_1} 
\wedge...\wedge\theta^{\alpha_k}\;\;,\;\;\nu =1,...,m'\;\;,\\
\eta^{\beta}&=&
\sum\limits_{\parbox{30pt}{\scriptsize $k=1$ \\
odd}}^n {1\over k!}
\sum\limits_{\alpha_1,...,\alpha_k =1}^n 
F^{\beta\wedge}_{\alpha_1...\alpha_k}(u)\;\theta^{\alpha_1} 
\wedge...\wedge\theta^{\alpha_k}\;\;,\;\;\beta=1,...,n'\;\;.
\end{eqnarray*}

\begin{PP}
Let $\{ x^I \}_{I=1}^{m + n}$, and
$\{ y^J \}_{J=1}^{m' + n'}$
be standard coordinate systems on 
${\cal S}^{m,n}_U$, and ${\cal S}^{m',n'}_V$,  respectively.
Let $F =(F^0,F_.) :
{\cal S}^{m,n}_U \rightarrow {\cal S}^{m',n'}_V$
be a ${\bf sm}$-morphism with 
 the coordinate representation 
 $ \{F^{J}(u,\theta)\}_{J=1}^{m' + n'}$.


Then for any superfunction $g\in {\cal S}^{m',n'}(V)$, and
any sequence ${\cal X}=\{ a_{1},...,a_{k} \}$ 
of homogeneous elements of $\Rm\! \oplus\! \Rn$ one has 
$
(F_Vg)^0(u) = g^0\circ F^0(u)\;\;\;,
$
and
\begin{eqnarray}
\label{schr}
\widetilde{D}^k(F_Vg)(u
\!\!\!\!\!\!&;&\!\!\!\!\!\!a_{1},...,a_{k})\;
= \\
&=&
\sum\limits_{l =1}^{k} {1\over l!}
\sum\limits_{ \parbox[t]{50pt}{ \scriptsize $
                                     \{ P_1,\ldots,P_l\} \\
                                    \makebox[10pt]{} P_i \neq \emptyset
                                   $  }}
\!\!\!\!\!\!\sigma({\cal X},{\cal P})
\widetilde{D}^l g(
F^0(u), \widetilde{D}^{|P_1|}F(u,a_{P_1}),\ldots,
\widetilde{D}^{|P_l|}F(u,a_{P_l}))\;\;\;,\nonumber
\end{eqnarray}
where the sum runs 
over all nonempty partitions of the
index set $\left\{1,...,k\right\}$, and for each $u\in U$,
$$
\widetilde{D}^{|P_1|}F(u,a_{P_1}) =
\sum\limits_{J=1}^{m'+n'} \widetilde{D}^{|P_1|}F^J(u,a_{P_1}) 
\overline{y}_J \in I\!\!R^{m'}\!\oplus\!I\!\!R^{n'}\;\;\; .
$$
\end{PP}

\noindent{\bf Proof.} 
For all even $a_{1},...,a_{k} \in \Rn\oplus \{o\}$ formula (\ref{schr})
is the multiple chain rule for Fr\'{e}chet manifolds (Prop.2.2.1)
applied to the function $g^0\circ F^0(u)$.

For all odd $a_{1},...,a_{k} \in \{o\} \oplus\Rm$ formula (\ref{schr})
is equivalent to the standard Taylor expansion
for the pull-back of a superfunction \cite{Leites,Schmitt,Berezin,Manin}.
Indeed, let
$$
F_Vg(u,\theta) = F_Vg^0(u) +
\sum\limits_{k=1}^n {1\over k!}
\sum\limits_{\alpha_1,...,\alpha_k =1}^n 
F_Vg^{\wedge}_{\alpha_1...\alpha_k}(u)\,\theta^{\alpha_1} 
\!\wedge...\wedge\theta^{\alpha_k}\;\;\;.
$$
be the coordinate expression for the superfunction $F_Vg$, and
 $\{ \overline{\theta}_{\a}\}_{\a=1}^{n}$  
the standard basis in $\Rn$.
Then  the coefficients of the representation above are given by 
\begin{eqnarray}
\label{pullback}
F_Vg^{\wedge}_{\alpha_k...\alpha_1}(u)&
=& \widetilde{D}^k F_Vg(u; \overline{\theta}_{\a_1},...,
\overline{\theta}_{\a_k} ) \\
&=&
\sum\limits_{l =1}^{k} {1\over l!}
\sum\limits_{ \parbox[t]{50pt}{ \scriptsize $
                                     \{ P_1,\ldots,P_l\} \\
                                    \makebox[10pt]{} P_i \neq \emptyset
                                   $  }}
\!\!\!\!\!\!\sigma({\cal X},{\cal P})
\widetilde{D}^l g(
F^0(u), F^{ \wedge}_{P_1}(u),\ldots,
F^{ \wedge}_{P_l}(u))\;\;\;,
\nonumber
\end{eqnarray}
where 
${\cal X}=\{ \overline{\theta}_{\a_1},...,
\overline{\theta}_{\a_k} \}$ and for each subset $P =\{\a_1,...,\a_i\}$
  ($\a_1<...<\a_i$) of the index set $\{1,...,k\}$
$$
F^{ \wedge}_{P}(u)\; = \;\sum\limits_{J=1}^{m'+n'}
\widetilde{D}^{|P|}F^{ J\wedge}(u,\overline{\theta}_{\a_1},...
,\overline{\theta}_{\a_i}) \overline{y}_J
\;= \;\sum\limits_{J=1}^{m'+n'}
F^{ J\wedge}_{\a_i,...,\a_1}(u) \overline{y}_J\;\;\;.
$$
The general  case can be derived from formula
 (\ref{pullback}) by  differentiating in even directions.
\hfill $\Box$
\bigskip

\noindent Formulae (\ref{schr}) and (\ref{pullback}) suggest
the following slightly modified  description
of an ${\bf sm}$-morphism.

\begin{DD}
Let  $ \{F^{J}(u,\theta)\}_{J=1}^{m' + n'}$ 
be  the coordinate representation of a ${\bf sm}$-morphism 
$F =(F^0,F_.) :
{\cal S}^{m,n}_U \rightarrow {\cal S}^{m',n'}_V$.
For every $k\geq1$ the {\it $k$-th infinitesimal component}
of $F =(F^0,F_.)$ is defined by
$$
F^+_k : U\times S^k(\Rm \!\oplus\! \Rn) \ni 
(u, a_1\cdot ... \cdot a_k) \lra
\sum\limits_{J=1}^{m'+n'}
\widetilde{D}^k F^{ J}(u;a_1,..., a_k) \overline{y}_J
\in {I\!\!R}^{m'}\!\oplus\! {I\!\!R}^{n'}\;\;\;.
$$
The restriction of the map $F^+_k$ to 
$U\times
\makebox(12,9)[t]{\footnotesize $\bigwedge$}^k {I\!\!R}^n
\subset U\times S^k(\Rm \!\oplus \!\Rn)$:
$$
F^{\wedge}_k : U\times
\makebox(12,9)[t]{\footnotesize $\bigwedge$}^k {I\!\!R}^n 
 \ni 
(u, \xi_1\cdot ... \cdot \xi_k) \lra
\sum\limits_{J=1}^{m'+n'}
\widetilde{D}^k F^{ J}(u;\xi_1,..., \xi_k) \overline{y}_J
\in {I\!\!R}^{m'}\!\oplus \!{I\!\!R}^{n'}\;\;\;,
$$
is called the {\it $k$-th exterior component } of the
${\bf sm}$-morphism  $F =(F^0,F_.)$.
\end{DD}

\noindent As a simple consequence of the definition and Prop.2.3.3
one has

\begin{PP}
Let $\{ x^I \}_{I=1}^{m + n}$, and
$\{ y^J \}_{J=1}^{m' + n'}$
be standard coordinate systems on 
${\cal S}^{m,n}_U$, and ${\cal S}^{m',n'}_V$,  respectively.
Let $\Phi^0: \Rm \supset U \ra V \subset I\!\!\R^{m'}$ be a smooth
map and $\{ \Phi^{\wedge}_k\}_{k=1}^n$ a family of smooth maps
$$
\Phi^{\wedge}_k : U\times
\makebox(12,9)[t]{\footnotesize $\bigwedge$}^k {I\!\!R}^n 
 \lra {I\!\!R}^{m'}\!\oplus \!{I\!\!R}^{n'}\;\;\;,
$$
linear and even in the second variable.

Then there exists a unique ${\bf sm}$-morphism
 $F=(F^0,F_.) :{\cal S}^{m,n}_U \rightarrow {\cal S}^{m',n'}_V$
 with underlying part $F^0 = \Phi^0$ and
exterior components $F^{\wedge}_k =\Phi^{\wedge}_k$;
$k=1,...,n$.
\end{PP}

\noindent{\bf Remark 2.3.2} The underlying part and the exterior
components of a ${\bf sm}$-morphism $F$ contain essentially the same 
data as the coordinate representation  of $F$.
The virtue of the description in terms of exterior components is that 
it is independent of the choice of basis in the model superspace and
as we shall see in the next section that it can be easily generalized
to the infinite-dimensional case. Here we shall  consider
composition of ${\bf sm}$-morphisms is this language.

Let $G =(G^0,G_.) :
{\cal S}^{m',n'}_V \rightarrow {\cal S}^{m'',n''}_W$
be another ${\bf sm}$-morphism, and 
$\{ z^K \}_{K=1}^{m'' + n''}$ the
standard coordinate system on ${\cal S}^{m'',n''}_W$.
For the underlying parts one has 
$$
(G\circ F)^{0} = G^{0}\circ F^{0}\;\;\;.
$$
The coordinate representation of the composition
$G\circ F :{\cal S}^{m,n}_U \rightarrow {\cal S}^{m'',n''}_W$
is given by
$$
(G\circ F)^K\;=\;
(G\circ F)_W(z^K) \; =\; F_V(G_W(z^K))\;=
\;G_V(F^K)\;\;.
$$
Calculating the r.h.s. by the formula (\ref{pullback}) and using 
Def.2.3.4 one gets the following expression for the exterior components
of the composition
\begin{eqnarray}
\label{comp} 
(G\circ F)^{\wedge}(u\!\!\!\!\!\!&,&\!\!\!\!\!\!\xi_{1},...,\xi_{k})\;
= \\
&=&
\sum\limits_{l =1}^{k} {1\over l!}
\sum\limits_{ \parbox[t]{50pt}{ \scriptsize $
                                     \{ P_1,\ldots,P_l\} \\
                                    \makebox[10pt]{} P_i \neq \emptyset
                                   $  } }
\!\!\!\!\!\!\sigma({\cal X},{\cal P})
 G^{+}(
F^0(u), F^{ \wedge}(u,\xi_{P_1}),\ldots,
F^{ \wedge}(u,\xi_{P_l}))\;\;\;.\nonumber
\end{eqnarray}
Let us note that the exterior components of the composition
depend on the infinitesimal components $G^+_k$ of $G$ and therefore
involve partial derivatives in even directions
of the exterior components $G^{\wedge}_k$. This property of composition
of {\bf sm}-morphisms is responsible for most of the
peculiar features of supergeometry. In particular
this is the reason for which smooth
structures and Fr\'{e}chet spaces are indispensable.

\subsection{BLK supermanifolds}

\begin{DD}
A supermanifold modelled on the superspace $\Rm \!\oplus \!\Rn$
is a sheaf ${\cal A}_M = (M,{\cal A}(\:.\:))$
 of $\ZZ$-graded algebras on a Hausdorff topological
space $M$ such that for each $p\in M$ there exist an open
neighbourhood $U$ of $p$, and an isomorphism of sheaves of
 $\ZZ$-graded algebras
$$
F = (F^0,F_{.}) :{\cal A}_U\lra 
{\cal S}^{m,n}_{\Phi^0(U)}\;\;\;,
$$
where ${\cal A}_U = (U,{\cal A}(\:.\:))$ is the restriction of 
 ${\cal A}_M$ to $U$, and  ${\cal S}^{m,n}_{\Phi^0(U)}$
is a superdomain i.e. $\Phi^0(U)$ is an open
subset of $\Rm$ and ${\cal S}^{m,n}_{\Phi^0(U)} =
(\Phi^0(U), {\cal C}^{\infty}(\:.\:)\otimes \!\!
          \makebox(12,9)[t]{ 
          \footnotesize $\bigwedge$} ({I\!\!R}^n)')$.
\end{DD}

\begin{DD}
The objects of the category ${\bf SM}$  are
supermanifolds modelled on superspaces  $\Rm \!\oplus \!\Rn$
where $m,n\geq 0$. 

For any two objects ${\cal A}_M, {\cal B}_N \in {\rm O}{\bf SM}$
the space of morphisms ${\rm M}{\bf SM}$ consists of
all morphisms of sheaves of $\ZZ$-graded algebras. The composition
of ${\bf SM}$-morphisms is defined as the composition of morphisms of
sheaves.
\end{DD}

\noindent
Let ${\cal A}_M$ be a supermanifold. 
The collection $\{ (U_{\a},F_{\a}) \}_{\a \in I}$ of isomorphisms 
of sheaves of $\ZZ$-graded algebras 
$$
F_{\a} = (F^0_{\a},F_{\a.}) :{\cal A}_{U_{\a}}\lra 
{\cal S}^{m,n}_{F^0_{\a}(U_{\a})}\;\;\;,
$$
such that $\{ U_{\a}\}_{\a\in I}$ is an open cover of $M$ is
called an {\it $(m,n)$-atlas} on ${\cal A}_M$.

Let $\{ (U_{\a},F_{\a}) \}_{\a \in I}$ be an atlas on a supermanifold
 ${\cal A}_M$. Then by Rem.2.3.1 the collection
$\{ F_{\a}^0 \}_{\a \in I}$ is a smooth atlas on $M$. By the same token
different atlases on ${\cal A}_M$ lead to compatible atlases 
 and $M$ acquires a unique smooth structure. $M$ with
this structure is called the {\it underlying manifold} of 
${\cal A}_M$.
\bigskip

\noindent
One has the following global version of Prop.2.3.1
\cite{Leites,Schmitt,Berezin,Manin}.

\begin{PP}Let ${\cal A}_M, {\cal B}_N$ be supermanifolds.
\begin{enumerate}
\item
For any ${\bf SM}$-morphism  $F =(F^0,F_.)
:{\cal A}_M \ra {\cal B}_N$ 
the underlying map $F^0 : M\ra N$ is smooth.
\item 
For any morphism $A : 
{\cal B}(N) \rightarrow {\cal A}(M)$ 
of  ${ Z\!\!\!Z}_2$-graded algebras there exists a unique
${\bf SM}$-morphism  $F = (F^0,F_.):
{\cal A}_M \ra {\cal B}_N$  
 such that
$A=F_N$.
\end{enumerate}
\end{PP} 

\noindent
 Let $F =(F^0,F_.):{\cal A}_M \ra {\cal B}_N$ 
be a  morphism of supermanifolds and
 $\left\{
(V_{\gamma}, G_{\gamma}) \right\}_{\gamma \in J} $  an 
atlas on ${\cal B}_N$. Then there exsists an atlas  
$\left\{( U_{\alpha}, F_{\alpha}) \right\}_{\alpha \in I} $ 
on ${\cal A}_M$, 
such that for all $\a \in I$ there exists (a non necessarily
unique) $\a' \in J$ for which 
$
F^0(U_{\alpha}) \subset  V_{\alpha'}$.
The family of ${\bf sm}$-morphisms
 $\left\{ F_{\a\a'} \right\}_{\a \in I}$
defined for each $\a\in I$ by  
$$
F_{\a\a'} = 
G_{\a'}\circ F\circ F_{\a}^{-1}\;\;\;
$$
is called {\it a representation of the morphism} 
$F =(F^0,F_.):{\cal A}_M \ra {\cal B}_N$ 
in the atlas  
$\left\{( V_{\gamma}, \Psi_{\gamma}) \right\}_{\gamma \in J} $ 
on ${\cal B}_N$. 

\begin{PP}
Let $\left\{( U_{\alpha}, F_{\alpha}) \right\}_{\alpha \in I}$
be an $(m,n)$-atlas on a supermanifold ${\cal A}_M$, and
  $\left\{( V_{\gamma}, G_{\gamma}) \right\}_{\gamma \in J}$ 
 an $(m',n')$-atlas  on a supermanifold ${\cal B}_N$. Let 
$\left\{ F_{\alpha\a'} \right\}_{\alpha \in I}$ be a family
of maps such that: 
\begin{enumerate}
\item for all $\a\in I$,
  $F_{\alpha\a'}: {\cal S}^{m,n}_{F^0_{\a}( U_{\a})} \ra
{\cal S}^{(m',n')}_{G^0_{\a'}( V_{\a'})}$
is a  ${\bf sm}$-morphism;
\item for all $\a,\b \in I$ such that $U_{\a} \cap U_{\b} \neq \emptyset$ 
one has
$$
F_{\alpha\b'}=  G_{\b'}\circ G_{\a'}^{-1}\circ
F_{\alpha\a'}\circ F_{\a} \circ F_{\b}^{-1} \;\;\;.
$$ 
\end{enumerate}  
Then there exists a unique smooth morphism 
$F =(F^0,F_.):{\cal A}_M \ra {\cal B}_N$
of supermanifolds such that 
$\left\{F_{\alpha\a'} \right\}_{\alpha \in I}$ is 
  a  representation of $F$ in the atlas 
$\left\{( V_{\gamma}, G_{\gamma}) \right\}_{\gamma\in J}$.
\end{PP}

\begin{DD}
Let 
 $\{ U_{\a},\vp_{\a} \}_{\a \in I}$ be
an admissible atlas on a smooth manifold $M$ of dimension
$m$. Let
 $\{ F_{\a\b} \}_{\a \in I}$ be a collection of maps
such that,
\begin{enumerate}
\item 
for all $\a \in I, \b\in I(\a)\equiv \{ \b \in I: U_{\a} \cap U_{\b}
\neq \emptyset \}$ 
$$
F_{\a\b}=(F^0,{F_{\a\b}}_{.})
: {\cal S}^{m,n}_{\vp_{\a}(U_{\a} \cap U_{\b})}
\lra
{\cal S}^{m,n}_{\vp_{\b}(U_{\a} \cap U_{\b})}
$$
is an isomorphism of superdomains such that $F^0_{\a\b} =
\vp_{\b}\circ\vp_{\a}^{-1}$;
\item 
for all  $\a \in I$, $ F_{\a\a} = 
{\rm id}_{ {\cal S}^{m,n}_{\vp_{\a}(U_{\a})} }$;
\item 
for all $\a,\b,\gamma \in I$ such that $U_{\a} \cap U_{\b}
 \cap U_{\gamma}\neq \emptyset$, 
$
F_{\b\gamma}\circ F_{\a\b} = F_{\a\gamma}
$
on 
${\cal S}^{m,n}_{\vp_{\a}(U_{\a} \cap U_{\b}\cap U_{\g})}$.
\end{enumerate}
A collection $\{ F_{\a\b} \}_{\a \in I}$ 
with the properties stated above is called an $(m,n)$-cocycle of
transition ${\bf sm}$-morphisms  over  the atlas 
$\{ U_{\a},\vp_{\a} \}_{\a \in I}$ on $M$. 
\end{DD} 

\noindent Two cocycles 
$\{F'_{\a'\b'} \}_{\a' \in I}$,
$\{ {F''}_{\a''\b''} \}_{\a'' \in I''}$ 
of transition ${\bf sm}$-morphisms on $M$ are said to be {\it compatible}
if there exists a third one
$\{F_{\a\b} \}_{\a \in I}$
such that 
\begin{eqnarray*}
\{ \vp'_{\a'} \}_{\a' \in I'}\cup\{ \vp''_{\a''} \}_{\a'' \in I''}
&\subset &\{ \vp_{\a} \}_{\a \in I}\;\;\;,\\
\{F'_{\a'\b'} \}_{\a' \in I'}\cup
\{F''_{\a''\b''} \}_{\a'' \in I''}
&\subset& \{F_{\a\b} \}_{\a \in I}\;\;\;,
\end{eqnarray*}
as sets of maps.

\begin{PP}
Let $\{ F_{\a\b} \}_{\a \in I}$ 
be  an $(m,n)$-cocycle of
transition ${\bf sm}$-morphisms on $M$. Then there exists a 
unique supermanifold ${\cal A}_M$ with the underlying manifold
$M$  and with the  $(m,n)$-atlas 
$\{ ( U_{\a},F_{\a}) \}_{\a \in I}$  such
that
$$
F_{\a\b}\;=\; F_{\b}\circ 
{F_{\a}^{-1}}_{| 
{\cal S}^{n,m}_{F^0_{\a}(U_{\a} \cap U_{\b})}}
\;\;\;,
$$
for all $\a\in I$, $\b\in I(\a)$. 

Compatible  $(m,n)$-cocycles of
transition ${\bf sm}$-morphisms on $M$ lead by the construction above 
to the same supermanifold ${\cal A}_M$.
\end{PP}

\noindent
The supermanifold ${\cal A}_M$ and the $(m,n)$-atlas 
$\{ ( U_{\a},F_{\a}) \}_{\a \in I}$ of the proposition above 
are said to be {\it generated } by  the $(m,n)$-cocycle 
$\{ F_{\a\b} \}_{\a \in I}$.\bigskip

\section{Model Category}

\subsection{Objects}

Let $X$ be a vector space. The {\it group-like Hopf algebra $G(X)$
of} $X$ is the free vector space $\R X$ of $X$ (i.e. the vector
space over $\R$ containing $X$ as a basis) endowed with the
trivial $\ZZ \oplus \Z_+$ grading
$$
\R X
 \; = \; \parbox[c]{35pt}{      \scriptsize $
                                    \makebox(5,22)[b]{} \bigoplus \\
                                     \makebox(2,2){}k \geq 0  \\
                                    i =   0,1 $  }  
                                 \!\!\!  \R X^k_i 
\hspace{20pt} , \hspace{20pt}
\R X^k_i \; = \; \left\{
                    \begin{array}{ll}
                    \R X & \mbox{for $k=i=0$} \\
                    \{o\} & \mbox{otherwise}
                    \end{array}  \right. \;\;\;,
$$
and with the Hopf algebra structure given by
$$
\makebox(150,40)[c]{ \parbox{140pt}{
\begin{eqnarray*}
M_G(x \otimes y) &=& x + y\;\;\;\\
u_G(1) &=& o\;\;\;
\end{eqnarray*}}}
\makebox(150,40)[c]{ \parbox{140pt}{
\begin{eqnarray*}
\Delta_G(x) &=& x \otimes x\;\;\;\\
\varepsilon_G(x) &=& 1\;\;\;\\
s_G(x) &=& -x\;\;\;
\end{eqnarray*}}}
$$
for all $x, y \in X$ ( $+$, $o$, $-$ stand for the addition, the zero
vector, and the inverse in the vector space $X$). One can easily verify
that $G(X)$ is a pointed bigraded commutative cocommutative
Hopf algebra. The direct sum decomposition into irreducible components
takes the form
$$
G(X) =
\parbox{30pt}{\scriptsize $    \makebox[4pt]{ } \bigoplus      \\
                              x \in X $} \!\!\!\R x \;\;\;,
$$
where $\R x$ is a 1-dimensional subcoalgebra generated by
the group-like element $x \in X \subset G(X)$.
For every subset $U \subset X$ the free vector space $\R U$
is a bigraded subcoalgebra of $\R X$,
$$
\R U =
\parbox{30pt}{\scriptsize $    \makebox[4pt]{ } \bigoplus      \\
                              x \in U $} \!\!\!\R x
                              \subset \;\R X \;\;\;.
$$
$\R U$ with the induced bigraded coalgebra structure is
called the {\it group-like coalgebra of} $U$.

\begin{DD}
Let $X = X_0 \oplus X_1$ be a graded space. The tensor product
$$
{\cal D}_X = G(X_0) \otimes S(X)
$$
of the group-like Hopf algebra $G(X_0)$ of $X_0$ and
the symmetric algebra $S(X)$ of $X$ is called the Hopf algebra of the
graded space $X$.
\end{DD}

\noindent The Hopf algebra structure on ${\cal D}_X$ is given by
\begin{eqnarray*}
M_H((u \otimes \alpha) \otimes ( w \otimes \beta))& =& 
(u + w)\otimes \alpha
\cdot \beta \;\;\;, \\
u_H(1)& =& o \otimes 1\;\;\;,\\
\Delta_H(u \otimes \alpha) & =& \sum_{(\alpha)}  (u \otimes \alpha_{(1)})
\otimes (u \otimes \alpha_{(2)}) \;\;\; , \\
\varepsilon_H(u \otimes \alpha)& =&
\varepsilon_G\otimes \varepsilon (u \otimes \alpha)\;=\;
\varepsilon(\alpha)\;\;\;,\\
s_H(u \otimes \alpha)& =&
s_G\otimes s (u \otimes \alpha)\;=\;
(-u )\otimes s(\alpha)\;\;\;,
\end{eqnarray*}
for all $u, w \in X_0; \alpha, \beta \in S(X)$, where
$\varepsilon$ is given by Prop.2.1.2, and $s$
by Rem.2.1.4.
By definition of $G(X_0)$ and Th.2.1.1,
${\cal D}_X$ is a pointed bigraded commutative cocommutative Hopf algebra.
Since $G(X_0)$ is generated by group-like elements and
$S(X)$ is pointed irreducible
one has the following decomposition into irreducible
components
$$
{\cal D}_X  =
\parbox{30pt}{\scriptsize $    \makebox[6pt]{ } \bigoplus      \\
                              u \in X_0 $} \!\!{{\cal D}_X}_u\;\;\;,
$$
where ${{\cal D}_X}_u = \R u \otimes S(X) \cong S(X)$ for all
$u \in X_0$.

The subcoalgebra ${{\cal D}_X}_o$ is a strictly bigraded Hopf subalgebra
of ${\cal D}_X$. It
acts on ${\cal D}_X$ by the left and right multiplication. 
In particular using the
identification ${{\cal D}_X}_o \cong S(X)$ one gets the right action
of $S(X)$ on ${\cal D}_X$,
$$
R : {\cal D}_X \otimes S(X) \ni (u \otimes \alpha) 
\otimes \beta \longrightarrow
(u \otimes \alpha) \cdot \beta = 
u \otimes \alpha \cdot \beta \in {\cal D}_X \;\;\;.
$$
By construction each irreducible component is stable with respect to $R$ and
for every $u \in X_0$ the map
$$
R_u : S(X) \ni \alpha \longrightarrow (u \otimes 1) 
\cdot \alpha \in {{\cal D}_X}_u
$$
is an isomorphism of bigraded coalgebras. For the sake of simplicity we will
write $u$ for the group like element $u \otimes 1 \in {\cal D}_X$. With this
convention $X_0 \subset {\cal D}_X$ and $u \otimes \alpha = u \cdot \alpha$ for
all $u \otimes \alpha \in {\cal D}_X$.

\begin{DD}
Let ${\cal D}_X$ be the Hopf algebra of a graded space 
$X = X_0 \oplus X_1$, and
$U$ a subset of $X_0$. The subcoalgebra 
${\cal D}_{X}(U) \subset {\cal D}_X$
$$
{\cal D}_{X}(U)  =
\parbox{30pt}{\scriptsize $    \makebox[4pt]{ } \bigoplus      \\
                              u \in U $} {{\cal D}_X}_u\;\;\;,
$$
is called the subcoalgebra over $U$.

In the case of a graded Fr\'{e}chet space i.e. a topological
direct sum $X=X_0 \oplus X_1$ of Fr\'{e}chet spaces,
 the subcoalgebra ${\cal D}_{X}(U)$ over
an open subset $U \subset X_0$ will be called an open subcoalgebra of
${\cal D}_X$.
\end{DD}
By definition, for any $U \subset X_0$  the subcoalgebra ${\cal D}_{X}(U)$ 
over $U$ is the tensor product of bigraded coalgebras
$
{\cal D}_{X}(U) = G(U) \otimes S(X).
$
As the direct sum of irreducible components of ${\cal D}_X$ it is stable 
uder the $S(X)$ right action.

\subsection{Morphisms}

Let $X, Y$ be graded spaces and ${\cal D}_{X}(U), {\cal D}_{Y}(V)$ 
subcoalgebras over $U \subset
X_0$ and $V \subset Y_0$, respectively. 
Let $\Phi : {\cal D}_{X}(U) \rightarrow {\cal D}_{Y}(V)$
be a morphism of graded coalgebras. $\Phi$ sends group-like elements into
group-like elements and irreducible components into irreducible ones.
It follows that $\Phi$ is uniquely determined by the map 
\begin{equation}
\label{map}
U\times S(X)  \ni u\otimes \alpha \longrightarrow
\Phi (u\otimes \alpha) \in V\times S(Y) \;\;\;,
\end{equation}
where the cartesian products $U\times S(X) , V\times S(Y) $ are
identified with the subsets of ${\cal D}_{X}(U), {\cal D}_{Y}(V)$ by the 
inclusions
\begin{eqnarray*}
U \times S(X) \ni (u,\alpha) & \longrightarrow & u\otimes \alpha =
u\cdot \alpha \in {\cal D}_{X}(U) \;\;\;, \\
U \times S(Y) \ni (v,\beta) & \longrightarrow & v\otimes \beta =
v\cdot \beta \in {\cal D}_{Y}(V) \;\;\;.
\end{eqnarray*}
Note that the map (\ref{map}) can be regarded as the family
$\left\{ {\Phi}_u \right\}_{u\in U}$ of $\ZZ$-graded
coalgebra morphisms
$$
{\Phi}_u : 
{{\cal D}_X}_u \ni u\cdot \alpha \longrightarrow {\Phi}(u\cdot
\alpha) \in {{\cal D}_X}_{\Phi(u)} \;\;\;.
$$
Since each irreducible component of ${\cal D}_{Y}(V)$ is isomorphic to  
the bigraded coalgebra $S(Y)$ the universal property of the
symmetric Hopf algebra can be used for a more detailed description
of $\Phi$. In order to study various properties of graded coalgebra 
morphisms it is covenient  to introduce the following definition.

\begin{DD}
Let $X, Y$ be $\ZZ$-graded spaces and ${\cal D}_{X}(U), {\cal D}_{Y}(V)$ 
subcoalgebras over $U \subset
X_0$ and $V \subset Y_0$, respectively. 
Denote by $\pi : V \times S(Y) \rightarrow
S(Y)$, resp. $\pi_Y : S(Y) \rightarrow Y$ the canonical projections on
the second factor, resp. on $S^1(Y) = Y$.
Let $\Phi : {\cal D}_{X}(U) \rightarrow {\cal D}_{Y}(V)$ be a
morphism of $\ZZ$-graded coalgberas. The maps
\begin{eqnarray*}
\Phi^0 
\!\!\!&: &\!\!\!
U \ni u  \longrightarrow  \Phi(u) \in V\;\;\;,\\
\Phi^+ 
\!\!\!\!&: &\!\!\!
 U \times S(X)^+ \ni (u,\beta)  \longrightarrow
 \pi_Y \circ \pi(\Phi(u \cdot \beta))
\in Y\;\;\;,\\
\Phi^{\wedge} 
\!\!\!\!&: &\!\!\!
 U \times
\makebox(12,9)[t]{\footnotesize $\bigwedge$}(X_1)^+ 
\ni (u,\gamma)  \longrightarrow
\pi_Y \circ \pi(\Phi(u \cdot \gamma)) \in Y \;\;\;,
\end{eqnarray*}
are called the underlying, the infinitesimal, and the exterior parts of
$\Phi$, respectively.

For every $k \geq 1$, the restrictions
\begin{eqnarray*}
\Phi^+_k : U \times 
\underbrace{X \times \ldots \times X}_{k}
 \ni (u,a_1,\ldots ,a_k) & \longrightarrow &
 \pi_Y \circ \pi(\Phi(u \cdot a_1 \cdot \ldots \cdot a_k))
\in Y\;\;\;,\\
\Phi^{\wedge}_k : U \times \underbrace{X_1 \times \ldots \times X_1}_{k}
 \ni (u,\xi_1,\ldots , \xi_k) & \longrightarrow &
\pi_Y \circ \pi(\Phi(u \cdot \xi_1 \cdot \ldots \cdot \xi_k)) \in Y
\end{eqnarray*}
are called the $k$-th infinitesimal and  the $k$-th exterior components 
of $\Phi$,
respectively.
\end{DD}

\noindent
For all $k>1$ the infinitesimal and the exterior components of a 
$\ZZ$-graded coalgebra morphism are totally symmetric and even.
By the universal property of the symmetric tensor product
 the infinitesimal $\Phi^+_k$ and  the exterior $\Phi^{\wedge}_k$ 
components uniquely extend to
maps on $U\times  S^k(X)$ and $U\times 
\makebox(12,9)[t]{\footnotesize $\bigwedge$}^k(X)$, respectively,
which are linear and even in the second variable.
For the sake of simplicity  the same symbols will be used
for the components and for their extensions above. Note that
the infinitesimal $\Phi^+$, and the exterior $\Phi^{\wedge}$ parts 
of $\Phi$ are uniquely determined by their components,
$\{ \Phi^+_k \}_{k\geq 1}$, and 
$\{ \Phi^{\wedge}_k \}_{k\geq 1}$, respectively. 
\bigskip 

\noindent
The following proposition is a  consequence of the universal property
of the symmetric algebra with respect to the coalgebra morphisms 
(Th.2.1.2).
It asserts
that a morphism $\Phi : {\cal D}_{X}(U) \rightarrow {\cal D}_{Y}(V)$ of 
$\ZZ$-graded coalgebras
is uniquely determined by its underlying and infinitesimal parts.


\begin{PP}
Let $X, Y$ be $\ZZ$-graded spaces and ${\cal D}_{X}(U), {\cal D}_{Y}(V)$ 
subcoalgebras  over $U \subset
X_0$ and $V \subset Y_0$, respectively. Let 
$\phi^+ : U \times S(X)^+ \rightarrow Y$ be a morphism of $\ZZ$-graded 
spaces in the second variable, and
$\phi : U \rightarrow V$  an
arbitrary map. 

Then there exists a unique morphism
$\Phi : {\cal D}_{X}(U) \rightarrow {\cal D}_{Y}(V)$ of 
$\ZZ$-graded coalgebras 
such that
$
\Phi^0 = \phi$, and $\Phi^+ = \phi^+$.

Moreover, for every $u \in U, \alpha \in S(X)$
\begin{equation}
\Phi(u,\alpha) = \phi(u) \cdot \ast\exp \phi^+_u(\alpha)\;\;\;,
\label{exp}
\end{equation}
where
\begin{eqnarray*}
\phi^+_u &\equiv & \phi^+(u,.) : S(X)^+ \longrightarrow Y\;\;\;,\\
\ast\exp \phi^+_u(\alpha) & \equiv & \sum_{k \geq 0} \frac{1}{k!}
\phi^{+k}_u(\alpha) \;\;,\\
\phi^{+0}_u(\alpha) & \equiv & u_{S(Y)} \circ \varepsilon_{S(X)}(\alpha) \;\;, \\
\phi^{+k}_u(\alpha) & \equiv &
\underbrace{\phi^+_u\circ \pi^+  \ast \ldots
\ast \phi^+_u \circ \pi^+}_{k}(\alpha)
\;\;\;,\;\;\; k \geq 1 \;\;.
\end{eqnarray*}
$\pi^+ : S(X)
\rightarrow S(X)^+$ denotes the canonical projection,
and $\ast$ is the convolution in ${\rm Hom}({\cal D}_{Xu},S(X))$.
\end{PP}

\noindent
It follows  that a $\ZZ$-graded coalgebra morphism 
$\Phi : {\cal D}_{X}(U) \rightarrow {\cal D}_{Y}(V)$ is uniquely determined by
 the underlying map $\Phi^0$ and the infinitesimal components
$\{ \Phi^+_k \}_{k\geq 1}$. In the following definition we
introduce the notion of smooth coalgebra morphism by imposing
some additional requirements on the maps 
$\Phi^0,\{ \Phi^+_k \}_{k\geq 1}$.

\begin{DD}
Let $X, Y$ be $\ZZ$-graded Fr\'{e}chet spaces and ${\cal D}_{X}(U),
{\cal D}_{Y}(V)$  open subcoalgebras of ${\cal D}_X$ and ${\cal D}_Y$, 
respectively. 

A morphism $\Phi : {\cal D}_{X}(U) \rightarrow {\cal D}_{Y}(V)$ of 
$\ZZ$-graded coalgebras is said
to be smooth if the following conditions are satisfied.
\begin{enumerate}
\item The underlying part $\Phi^0 : U \rightarrow V$ is a continuous map and
for all $x \in X_0, u \in U$, the directional derivative $D^1\Phi^0(u;x)$
exists.
\item
For every $k \geq 1$, the $k$-th infinitesimal component
$$
\Phi^+_k : U \times \underbrace{X \times \ldots \times X}_{k} \longrightarrow Y
$$
is jointly continuous with respect to the cartesian product
topology on $U \times X^{\times k}$ and the directional derivatives 
$D^1\Phi^+_k(u,a_1,\ldots,a_k;x)$ with respect to the first variable
exist for all $x \in X_0, u \in U, a_i \in X$.
\item
For every  $u \in U, x \in X_0, a_i \in X$ the following
relations hold
\begin{eqnarray}
D^1\Phi^0(u;x) & = & \Phi^+_1(u,x)\;\;\;, \label{defu}\\
D^1\Phi^+_k(u,a_1,\ldots,a_k;x) & = & \Phi^+_k(u,a_1,\ldots,a_k,x)\;\;\;.
\label{defc}
\end{eqnarray}
\end{enumerate}
\end{DD}


\begin{PP}
Let $\Phi : {\cal D}_{X}(U) \rightarrow {\cal D}_{Y}(V)$ be a smooth morphism 
of $\ZZ$-graded
coalgebras and $\left\{ x_j \right\}_{j = 1}^l$ an arbitrary sequence
of elements of $X_0$.
\begin{enumerate}
\item
The underlying part $\Phi^0: U \rightarrow V$ of $\Phi$ is a smooth map and
for every $u \in U$ the $l$-th order partial derivatives satisfy
the relation
\begin{equation} \label{eq:pdphiu}
D^{l}\Phi^0(u;x_1 , \ldots , x_l) =\Phi^+_l(u,x_1,\ldots, x_l)
\;\;\;.
\end{equation}
\item
For every $k \geq 1$ the $k$-th  exterior component $\Phi^{\wedge}_k$ 
of $\Phi$ is a smooth map. For evry $u \in U, \xi_i \in X_1 $
 the $l$-th order partial
derivatives with respect to the first variable are given by the formula
\begin{equation} \label{eq:pdphiw}
D^l\Phi^{\wedge}(u,\xi_1,\ldots,\xi_k;x_1 , \ldots , x_l) =
\Phi^+_{k+l}(u,\xi_1,\ldots,\xi_k,x_1, \ldots, x_l)
\;\;\;.
\end{equation}
\item
For every $k \geq 1$ the $k$-th infinitesimal component $\Phi^+_k$ of 
$\Phi$ is a smooth map. For evry $u \in U$, $a_i \in X$ the $l$-th 
order partial derivatives satisfy the relation

\begin{equation} \label{pdphip}
D^l\Phi^+_k(u,a_1,\ldots,a_k;x_1 , \ldots , x_l) =
\Phi^+_{k+l}(u,a_1,\ldots,a_k,x_1, \ldots, x_l) 
\;\;\;.
\end{equation}
\end{enumerate}
\end{PP}

\noindent {\bf Proof.}
{\it 1.} By condition 1) of Def.3.2.2 the directional derivative
$D^1\Phi^0(u;x)$ exists for all $u \in U$ and $ x \in X_0$.
By condition 3) one gets the relation (\ref{eq:pdphiu}) for
$l=1$. Since by 2) the first component of $\Phi^+$ is jointly continuous
on $U \times X$ so is $D^1\Phi^0$ on $U \times X_0$, hence
$\Phi^0$ is $C^1$.  The $C^l$ smoothness and the relation (\ref{eq:pdphiu})
for arbitrary $l$ follow from induction on $l$.

{\it 2.} Let us fix $k\geq 1$.
Repeating the reasoning above one gets that $\Phi^{\wedge}_k$ is $C^1$
separately in the first variable and the relation (\ref{eq:pdphiw})
holds for $l=1$. But $\Phi^{\wedge}_k$ is linear in the second variable and
by the condition 2) of Def.3.2.2, jointly continuous. It follows
\cite{Hamilton} that $\Phi^{\wedge}_k$ is  jointly $C^1$.
The induction
on $l$ yields the $C^l$-smoothness and the relation (\ref{eq:pdphiw}) for
arbitrary $x_i \in X_0$ and for all $l$.

{\it 3.} It is a straightforward consequence of {\it 1.} and 
{\it 2.}. \hfill $\;\;\;\Box$
\bigskip

\noindent
Let ${\cal D}_{X}(U),{\cal D}_{Y}(V)$ be open subcoalgebras. For $k\geq 1$ and
$X_1\neq \{o\}$
we introduce the space $C^{\wedge}_k({\cal D}_{X}(U),{\cal D}_{Y}(V))$ 
of all smooth maps
$$
\phi_k :  U\times \underbrace{X_1 \times \ldots \times X_1 }_{k}
\longrightarrow Y\;\;\;,
$$
which are $k$-linear, totally symmetric, and even in the second variable.
Note that in the definition above $X_1$ is regarded as a purely
odd graded Fr\'{e}chet space $\{o\} \oplus X_1$. Thus the maps $\phi_k$ are
totally antisymmetric in the usual sense.
Let  $C^{\wedge}_0({\cal D}_{X}(U),{\cal D}_{Y}(V)) \equiv
C^{\infty}(U,V)$ be the space of all smooth maps from the
open subset $U \subset X_0$ to $V \subset Y_0$ and
$$
C^{\wedge}({\cal D}_{X}(U),{\cal D}_{Y}(V)) \equiv 
\parbox{25pt}{  { \scriptsize $\makebox[6pt]{} \infty $} \\
                      $ \makebox[6pt]{} \times $      \\
                     {\scriptsize $ k=0$}
                                            }
C^{\wedge}_k({\cal D}_{X}(U),{\cal D}_{Y}(V))\;\;\;.
$$
For $X_1 = \{o\}$ we put
$
C^{\wedge}({\cal D}_{X}(U),{\cal D}_{Y}(V)) \equiv C^{\infty}(U,V)
$.

\noindent
As a consequence of Propositions 3.2.1 and  3.2.2 one gets the following 
characterization
of the space ${\rm Mor}({\cal D}_{X}(U),{\cal D}_{Y}(V))$ of
all smooth morphisms $\Phi : {\cal D}_{X}(U) \rightarrow {\cal D}_{Y}(V)$
of open subcoalgebras
\begin{TT}
The map
$$
{\rm Mor}({\cal D}_{X}(U),{\cal D}_{Y}(V)) \ni \Phi \longrightarrow 
(\Phi^0,\Phi^{\wedge}_1,\Phi^{\wedge}_2,\ldots)
\in C^{\wedge}({\cal D}_{X}(U),{\cal D}_{Y}(V))
$$
is bijective.
\end{TT}

\noindent 
In order to complete the construction of the model category we shall show 
that the composition of smooth coalgebra morphisms is smooth. For this 
purpose we start with the description of the compositon of 
$\ZZ$-graded coalgebra morphisms in terms of components.

\begin{PP}
Let $\Phi : {\cal D}_{X}(U) \rightarrow {\cal D}_{Y}(V)$, 
$\Psi : {\cal D}_{Y}(V) \rightarrow {\cal D}_{Z}(W)$ be morphisms of 
$\ZZ$-graded coalgebras. The underlying part $(\Psi \circ \Phi)^0$ and
the $k$-th infintesimal components $(\Psi \circ \Phi)^+_k$ of the 
$\ZZ$-graded coalgebra morphism 
$\Psi \circ \Phi : {\cal D}_{X}(U) \rightarrow {\cal D}_{Z}(W)$
are given by
$$
(\Psi \circ \Phi)^0 = \Psi^0 \circ \Phi^0 \;\;\;,
$$
and
\begin{eqnarray}
(\Psi \circ
 \Phi)^+_k(u,a_1,\ldots,a_k) &= &
\sum\limits_{i=1}^k {1\over i !}
\sum\limits_{
               \parbox[t]{50pt}{ \scriptsize $
                                     \{ P_1,\ldots,P_i\} \\
                                    \makebox[10pt]{}| P_i| > 0 
                                   $  }}
{\sigma({\cal X},{\cal P})}
\label{rep} \\
& & \times \;
\Psi^+_i(\Phi^0(u),\Phi^+_{|P_1|}(u,a_{P_1}),\ldots,
\Phi^+_{|P_i|}(u,a_{P_i}))
\;,
\nonumber
\end{eqnarray}
where the sum is over all nonempty partitions 
${\cal P} =\{ P_1,\ldots,P_i\}$  of the index set $\{ 1,\ldots,k\}$ and 
 the notation of {\rm Rem.2.1.2.} is used.
\end{PP}

\noindent {\bf Proof.}
According to Def.3.2.1 the $k$-th infinitesimal compnent of 
$\Psi \circ \Phi$ is given by
$$
(\Psi \circ \Phi)^+_k(u,a_1,\ldots,a_k) =
 \pi_Y \circ \pi\circ \Psi \circ \Phi(u \cdot a_1 \cdot \ldots \cdot a_k)\;\;\;.
$$
Representing $\Phi$ in the formula above in terms of its infinitesimal
components (Prop.3.2.1) and using explicit form of comultiplication in 
$S(X)$ given in Prop.2.1.3 one gets the result required. \hfill $\;\;\;\Box$
\bigskip

\begin{TT}
Let $X, Y, Z$ be $\ZZ$-graded Fr\'{e}chet spaces and
${\cal D}_{X}(U), {\cal D}_{Y}(V), {\cal D}_{Z}(W)$
open subcoalgebras of ${\cal D}_X, {\cal D}_Y$ and ${\cal D}_Z$, 
respectively.

If $\Phi : {\cal D}_{X}(U) \rightarrow {\cal D}_{Y}(V)$, 
$\Psi : {\cal D}_{Y}(V) \rightarrow {\cal D}_{Z}(W)$
are smooth morphisms of $\ZZ$-graded coalgebras so is their composition
$\Psi \circ \Phi : {\cal D}_{X}(U) \rightarrow {\cal D}_{Z}(W)$.

The underlying part of $\Psi\circ\Phi$ is the composition of underlying 
parts $(\Psi\circ\Phi)^0 = \Psi^0\circ\Phi^0$.

The $k$-th exterior component of $\Psi \circ \Phi$ is given by 
\begin{eqnarray}
(\Psi 
\circ
\Phi)^{\wedge}_k(u,\xi_1,\ldots,\xi_k) & = &   
\sum\limits_{i=1}^k {1\over i !}
\sum\limits_{
               \parbox[t]{50pt}{ \scriptsize $
                                     \{ P_1,\ldots,P_i\} \\
                                    \makebox[10pt]{}| P_i| > 0 
                                   $  }}
{\sigma({\cal X},{\cal P})}
\label{irep} \\
& & \makebox[40pt]{}\times \;
\Psi^+_i(\Phi^0(u),\Phi^{\wedge}_{|P_1|}
(u,\xi_{P_1}),\ldots,\Phi^{\wedge}_{|P_i|}(u,\xi_{P_i}))
\;, \nonumber
\end{eqnarray}
where the sum is over all nonempty  partitions 
$\{ P_1,\ldots,P_i\}$ of the index set $\{ 1,\ldots,k\}$.
\end{TT}

\noindent {\bf Proof.}
By Prop.3.2.2 $\Psi^0, \Phi^0$ are smooth maps so is their composition 
$(\Psi \circ \Phi)^0 = \Psi^0 \circ \Phi^0$.
By Prop.3.2.3 the $k$-th infinitesimal component
$(\Psi \circ \Phi)^+_k$ can be expressed as a finite sum of compositions
$
\Psi^+_i \circ ( \Phi^+_{|P_1|} \times \ldots \times \Phi^+_{|P_i|} ) 
\circ \widetilde{{\cal P}}$,
where
$$
\widetilde{{\cal P}} : U\times X^{\times k} \ni (u,a_1,\ldots,a_k)
\longrightarrow (u,a_{|P_1|})\times \ldots\times (u,a_{|P_i|}) \in
\parbox{25pt}{  { \scriptsize $\makebox(8,2){} i$} \\
                      $ \makebox(6,2){} \times $      \\
                     {\scriptsize $ j=1$}
                                            }
( U \times X^{\times |P_j|})\;\;\;.
$$
Since $\widetilde{{\cal P}}$ is smooth with respect to the cartesian 
product topology
the smoothness of $(\Psi \circ \Phi)^+_k$ follows from the smoothness of
$\Psi^+_i$ and $\Phi^+_j$. 

It remains to check the condition 3) of Def.3.2.2. Using  relation 
(\ref{defu}) for
$\Psi^0$ and $\Phi^0$ and  (\ref{rep}) for $k=1$
one gets
\begin{eqnarray*}
D^1(\Psi \circ \Phi)^0(u;x) & =&  D^1\Psi^0 
( \Phi^0(u); D^1\Phi^0 (u;x))\\
& = & \Psi^+_1(\Phi^0(u),\Phi^+_1(u,x)) = 
(\Psi \circ \Phi)^+_1(u,x)\;\;\;,
\end{eqnarray*}
for all $u \in U, x \in X_0$. Hence the  relation (\ref{defu}) of
Def.3.2.2 is satisfied.

Differentiating  expression (\ref{rep}) for the $k$-th infinitesimal
component $(\Psi \circ \Phi)^+_k$ and using  relations
(\ref{defu},\ref{defc}) for  components of
$\Psi$ and $\Phi$ one has
\begin{eqnarray*}
D^1(\Psi\!\!\!\!\! &\circ &\!\!\!\!\!\Phi)^+_k(u,a_1,\ldots,a_k;x)\;= \\
&=&
\sum\limits_{i=1}^k {1\over i !}
\sum\limits_{
               \parbox[t]{50pt}{ \scriptsize $
                                     \{ P_1,\ldots,P_i\} \\
                                     \makebox[10pt]{}| P_i| > 0 
                                   $  }}
{\sigma({\cal X},{\cal P})} \\
& &\times \; \left[ 
D^1\Psi^+_i
(\Phi^0(u),\Phi^+_{|P_1|}(u,a_{P_1}),\ldots,\Phi^+_{|P_i|}(u,a_{P_i});
D^1\Phi^0(u;x)) \right.\\
& & \;\;\;+\;   \sum\limits_{j=1}^i \left.
\Psi^+_i(\Phi^0(u),\Phi^+_{|P_1|}(u,a_{P_1}),\ldots,
D^1\Phi^+_{|P_j|}(u,a_{P_j};x),
\ldots, 
\Phi^+_{|P_i|}(u,a_{P_i})) \right] \\
&=&
\sum\limits_{i=2}^{k+1} {1\over( i -1)!}
\sum\limits_{
               \parbox[t]{55pt}{ \scriptsize $
                                    \{ P_1,\ldots,P_{i-1}\} \\
                                    \makebox[10pt]{}| P_i| > 0 
                                   $  }}
{\sigma({\cal X},{\cal P})} \\
& &\times\;
\Psi^+_i(\Phi^0(u),\Phi^+_{|P_1|}(u,a_{P_1}),\ldots,
\Phi^+_{|P_i|}(u,a_{P_i}),
\Phi^+_1(u,x)) \\
&+&
\sum\limits_{i=1}^k {1\over i !}
\sum\limits_{
               \parbox[t]{50pt}{ \scriptsize $
                                     \{ P_1,\ldots,P_i\} \\
                                    \makebox[10pt]{}| P_i| > 0 
                                   $  }}
{\sigma({\cal X},{\cal P})} \\
& &\times \;
\sum\limits_{j=1}^i
\Psi^+_i(\Phi^0(u),\Phi^+_{|P_1|}(u,a_{P_1}),\ldots,
\Phi^+_{|P_j|+1}(u,a_{P_j},x),
\ldots, 
\Phi^+_{|P_i|}(u,a_{P_i})) \\
&=& 
\sum\limits_{i=1}^{k +1}{1\over i !}
\sum\limits_{
               \parbox[t]{50pt}{ \scriptsize $
                                     \{ Q_1,\ldots,Q_i\} \\
                                    \makebox[10pt]{}| Q_i| > 0 
                                  $  }}
{\sigma({\cal X}',{\cal Q})} 
\Psi^+_i(\Phi^0(u),\Phi^+_{|Q_1|}(u,a_{Q_1}),\ldots,
\Phi^+_{|Q_i|}(u,a_{Q_i}))\\
&=& (\Psi \circ \Phi)^+_{k+1}(u,a_1,\ldots,a_k,x)\;\;\;,
\end{eqnarray*}
where $ {\cal P} =\{ P_1,\ldots,P_i\}$ and 
${\cal Q} = \{ Q_1,\ldots,Q_i\}$ are  partitions of
the index sets  $\{1,\ldots,k\}$ and $\{1,\ldots,k+1\}$ respectively, and 
${\cal X}' = \{ a_1,\ldots,a_{k+1}\}, a_{k+1} = x$.

Formula (\ref{irep}) is a special case of (\ref{rep}).\hfill $\;\;\;\Box$
\bigskip

\subsection{Direct product}

The considerations of last two subsections leads to the following 
definition of the model category.
\begin{DD}
The objects of the model category ${\bf sc}$ of smooth graded coalgebras 
are open subcoalgebras ${\cal D}_{X}(U)$ where $X$ runs over 
the category of graded Fr\'{e}chet spaces and $U$ over all open subsets
of the even part $X_0$ of a graded Fr\'{e}chet space $X$.

For any two objects ${\cal D}_{X}(U),{\cal D}_{Y}(V) \in {\rm O}{\bf sc}$
the space of morphisms ${\rm M}{\bf sc}({\cal D}_{X}(U),{\cal D}_{Y}(V))$ 
consists of all smooth coalgebra morphisms 
$\Phi : {\cal D}_{X}(U)\rightarrow {\cal D}_{Y}(V)$.

The composition of morphisms in ${\bf sc}$ is defined as a composition 
of coalgebra maps.

An isomorphism in the category ${\bf sc}$ is called a diffeomorphism of 
open subcoalgebras.
\end{DD}

\noindent
We shall show that in the model category defined above
 the direct product exists. We start with the corresponding result for
the category of $\ZZ$-graded cocommutative coalgebras \cite{sweedler}.

\begin{TT}
Let $({\cal C},\Delta_C,\varepsilon_C), ({\cal D},\Delta_D,\varepsilon_D)$ be
$\ZZ$-graded
cocommutative coalgebras. 

\noindent Define the maps
\begin{eqnarray*}
\pi_C : {\cal C}\otimes {\cal D} \ni c\otimes d \longrightarrow \varepsilon_D(d)\cdot c
\in {\cal C}\;\;\;,\\
\pi_D : {\cal C}\otimes {\cal D} \ni c\otimes d \longrightarrow \varepsilon_C(c)\cdot d
\in {\cal D}\;\;\;.
\end{eqnarray*}
Then $\pi_C, \pi_D$ are $\ZZ$-graded coalgebra morphisms and for every 
$\ZZ$-graded cocommutative coalgebra $({\cal E},\Delta_E,\varepsilon_D)$
and morphisms of $\ZZ$-graded coalgebras $\Phi_C :  {\cal E} \rightarrow {\cal C}$,
and $ \Phi_D :  {\cal E} \rightarrow {\cal D}$,
there exists a unique morphism of $\ZZ$-graded coalgebras
$\Phi :  {\cal E} \rightarrow {\cal C}\otimes {\cal D}$
making the diagram
\begin{equation}
\label{diag}
\begin{picture}(120,100)(0,10)
\put(40,93){\makebox(40,10){$
%
   {\cal C}\otimes {\cal D}
$}}
\put(5,75){\makebox(20,10)[r]{\small $
%
 \pi_C
$}}
\put(95,75){\makebox(20,10)[l]{$
%
 \pi_D
$}}
\put(-20,48){\makebox(20,10)[br]{\small$
%
 {\cal C}
$}}
\put(120,48){\makebox(20,10)[bl]{$
%
 {\cal D}
$}}
\put(5,18){\makebox(20,10)[tr]{\small$
%
 \Phi_C
$}}
\put(95,18){\makebox(20,10)[tl]{\small$
%
 \Phi_D
$}}
\put(65,50){\makebox(20,10)[l]{\small$
%
 \Phi
$}}
\put(50,2){\makebox(20,10)[t]{\small$
%
 {\cal E}
$}}
\put(50,90){\vector(-3,-2){48}}
\put(70,90){\vector(3,-2){48}}
\put(50,15){\vector(-3,2){48}}
\put(70,15){\vector(3,2){48}}
\put(60,81){\vector(0,1){8}}
\multiput(60,17)(0,8){8}{\line(0,1){4}}
\end{picture}
\end{equation}
commute. The $\ZZ$-graded coalgebra morphism $\Phi$ is given by
$
\Phi = (\Phi_C \otimes \Phi_D) \circ \Delta_E
$.
\end{TT}

\begin{DD}
Let ${\cal D}_{X}(U), {\cal D}_{Y}(V) \in {\rm O}{\bf sc}$. The tensor product
${\cal D}_{X}(U)\otimes {\cal D}_{Y}(V) \in {\rm O}{\bf sc}$ is defined
as follows:

\noindent 1) With respect to the coalgebra structure
${\cal D}_{X}(U)\otimes {\cal D}_{Y}(V) $ is the tensor product of 
graded cocommutative coalgebras.

\noindent 2) With respect to the topological structure 
${\cal D}_{X}(U)\otimes {\cal D}_{Y}(V) $ is identified with an
open subcoalgebra of ${\cal D}_{X \oplus Y}$ by the
canonical isomorphism
$$
{\cal D}_{X}(U)\otimes {\cal D}_{Y}(V) \longrightarrow
{\cal D}_{X \oplus Y}(U\times V)
$$
where the direct sum topology on $X\oplus Y = (X_0 \oplus Y_0)
\oplus (X_1 \oplus Y_1)$ is assumed.
\end{DD}

\noindent Calculating infinitesimal components and checking the
conditions of Def.3.2.2 by explicit calculation of directional 
derivatives one gets

\begin{PP}\parbox{10pt}{}

\begin{enumerate}
\item
Let $\Phi:{\cal D}_{X}(U)\rightarrow {\cal D}_{Y}(V) $, 
$\Phi':{\cal D}_{X'}(U')\rightarrow {\cal D}_{Y'}(V')$ be smooth morphisms
of open subcoalgebras. Then the tensor product of $\ZZ$-graded coalgebra
morphisms
$$
\Phi\otimes \Phi':{\cal D}_{X}(U)\otimes {\cal D}_{X'}(U')
\ni m\otimes m'\rightarrow 
\Phi(m)\otimes \Phi'(m') \in  {\cal D}_{Y}(V) \otimes {\cal D}_{Y'}(V') \;\;
$$
is smooth.
\item
Let $\varepsilon_U, \varepsilon_V$ be counits of open subcoalgebras 
${\cal D}_{X}(U)$, and  ${\cal D}_{Y}(V)$, respectively.
Then the maps
\begin{eqnarray*}
P_U & \equiv & I\otimes \varepsilon_V : {\cal D}_{X}(U)
\otimes {\cal D}_{Y}(V)
\longrightarrow {\cal D}_{X}(U) \;\;\;,\\
P_V & \equiv &  \varepsilon_U\otimes I : {\cal D}_{X}(U)
\otimes {\cal D}_{Y}(V)
\longrightarrow {\cal D}_{Y}(V)\;\;\;
\end{eqnarray*}
are smooth morphisms of $\ZZ$-graded coalgebras.
\item
 The comultiplication 
$\Delta: {\cal D}_{X}(U) \ra {\cal D}_{X}(U)\otimes {\cal D}_{X}(U)$
of an open subcoalgebra ${\cal D}_{X}(U)$ 
is a smooth morphism of $\ZZ$-graded coalgebras.
\end{enumerate}
\end{PP}

\noindent {\bf Remark 3.3.1}
The underlying parts and the exterior components of
smooth $\ZZ$-graded coalgebra morphisms of the 
proposition above are given by 
\begin{eqnarray*}
(\Phi\otimes \Phi')^0(u,u') &= &(\Phi^0(u),\Phi'^0(u')) \;\;\;,\\
(\Phi\otimes \Phi')^{\wedge}_k
((u,u'),\theta_1 \oplus \theta'_1,...,\theta_k\oplus \theta'_k) &=& 
\Phi^{\wedge}_k(u,\theta_1,...,\theta_k) \oplus
\Phi'^{\wedge}_k(u',\theta'_1,...,\theta'_k)\;\;\;;
\end{eqnarray*}
\begin{eqnarray*}
(P_U)^0(u,v) &= &u \;\;\;,\\
(P_U)^{\wedge}_1((u,v),\theta \oplus \eta) &=& \theta\;\;\;,\\
(P_U)^{\wedge}_k((u,v),\theta_1 \oplus \eta_1,\ldots,
\theta_k \oplus \eta_k) &= & 0\;\;\;,\; \;\;k\geq 2\;\;\;;
\end{eqnarray*}
\begin{eqnarray*}
\Delta^0(u) &=& (u,u)\;\;\;,\\
\Delta^{\wedge}_1(u,\theta) &=& \theta \oplus \theta\;\;\;,\\
\Delta^{\wedge}_k(u,\theta_1,\ldots,\theta_k) &=&0\;\;\;
\; \;\;\;,\;\;\;k\geq 2\;\;\;.
\end{eqnarray*}

\begin{TT}
$({\cal D}_{X}(U)\otimes {\cal D}_{Y}(V),P_U,P_V)$ is the direct product
in the model category of smooth $\ZZ$-graded coalgebras.
\end{TT}

\noindent {\bf Proof.}
By Prop.3.3.1 $P_U, P_V$ are ${\bf sc}$-morphisms. 
By Th.3.3.1 for any pair of smooth morphisms of  $\ZZ$-graded coalgebras
$\Phi_U  : {\cal D}_{Z}(W)\rightarrow {\cal D}_{X}(U)$, and 
$\Phi_V  : {\cal D}_{Z}(W)\rightarrow {\cal D}_{Y}(V)$,
the map
$
\Phi = \Phi_U\otimes \Phi_V \circ \Delta_W
$
is the unique  $\ZZ$-graded coalgebra morphism for which the diagram
\begin{center}
\begin{picture}(120,100)(0,5)
\put(40,93){\makebox(40,10){$
%
   {\cal D}_{X}(U)\otimes {\cal D}_{Y}(V)
$}}
\put(-5,75){\makebox(20,10)[r]{\small $
%
 P_U
$}}
\put(105,75){\makebox(20,10)[l]{$
%
 P_V
$}}
\put(-30,48){\makebox(20,10)[br]{\small$
%
 {\cal D}_{X}(U)
$}}
\put(130,48){\makebox(20,10)[bl]{$
%
 {\cal D}_{Y}(V)
$}}
\put(-5,18){\makebox(20,10)[tr]{\small$
%
 \Phi_U
$}}
\put(105,18){\makebox(20,10)[tl]{\small$
%
 \Phi_V
$}}
\put(65,50){\makebox(20,10)[l]{\small$
%
 \Phi
$}}
\put(50,2){\makebox(20,10)[t]{\small$
%
  {\cal D}_{Z}(W)
$}}
\put(40,90){\vector(-3,-2){48}}
\put(80,90){\vector(3,-2){48}}
\put(40,15){\vector(-3,2){48}}
\put(80,15){\vector(3,2){48}}
\put(60,81){\vector(0,1){8}}
\multiput(60,17)(0,8){8}{\line(0,1){4}}
\end{picture}
\end{center}
commutes. 
By Prop.3.3.1 $\Phi$ is a composition of smooth morphisms of  $\ZZ$-graded
coalgebras. Hence $\Phi$ is smooth by Th.3.2.2.\hfill$\;\;\;\Box$
\bigskip

\noindent {\bf Remark 3.3.2}
The underlying part and the exterior components
of the smooth morphism $\Phi = \Phi_U\otimes \Phi_V \circ \Delta_W$ are given by
\begin{eqnarray*}
\Phi^0(w) &= &(\Phi^0_U(w),\Phi^0_V(w)) \;\;\;,\\
\Phi^{\wedge}_k
(w,\theta_1 ,...,\theta_k) &=& 
(\Phi_U)^{\wedge}_k(w,\theta_1,...,\theta_k) \oplus
(\Phi_V)^{\wedge}_k(w,\theta_1,...,\theta_k)\;\;.
\end{eqnarray*}

\subsection{Subcategories  ${\bf sc }_0$ and ${\bf sc }^{<}$}

In this subsection we shall show that the model category ${\bf fm}$
of Fr\'{e}chet manifolds and the model category ${\bf sm}$ of
BLK supermanifolds are both full subcategories of the model
category ${\bf sc}$ of smooth coalgebras. This justifies
introduction of ${\bf sc}$ as an  extension of ${\bf fm}$ and
${\bf sm}$. \bigskip

\noindent
We define the category  ${\bf sc }_0$ of even open coalgebras as 
the full subcategory of ${\bf sc }$
consisting of all objects of the form ${\cal D}_{X_0 \oplus \{o\}}(U)$
and all ${\bf sc }$-morphisms between them. Similarly, the category
${\bf sc }^{<}$ of finite-dimensional open coalgebras is defined as
the full subcategory of  ${\bf sc }$
consisting of all objects of the form ${\cal D}_{\Rm \oplus\Rn}(U)$
with arbitrary $m,n \geq 0$, 
and all ${\bf sc }$-morphisms between them.
For notational convenience we shall introduce simplified symbols
${\cal D}(U) \equiv {\cal D}_{X_0 \oplus \{o\}}(U)$ and
${\cal D}_{m,n}(U) \equiv {\cal D}_{\Rm \oplus\Rn}(U)$ for objects of
${\bf sc }_0$ and ${\bf sc }^{<}$, respectively.
Note that by definition, for all ${\cal D}(U),{\cal D}(V) \in
{\rm O}{\bf sc}_0$ and ${\cal D}_{m,n}(U),{\cal D}_{m',n'}(V)
\in {\rm O}{\bf sc}^{<}$ one has 
\begin{eqnarray*}
{\rm M}{\bf sc}_0({\cal D}(U),{\cal D}(V)) &=&
{\rm M}{\bf sc}({\cal D}(U),{\cal D}(V))\;\;\;,\\
{\rm M}{\bf sc}^{<}({\cal D}_{m,n}(U),{\cal D}_{m',n'}(V))&=&
{\rm M}{\bf sc}({\cal D}_{m,n}(U),{\cal D}_{m',n'}(V))\;\;\;.
\end{eqnarray*}
By Th.3.3.2 both ${\bf sc}_0$ and ${\bf sc }^{<}$  inherit
 the direct product from ${\bf sc}$. \bigskip

\noindent
As a consequence of Th.3.2.1 and Th.3.2.2 one has
\begin{PP} 
The correspondence
\begin{eqnarray}
 {\rm O}{\bf sc}_0\ni {\cal D}(U) &\longrightarrow & 
U \in {\rm O}{\bf fm}\;\;\;,\label{funct1}\\
{\rm M}{\bf sc}_0({\cal D}(U),{\cal D}(V)) \ni \Phi &
\longrightarrow &  \Phi^0 
 \in 
{\rm M}{\bf fm}(U,V)\;\;\;\nonumber
\end{eqnarray}
is  an equivalence of categories.
\end{PP}
It follows that the model category ${\bf fm}$ of Fr\'{e}chet manifolds
 can be identified with
the full subcategory ${\bf sc}_0$ of the 
category ${\bf sc}$ of open coalgebras. \bigskip

\noindent{\bf Remark 3.4.1}
The inverse to the correspondence (\ref{funct1}) is given by
\begin{eqnarray*}
{\rm O}{\bf fm} \ni U &\longrightarrow & 
{\cal D}(U) \in  {\rm O}{\bf sc}_0 \;\;\;, \\
{\rm M}{\bf fm}(U,V) \ni \phi &
\longrightarrow &  \phi_* 
 \in {\rm M}{\bf sc}_0({\cal D}(U),{\cal D}(V))
\;\;\;
\end{eqnarray*}
where 
$\phi_* $ is a unique smooth morphism of coalgebras 
such that $(\phi_*)^0 = \phi$. 
Using formula (\ref{delta}) of Prop.2.1.3
for comultiplication in $S(X)$  and  formulae
 (\ref{exp}) of Prop.3.2.1 and 
(\ref{eq:pdphiu}) of Prop.3.2.2 
one gets for $u\in U, x_1,\ldots,x_k \in X$,
\begin{eqnarray}
\label{phistar}
\phi_*(u\!\!\!\!&\cdot & \!\!\!\!  x_1\cdot\ldots\cdot x_k) = \\
 &=&\sum\limits_{i=1}^k {1\over i!} \!
\sum\limits_{
               \parbox[t]{50pt}{ \scriptsize $
                                     \{ P_1,\ldots,P_i\} \\
                                    \makebox[10pt]{}| P_i| > 0 
                                   $  }}
\!\!\! \phi(u)\cdot D^{|P_1|}\phi(u;x_{P_1})\cdot\ldots\cdot
D^{|P_i|}\phi(u;x_{P_i}))
\;, \nonumber
\end{eqnarray}
where the sum runs over all  nonempty $i$-partitions of the index
set $\{1,\ldots,k\}$. \bigskip

\noindent  Let us now compare ${\cal D}(U)$ to the dual coalgebra
(see Appendix A.5 )
${\cal C}^{\infty}(U)^{\circ}$ of the algebra
${\cal C}^{\infty}(U)$ of
 smooth real-valued functions on an open subset
 $U$ of a Fr\'{e}chet space$X$.
For this purpose we introduce the pairing
$$
\langle \;.\;,\;.\; \rangle_U :
{\cal D}(U) \times {\cal C}^{\infty}(U) \longrightarrow \R
$$
defined, for all $u\in U, x_1,\ldots,x_k \in X$, and 
$f \in {\cal C}^{\infty}(U)$ by
\begin{eqnarray}
\label{pair}
\langle u , f \rangle_U & \equiv & f (u) \;\;\;,\\
\langle u\cdot x_1\cdot\ldots \cdot x_k, f\rangle_U & \equiv & 
D^k f(u;x_1,\ldots,x_k)\;\;\;. \nonumber
\end{eqnarray}

\begin{PP}
Let $U$ be an open subset of a Fr\'{e}chet space $X$,
$({\cal D}(U),\Delta_U, \varepsilon_U )$ the even open
coalgebra, and
${\cal C}^{\infty}(U)$ the algebra of smooth functions on $U$.
\begin{enumerate}
\item
For all $f,g \in {\cal C}^{\infty}(U)$, $\omega \in {\cal D}(U)$, 
\begin{eqnarray*}
\langle \omega,  f\cdot g\rangle_U &=& 
\sum\limits_{(\omega)} \langle \omega_{(1)},  f\rangle_U
\langle \omega_{(2)},  g\rangle_U
\;\;\;,\\
\langle \omega,1 \rangle_U & = & \varepsilon_U(\omega )\;\;\;,
\end{eqnarray*}
where $\Delta_U\omega = \sum\limits_{(\omega)} \omega_{(1)}\otimes 
\omega_{(2)}$.
\item
Let $\phi :U\rightarrow V$ be a smooth map of open subsets of Fr\'{e}chet
spaces. Then
$$
\langle \phi_*\omega, f\rangle_V = \langle \omega, f\circ 
\phi\rangle_U\;\;\;,
$$
for all $\omega \in {\cal D}(U), f \in {\cal C}^{\infty}(V)$.
\end{enumerate}
\end{PP}

\noindent 
The first part follows from the mutiple Leibnitz rule (Prop.2.2.2) 
and the explicit 
formula for the comultiplication in $S(X)$ (Prop.2.1.3). 
The second part is a  straightforward
consequence of the multiple chain rule (Prop.2.2.1)
and formula (\ref{phistar})  for $\phi_*$.

\begin{PP} With the notation of {\rm Prop.3.4.2} the
map 
\begin{equation}
\label{sub}
{\cal D}(U) \ni \a \lra \langle \a, \;.\; \rangle \in 
{\cal C}^{\infty}(U)^{\circ}
\end{equation}
is an injective morphism of $\ZZ$-graded coalgebras.
\end{PP}

\noindent {\bf Proof.} 
By Prop.3.4.2 and Prop.2.5.3 for all $\omega \in {\cal D}(U)$,
$\langle \omega, \;.\; \rangle_U \in {\cal C}^{\infty}(U)^{\circ}$.
By Definition 2.5.1 of dual coalgebra and Prop.3.4.2 (\ref{sub})
is a morphism of $\ZZ$-graded coalgebras. 
Using the Hahn-Banach theorem for Fr\'{e}chet spaces one can show that
the  pairing (\ref{pair}) is nonsingular which implies the  injectivity of
the map (\ref{sub}). 
\hfill  $\;\;\;\Box$ \bigskip

\noindent{\bf Remark 3.4.2}
The image of the map (\ref{sub}) is a subcoalgebra of 
 ${\cal C}^{\infty}(U)^{\circ}$ consisting of all finite linear
combinations of evaluations of directional derivatives of
arbitrary order. This subcoalgebra will be called
the {\it coalgebra of Dirac distributions on $U$}.
For infinite-dimensional Fr\'{e}chet spaces this is a proper subcoalgebra
of ${\cal C}^{\infty}(U)^{\circ}$.\bigskip

\noindent
We shall proceed to the model category of BLK supermanifolds.
Let
$F =(F^0,F_.) :
{\cal S}^{m,n}_U \rightarrow {\cal S}^{m',n'}_V$ be a
${\bf sm}$-morphism and $\{ F^{\wedge}_k\}_{k=1}^{n}$ its
exterior components. The collection 
$\{ F^0, F^{\wedge}_1,..., F^{\wedge}_n \}$ can be regarded
as a point in the space 
${\cal C}^{\wedge}({\cal D}_{m,n}(U),{\cal D}_{m',n'}(V))$.
By  Th.3.2.1 there exists a unique smooth $\ZZ$-graded 
 coalgebra map  with the underlying
part $F^0$ and with the exterior components 
$\{ F^{\wedge}_k\}_{k=1}^{n}$. We denote this map
by $F_*$. 

Let $G =(G^0,G_.) :
 {\cal S}^{m',n'}_V\rightarrow {\cal S}^{m'',n''}_W$ be another
${\bf sm}$-morphism. Then by the above construction one has
a smooth $\ZZ$-graded  coalgebra map 
$G_*: {\cal D}_{m',n'}(V)\ra {\cal D}_{m'',n''}(W)$.
Comparing formula (\ref{comp}) for the exterior components
of composition of ${\bf sm}$-morphisms (Rem.2.3.2) with the corresponding
one  (\ref{irep}) for the composition of
smooth  $\ZZ$-graded  coalgebra morphisms (Th.3.2.2) one gets
$
(G \circ F)_* = G_* \circ F_*$. 
\begin{PP}
The correspondence
\begin{eqnarray}
{\rm O}{\bf sm} \ni {\cal S}^{m,n}_U &\longrightarrow & 
{\cal D}_{m,n}(U) \in  {\rm O}{\bf sc}^{<} \;\;\;, \label{funct3}\\
{\rm M}{\bf sm}({\cal S}^{m,n}_U,{\cal S}^{m',n'}_V) 
\ni F = (F^0,F_.) &
\longrightarrow &  F_* 
 \in {\rm M}{\bf sc}^{<}({\cal D}_{m,n}(U),{\cal D}_{m',n'}(V))
\;\;\;\nonumber
\end{eqnarray}
is an equivalence of categories.
\end{PP}

\noindent
Let ${\cal S}^{m,n}(U)$ be the $\ZZ$-graded algebra of superfunctions
of a superdomain ${\cal S}^{m,n}_U$, and
$$
\langle \;.\;,\;.\; \rangle_U :
{\cal D}_{m,n}(U) \times {\cal S}^{m,n}(U) \longrightarrow \R
$$
a pairing defined by
\begin{eqnarray}
\label{spair}
\langle u , f \rangle_U & \equiv & f^0(u) \;\;\;,\\
\langle u\cdot a_1\cdot\ldots \cdot a_k, f\rangle_U & \equiv & 
\widetilde{D}^k f(u;a_1,\ldots,a_k)\;\;\;. \nonumber
\end{eqnarray}
 for all $u\in U, a_1,\ldots,a_k \in \Rm\!\oplus \!\Rn$, and 
$f \in {\cal S}^{m,n}(U)$.

As a consequence of the multiple Leibnitz rule (Prop.2.3.3) and
the multiple chain rule (Prop.2.3.5) for superfunctions one has
the following ${\bf sm}$-counterpart of Prop.3.4.2.
\begin{PP}
Let ${\cal S}^{m,n}(U)$ be the $\ZZ$-graded algebra of superfunctions
of a superdomain ${\cal S}^{m,n}_U$, and 
$({\cal D}_{m,n}(U),\Delta_U, \varepsilon_U )$ the open
coalgebra ${\cal D}_{m,n}(U) = {\cal H}_U(\Rm\!\oplus \!\Rn)$. 
\begin{enumerate}
\item
For all $f,g \in {\cal S}^{m,n}(U)_0 \cup
{\cal S}^{m,n}(U)_1$, $\omega \in {\cal D}_{m,n}(U)$, 
\begin{eqnarray*}
\langle \omega,  f\cdot g\rangle_U &=& 
\sum\limits_{(\omega)} (-1)^{|f||\omega_{(2)}|}
\langle \omega_{(1)},  f\rangle_U
\langle \omega_{(2)},  g\rangle_U
\;\;\;,\\
\langle \omega,1 \rangle_U & = & \varepsilon_U(\omega )\;\;\;,
\end{eqnarray*}
where $\Delta_U\omega = \sum\limits_{(\omega)} \omega_{(1)}\otimes \omega_{(2)}$.
\item
Let  $F =(F^0,F_.) : {\cal S}^{m,n}_U \rightarrow {\cal S}^{m',n'}_V$ 
be a ${\bf sm}$-morphism. Then
$$
\langle F_*\omega, f\rangle_V = 
\langle \omega, F_Vf\rangle_U\;\;\;,
$$
for all $\omega \in {\cal D}_{m,n}(U) , f \in {\cal S}^{m',n'}(V)  $.
\end{enumerate}
\end{PP}
\begin{PP}
With the notation of {\rm Prop.3.4.5} the map
\begin{equation}
\label{idd}
{\cal D}_{m,n}(U) \ni \omega \lra \langle \omega, \;. \;\rangle \in 
{\cal S}^{m,n}(U)^{\circ}
\end{equation}
is an isomorphism of $\ZZ$-graded coalgebras.
\end{PP}

\noindent{\bf Proof.} Since the pairing (\ref{spair}) is nonsingular
the map (\ref{idd}) is injective. By Prop.3.4.5 it is a morphism of
$\ZZ$-graded coalgebras. In particular it preserves the coradical
filtration 
$$
{\cal D}_{m,n}(U)_u = 
 \parbox[c]{35pt}{   \scriptsize $
                                    \makebox(7,12)[b]{} \bigcup \\
                                     \makebox(2,2){}k\geq 0 $  }
\!\!\!\!{\cal D}_{m,n}(U)^{(k)}_u
$$
 of the irreducible components ${\cal D}_{m,n}(U)_u$,
$u\in U$. Hence, for all $u\in U$, $k\geq 0$
one has  the injective  maps
\begin{equation}
\label{small}
{\cal D}_{m,n}(U)^{(k)}_u
 \ni \a \lra \langle \a, \;. \;\rangle \in 
{{\cal S}^{m,n}(U)^{\circ}}^{(k)}_u\;\;\;.
\end{equation}
By Prop.A.5.6 $ {{\cal S}^{m,n}(U)^{\circ}}^{(k)}_u=
({\cal S}^{m,n}(U) / I^{k+1}_u)'$ and therefore
$$
{\rm dim}({\cal D}_{m,n}(U)^{(k)}_u) \;=\; 
{\rm dim}({{\cal S}^{m,n}(U)^{\circ}}^{(k)}_u) \;<\; +\infty\;\;\;.
$$
It follows that the maps (\ref{small}) are surjective for all 
$k\geq 0$, and $u\in U$, and so is the map (\ref{idd}).
\hfill $\Box$ \bigskip

\noindent{\bf Remark 3.4.3} 
By Prop.3.4.6 the open coalgebra ${\cal D}_{m,n}(U)$ can be identified
with the dual coalgebra of the algebra of superfunctions
${\cal S}^{m,n}(U)$.
The  {\it coalgebra of Dirac distributions on the  superdomain } 
${\cal S}^{m,n}_U$ is defined as
coalgebra of all finite linear combinations of evaluations of 
differentiations  of superfunctions. This coalgebra
coincides with the 
dual coalgebra ${\cal S}^{m,n}(U)^{\circ}$ \cite{Kostant}.

\subsection{Superfunctions}

Prop.3.4.1 and Prop.3.4.3
of the previous subsection provide characterization of even open
coalgebras as objects dual to algebras of smooth functions on 
open subsets of Fr\'{e}chet spaces. Similarly by Prop.3.4.4 and Prop.3.4.6
finite-dimensional open coalgebras are dual coalgebras to
alegbras of superfunctions on superdomains. Using this duality
smooth functions in both cases  can be regarded as linear functionals
on corresponding coalgebras. This leads to the following definition
of superfunction in general case.
\begin{DD}
Let ${\cal D}_{X}(U)$ be an open coalgebra. A linear functional 
$f \in {\cal D}_{X}(U)'$ is called a superfunction on 
${\cal D}_{X}(U)$ if the following conditions are satisfied.
\begin{enumerate}
\item For each $k\geq 0$ the function
$$
f_k : U\times \underbrace{X\times...\times X}_{k} \ni
(u,a_1,...,a_k)
 \longrightarrow 
\langle f ,  u \cdot a_1 \cdot ... \cdot a_k\rangle \in \R 
$$
is jointly continuous with respect to the Cartesian product topology
on $U\times X^{\times k}$.
\item For each $k\geq 0$ and for all $x\in X_0, u\in U$, and
$a_1,...,a_k\in X$ 
the partial derivative 
$$
D_x  \langle f , u \cdot a_1 \cdot ... \cdot a_k \rangle =
\lim_{\epsilon \rightarrow 0}
{  \langle f , ( u +\epsilon x) \cdot a_1 \cdot ... \cdot a_k\rangle- 
 \langle f ,u \cdot a_1 \cdot ... \cdot a_k \rangle \over \epsilon}
$$
exists.
\item For each $k\geq 0$ and for all $x\in X_0, u\in U$, and
$a_1,...,a_k\in X$ 
$$
D_x  \langle f , u \cdot a_1 \cdot ... \cdot a_k\rangle =
 \langle f , u \cdot a_1 \cdot ... \cdot a_k \cdot x  \rangle \;\;\;.
$$
\end{enumerate}
\end{DD}

\noindent
{\bf Remark 3.5.1}
Let ${\cal D}_{X}(U)'$ be the full algebraic dual of
an open coalgebra ${\cal D}_{X}(U)$. Let us consider 
 the inclusion of $\ZZ$-graded spaces 
$$
i : {\cal D}_{X}(U)'\otimes {\cal D}_{X}(U)' \lra
({\cal D}_{X}(U) \otimes {\cal D}_{X}(U))'\;\;\;,
$$
defined for each $f,g \in {\cal D}_{X}(U)'_0 \cup {\cal D}_{X}(U)'_1$, and
$\alpha, \beta \in {\cal D}_{X}(U)_0 \cup {\cal D}_{X}(U)_1 $ by
$$
\langle i(f \otimes g ), \a \otimes \b\rangle = 
(-1)^{|g||\a|} \langle f,\a\rangle \langle g,\b\rangle\;\;\;.
$$
${\cal D}_{X}(U)'$ has the structure of 
$\ZZ$-graded commutative algebra with multiplication
$$
M: {\cal D}_{X}(U)'\otimes {\cal D}_{X}(U)' 
\stackrel{i}{\lra}
({\cal D}_{X}(U) \otimes {\cal D}_{X}(U))' \stackrel{\Delta'}{\lra}
{\cal D}_{X}(U)'\;\;\;,
$$
and  unit
$
u: \R \stackrel{\ve'}{\lra} {\cal D}_{X}(U)'$.

The explicit formula for the product $f \cdot g$ reads
\begin{equation}
\label{product}
\langle f\cdot g,u \cdot a_1 \cdot ... \cdot a_k\rangle 
 =
\sum\limits_{{\cal P}=\{P_1,P_2\}} \sigma({\cal X},{\cal P}) 
(-1)^{|g||a_{P_1}|}
\langle f,u\cdot a_{P_1} \rangle 
\langle g,u\cdot a_{P_2} \rangle \;\;\;,
\end{equation}
where $u\in U$, $a_1 , ... , a_k \in X$, 
the sum runs over all 2-partitions of the index set
$\{1,...,k\}$, and the notation of Rem.2.1.2 is used.
 The unit $1\!\!1 = u(1)$ is given by
$$
\langle 1\!\!1, u \rangle = 1 \;\;\;,
\;\;\;\langle 1\!\!1, u \cdot a_1 \cdot ... \cdot a_k\rangle \;=\; 0\;\;\;,
$$
for all $u \in U$, $ a_1, ... ,a_k \in X$. \bigskip

\noindent
Differentiating the formulae above with respect to the variable $u$
and checking the conditions of Def.3.5.1 one  gets

\begin{PP}
The subspace ${\cal D}_{X}(U)^{\wedge} \subset {\cal D}_{X}(U)'$ 
consisting of all superfunctions on an open coalgebra 
${\cal D}_{X}(U)$ is a $\ZZ$-graded subalgebra of
${\cal D}_{X}(U)'$.
\end{PP}

\begin{DD}
Let $f$ be a superfunction on an open coalgebra ${\cal D}_{X}(U)$.
The map
$$
f^0: U \ni u \lra \langle f, u\rangle \in \R
$$
is called the underlying part of the superfunction $f$.
For each $k\geq 1$ the $k$-th exterior component $f_k^{\wedge}$
of the superfunction $f$ is defined as a map
$$
f_k^{\wedge} : U\times \underbrace{X_1\times...\times X_1}_{k} \ni
(u,\xi_1,...,\xi_k)
 \longrightarrow 
\langle f , u \cdot \xi_1 \cdot ... \cdot \xi_k \rangle \in \R\;\;\;. 
$$
\end{DD}
It follows from Def.3.5.1 that the underlying part and the
exterior components of a superfunction $f$ on an open coalgebra 
${\cal D}_{X}(U)$ are
smooth functions on $U$ and $U\times X_1^{\times k}$, respectively.
  For all $u\in U$; $x_1, ... ,x_k \in X_0$; $\xi_1, ... ,\xi_l \in X_1$ 
one has
$$
\langle f ,u \cdot \xi_1 \cdot ... \cdot \xi_l \cdot 
x_k \cdot ... \cdot x_1\rangle \
= D_{x_1}... D_{x_k}\langle f ,u \cdot \xi_1 \cdot ... \cdot \xi_l  \rangle
$$
where $ D_{x}$ denotes the partial derivative with respect to 
 $u$-variable in the
direction $x\in X_0$. This implies that 
the superfunction is uniquely determined by
$f^0$, and $\{f^{\wedge}_k\}_{k\geq1}$.\bigskip

\noindent
Let ${\cal D}_{X}(U)$ be an open coalgebra and 
$$
R : {\cal D}_{X}(U)\times S(X) \ni (\omega,  u\cdot\a) 
\lra u\cdot \a \cdot \omega \in {\cal D}_{X}(U)
$$
the right action of the Hopf algebra $S(X)$ on ${\cal D}_{X}(U)$ introduced
in Subsect.3.1. For each $\omega \in S(X)$ one has a linear
map $R'_{\omega} : {\cal D}_{X}(U)' \ra {\cal D}_{X}(U)'$ given by
$$ 
\langle R'_{\omega} f , u\cdot\a \rangle =
\langle f , R_{\omega} ( u\cdot\a ) \rangle 
\;=\; \langle f , u\cdot\a \cdot\omega  \rangle \;\;\;,
$$         
for all $u\in U$, $\a \in S(X)$. One can easily check that if
$f\in {\cal D}_{X}(U)'$ is a superfunction so is $R'_{\omega} f$.

\begin{DD}
Let $f$ be a superfunction on an open coalgebra ${\cal D}_{X}(U)$.
For each $\omega \in S(X)$ the superfunction 
$D_{\omega}f = R'_{\omega}f$
is called the  derivative of $f$ in the direction
$\omega$.
\end{DD}

\noindent
The following proposition says that the notions of superfunction on open
coalgebra (Def.3.5.1) and its derivative (Def.3.5.3) are generalizations
of the corresponding notions both in the smooth Fr\'{e}chet geometry
and in the  finite-dimensional BLK supergeometry.
\begin{PP}\parbox{10pt}{}
\begin{enumerate}
\item Let $X = X_0\oplus \{o\}$ be a purely even $\ZZ$-graded
Fr\'{e}chet space. Let $U$ be an open subset of $X_0$ and let
$$
\langle \;.\;,\;.\; \rangle_U :
{\cal D}(U) \times {\cal C}^{\infty}(U) \longrightarrow \R
$$
be the pairing of {\rm Prop.3.4.2}. Then the map
$$
{\cal C}^{\infty}(U)\ni f \lra \overline{f} =
\langle \;.\;, f \rangle_U \in {\cal D}_{X}(U)^{\wedge} 
$$
is an isomorphism of $\ZZ$-graded algebras. Moreover, for
all $u\in U$, $x_1,...,x_k \in X$ one has
$$
D_{x_1...x_k}\overline{f}(u) = \overline{ D^k f}(u;x_1,...,x_k)\;\;\;.
$$
\item Let $X = \Rm\!\oplus\! \Rn$ be a finite-dimensional $\ZZ$-graded
Fr\'{e}chet space. Let $U$ be an open subset of $\Rm$ and
$$
\langle \;.\;,\;.\; \rangle_U :
{\cal D}_{m,n}(U) \times {\cal S}^{m,n}(U) \longrightarrow \R\;\;\;.
$$
the pairing of {\rm Prop.3.4.5}. Then the map
$$
{\cal S}^{m,n}(U)\ni f \lra \overline{f} =
\langle \;.\;, f \rangle_U \in {\cal D}_{X}(U)^{\wedge} 
$$
is an isomorphism of $\ZZ$-graded algebras. Moreover, for
all $u\in U$, $a_1,...,a_k \in \Rm\!\oplus\! \Rn$ one has
$$
D_{a_1\cdot ...\cdot a_k}\overline{f}(u,\theta) =
 \overline{D^k f}(u,\theta;a_1,...,a_k)\;\;\;.
$$
\end{enumerate}
\end{PP}

\noindent{\bf Remark 3.5.2}
Let $\Phi: {\cal D}_{X}(U) \ra {\cal D}_{Y}(V)$ be a smooth morphism of
open coalgebras. Then the dual map $\Phi':{\cal D}_{Y}(V)'\ra
{\cal D}_{X}(U)'$ is a morphism of $\ZZ$-graded algebras. 
Let $f$ be a superfunction on ${\cal D}_{Y}(V)$. For all  $u\in U$, 
$a_1,...,a_k \in X$ one has
\begin{eqnarray}
\label{spullback}
\langle \Phi' f 
\!\!\!\!\!\!&,& \!\!\!\!\!\! u \cdot a_1\cdot ... \cdot a_k \rangle
\;=\; \langle  f , \Phi( u \cdot a_1\cdot ... \cdot a_k) \rangle\\
&=& 
\sum\limits_{i=1}^k {1\over i !}
\sum\limits_{
               \parbox[t]{50pt}{ \scriptsize $
                                     \{ P_1,\ldots,P_i\} \\
                                    \makebox[10pt]{}| P_i| > 0 
                                   $  }}
{\sigma({\cal X},{\cal P})}
\langle  f ,
\Phi^0(u)\cdot \Phi^+_{|P_1|}(u,a_{P_1})
\cdot ... \cdot\Phi^+_{|P_i|}(u,a_{P_i}))
\;\;\;,\nonumber
\end{eqnarray}
where the sum runs over all nonempty partitions of the index set
$\{1,...,k\}$. 
Following the  proof of Th.3.2.2 one
can show that the functional $\Phi' f$ is a superfunction on 
${\cal D}_{X}(U)$. It follows that $\Phi'$ defines
a morphism of $\ZZ$-graded algebras   
$$
\Phi^*:{\cal D}_{Y}(V)^{\wedge} \ni f \lra \Phi' f \in 
{\cal D}_{X}(U)^{\wedge}\;\;\;.
$$
The superfunction $\Phi^*f = \Phi' f$ is called  the
{\it pull-back} of the superfunction $f$.
\bigskip

\noindent Differentiating formulae (\ref{product}) and
(\ref{spullback}) according to Def.3.5.3 one gets the
multiple Leibnitz and the chain rules for superfunctions
\begin{PP}
Let ${\cal D}_{X}(U)^{\wedge}, {\cal D}_{Y}(V)^{\wedge}$ be 
algebras of superfunctions on open coalgebras
${\cal D}_{X}(U)$, and ${\cal D}_{Y}(V)$, respectively,
Let $\Phi : {\cal D}_{X}(U)\ra {\cal D}_{Y}(V)$ be a 
smooth morphisms of open coalgebras. 
\begin{enumerate}
\item
For each $f,g \in {\cal D}_{X}(U)^{\wedge}$ and $a_1, ... , a_k \in
X$ 
\begin{equation}
\label{leibnitz}
D_{a_1\cdot ... \cdot a_k} (f\cdot g)=
\sum\limits_{{\cal P}=\{P_1,P_2\}} \sigma({\cal X},{\cal P}) 
(-1)^{|f||a_{P_2}|}
D_{a_{P_1}}f\cdot D_{a_{P_2}}g
\;\;\;,
\end{equation}
where the sum runs over all $2$-partitions of the index set
$\{1,...,k\}$ and the convention $D_{a_{\emptyset}}f = D_1f =f$ is
used.
\item
For each $g \in {\cal D}_{Y}(V)^{\wedge}$ and $a_1, ... , a_k \in
X$ 
\begin{eqnarray}
\label{chain}
(D_{a_{1}\cdot...\cdot a_{k}}\Phi^*
\!\!\!\!\!\!&g&\!\!\!\!\!\!)^0(u) \;=\\
&=&\sum\limits_{l =1}^{k} {1\over l!}
\sum\limits_{ \parbox[t]{50pt}{ \scriptsize $
                                     \{ P_1,\ldots,P_l\} \\
                                    \makebox[10pt]{} P_i \neq \emptyset
                                   $  }}
\!\!\!\!\!\!\sigma({\cal X},{\cal P})
\langle g,
\Phi^0(u)\cdot  \Phi^+(u,a_{P_1})\cdot ...
\cdot \Phi^+(u,a_{P_l})\rangle \;,\nonumber
\end{eqnarray}
where the sum runs 
over all nonempty partitions of the
index set $\left\{1,...,k\right\}$.
\end{enumerate}
\end{PP}

\noindent Let ${\cal D}_X(U)$ be an open coalgebra. For each pair of open 
subsets $U''\subset U' \subset U$ we define the restriction map
$$
\varrho_{U'U''} : {\cal D}_X(U')^{\wedge} \lra {\cal D}_X(U'')^{\wedge}
$$
as  dual to the inclusion ${\cal D}_X(U'') \subset {\cal D}_X(U')$.
The assignment for each open
subset $U'\subset U$  the $\ZZ$-graded algebra ${\cal D}_X(U')^{\wedge}$
of superfunctions with the restriction maps above 
 defines a sheaf 
${\cal D}_U^{\wedge} =(U,{\cal D}_X(\: .\:)^{\wedge})$
of $\ZZ$-graded algebras.

Let $\Phi:{\cal D}_X(U)\ra{\cal D}_Y(V)$ be a smooth morphism of open
coalgebras. For each open $V'\subset V$  the restriction
$$
\Phi_{V'}:{\cal D}_X({\Phi^0}^{-1}(V'))\ni
 \mu \lra \Phi(\mu)\in {\cal D}_Y(V')
$$
is a smooth morphism of open coalgebras.
Then the family of dual maps 
$$
\Phi^*_{V'}:{\cal D}_Y(V')^{\wedge}
\lra {\cal D}_X({\Phi^0}^{-1}(V'))^{\wedge}
$$ 
defines the morphism of
sheaves of $\ZZ$-graded algebras
$
\Phi^* = (\Phi^0,\Phi^*_.) : {\cal D}_U^{\wedge} \lra
{\cal D}_V^{\wedge}.$  
One has the following
\begin{PP}
The correspondence
\begin{eqnarray*}
{\cal D}_X(U) &\lra & {\cal D}_U^{\wedge}=(U,{\cal D}_X(\: .\:)^{\wedge})
\;\;\;,\\
\Phi &\lra & \Phi^* = (\Phi^0,\Phi^*_.)\;\;\;,
\end{eqnarray*}
is a covariant functor from the model category ${\bf sc}$ of open
coalgebras to the category of sheavs of $\ZZ$-graded algebras.

The functor above restricted to the subcategory ${\bf sc}^{<}$
of finite-dimensional open coalgebras yields an inverse to
the functor of {\rm Prop.3.4.4}.
\end{PP}

\section{Smooth  coalgebras}

\subsection{Category of smooth coalgebras}

\begin{DD}
Let ${\cal M}$ be a pointed $\ZZ$-graded cocommutative coalgebra
and $X = X_0 \oplus X_1$ a $\ZZ$-graded Fr\'{e}chet space. An
$X$-atlas on ${\cal M}$ is a collection $\left\{
(U_{\alpha}, \Phi_{\alpha}) \right\}_{\alpha \in I} $  of charts
$( U_{\alpha}, \Phi_{\alpha})$
satisfying the following conditions;
\begin{enumerate}
\item  The collection $\left\{ U_{\alpha} \right\}_{\alpha \in I}$ is
a covering of the set $M$ of group-like elements of ${\cal M}$
$$
 M = \parbox{35pt}{\scriptsize $
                               \hspace{7pt} \bigcup \\
                               \alpha \in I $}
\!\!\!\!\!U_{\alpha}\;\;\;.
$$
\item For each $\a\in I$, let ${\cal M}(U_{\alpha})$ be
a subcoalgebra of ${\cal M}$ given by 
$$
{\cal M}(U_{\alpha}) = \parbox{35pt}{\scriptsize $
                               \hspace{7pt} \bigoplus \\
                               p \in U_{\alpha} $}
\!\!\!{\cal M}_{p}\;\;\;,
$$
where  ${\cal M}_{p}$ denotes the  irreducible components   of ${\cal M}$
containing $p\in U_{\a}$.

Each $\Phi_{\alpha}$ is an isomorphisms of $\ZZ$-graded coalgebras
$$
\Phi_{\alpha}: {\cal M}(U_{\alpha}) \lra {\cal D}_X(\Phi^0(U_{\alpha}))
\;\;\;,
$$
where ${\cal D}_X(\Phi^0(U_{\alpha}))$ is an open subcoalgebra of
${\cal D}_X$. 
\item
For any $\alpha, \beta \in I$ such that $ U_{\alpha} \cap
U_{\beta} \neq \emptyset$, $\Phi^0_{\alpha}(U_{\alpha} \cap
U_{\beta})$ is an open subset of $X_0$,
and
$$
\Phi_{\alpha} \circ \Phi_{\beta}^{-1} :
{\cal D}_X(\Phi^0_{\beta}( U_{\a} \cap U_{\beta}) )\longrightarrow
{\cal D}_X(\Phi^0_{\a}(U_{\alpha} \cap  U_{\b}))
$$
is a diffeomorphism of open subcoalgebras of ${\cal D}_X$.
\end{enumerate}
\end{DD}

\noindent
Let $\left\{
( U_{\alpha}, \Phi_{\alpha}) \right\}_{\alpha \in I} $
be an $X$-atlas of ${\cal M}$. For each $\alpha \in I$ the underlying 
part of $\Phi_{\alpha}$
$$
\Phi^0_{\alpha} : U_{\alpha} \longrightarrow X_0
$$
is a bijective map onto an open subset of $X_0$, and the compositions
$$
\Phi_{\alpha}^0 \circ (\Phi_{\beta}^0)^{-1} :
\Phi_{\beta}^0(U_{\alpha}) \longrightarrow
\Phi_{\alpha}^0(U_{\alpha})
$$
are homeomorphisms. As in the standard theory of manifolds \cite{Lang}
one easily
shows that there exists a unique topology ${\cal T}_M$ on $M$ 
such that all $U_{\alpha}$ are open and all $\Phi_{\alpha}^0$ are
homeomorphisms onto open subsets of $X_0$. 

\begin{DD}
Two $X$-atlases
$\left\{
(U_{\alpha}, \Phi_{\alpha}) \right\}_{\alpha \in I} $,
$\left\{
( U_{\beta}, \Phi_{\beta}) \right\}_{\beta \in J} $ on 
${\cal M}$ are {\it compatible} if the union of them
$\left\{
( U_{\gamma}, \Phi_{\gamma}) \right\}_{\gamma \in I \cup J} $
is an $X$-atlas on ${\cal M}$. 
\end{DD}

\noindent
One easily verifies that compatible
$X$-atlases on ${\cal M}$ determine the same topology 
${\cal T}_M$ on $M$.
Since the smoothness of a graded coalgebra morphism in ${\bf sc}$
is a local property one also has:

\begin{PP}
The relation of compatibility of $X$-atlases is an equivalence 
relation.
\end{PP}

\begin{DD}
Let ${\cal M}$ be a pointed $\ZZ$-graded cocommutative coalgebra and
$X$ a $\ZZ$-graded Fr\'{e}chet space. An equivalence class of compatible
$X$-atlases on ${\cal M}$ is called an $X$-smooth structure on 
${\cal M}$.
A coalgebra ${\cal M}$ with an $X$-smooth structure inducing a
Hausdorff topology ${\cal T}_M$ on $M$  is called a smooth
coalgebra modelled on the $\ZZ$-graded Fr\'{e}chet space $X$, or simply an
$X$-smooth coalgebra. 
\end{DD}

\noindent
Note that by Prop.4.1.1 an $X$-atlas on ${\cal M}$ uniquely defines
a smooth structure on ${\cal M}$.

\begin{DD}
Let ${\cal M}$ be an $X$-smooth coalgebra. An $X$-atlas on ${\cal M}$
is said to be admissible if it defines an original $X$-smooth structure
on ${\cal M}$. A chart $( U_{\alpha}, \Phi_{\alpha})$ on ${\cal M}$
is called admissible if it belongs to an admissible atlas on ${\cal M}$.
\end{DD}

\noindent {\bf Remark.4.1.1} Let $\left\{
({\cal U}_{\alpha}, \Phi_{\alpha}) \right\}_{\alpha \in I} $ be an 
admissible $X$-atlas on a smooth coalgebra ${\cal M}$. By Prop.3.2.2
and Th.3.2.2,
$\left\{
(U_{\alpha}, \Phi_{\alpha}^0) \right\}_{\alpha \in I} $ is a 
smooth $X_0$-atlas on $M$. By the same token compatible
atlases on  ${\cal M}$ induce compatible smooth atlases on $M$.
The space $M$
of group-like elements of ${\cal M}$ with the smooth structure determined
by the atlas $\left\{
(U_{\alpha}, \Phi_{\alpha}^0) \right\}_{\alpha \in I} $ is called 
{\it the underlying manifold of ${\cal M}$}.

\begin{DD}
Let ${\cal M}, {\cal N}$ be smooth coalgebras. A morphism $\Phi:
{\cal M} \rightarrow {\cal N}$ of $\ZZ$-graded coalgebras is said to be
smooth if for each $p\in M$ there exist  admissible charts 
$(U_{\alpha}, \Phi_{\alpha})$ on ${\cal M}$ and 
$(V_{\gamma}, \Psi_{\gamma})$ on ${\cal N}$ such that
$p\in U_{\alpha}$, 
$\Phi^0(U_{\alpha})\subset V_{\gamma}$, and the
map
$$
\Psi_{\gamma}\circ\Phi\circ\Phi_{\alpha}^{-1} : {\cal D}_X(
 U_{\alpha})
\longrightarrow {\cal D}_X(V_{\gamma})\;\;\;
$$
is a smooth morphism of open subcoalgebras.
\end{DD}

\noindent
As a simple consequence of Th.3.2.2 one gets

\begin{PP}
The composition of smooth morphisms of smooth coalgebras is smooth.
\end{PP}

\begin{DD}
The objects of the category ${\bf SC}$ of smooth coalgebras are $X$-smooth
coalgebras, where $X$ runs over the category of graded Fr\'{e}chet spaces.

For any two objects ${\cal M}, {\cal N} \in {\rm O}{\bf SC}$ the space of
morphisms ${\rm M}{\bf SC}({\cal M},{\cal N})$ consists of all smooth morphisms of
graded coalgebras.

The composition of morphisms in ${\bf SC}$ is defined as a composition of
graded coalgebra morphisms.

An isomorphism in the category ${\bf SC}$ is called a diffeomorphism of
smooth coalgebras.
\end{DD}

\noindent{\bf Remark 4.1.2.}
In order to simplify further considerations we assume that objects and morphisms
of the category ${\bf SC}$ are defined up to diffeomorphisms of smooth
coalgebras with underlying parts being identitical maps in ${\bf FM}$.
\bigskip

\noindent {\bf Remark 4.1.3.} 
Let ${\cal M}, {\cal N}$ be smooth coalgebras, and $\Phi:
{\cal M} \rightarrow {\cal N}$ a morphism of $\ZZ$-graded coalgebras. 
The restriction of $\Phi$ to the set $M$ of group-like elements
of ${\cal M}$, $\Phi^0 : M \rightarrow N$
 is called  {\it the underlying part} of the morphism $\Phi$.  
For composition of two $\ZZ$-graded coalgebra morphisms one has 
$(\Psi\circ\Phi)^0 =\Psi^0\circ\Phi^0$.
By Rem.4.1.1 if $(U_{\alpha}, \Phi_{\alpha})$, 
$(V_{\gamma}, \Psi_{\gamma})$ are admissible charts on
 ${\cal M}$ and  ${\cal N}$, then $(U_{\alpha}, \Phi^0_{\alpha})$, 
$(V_{\gamma}, \Psi^0_{\gamma})$ are admissible charts on
 $M$ and  $N$, and the map
$$
\Psi^0_{\gamma}\circ\Phi^0\circ\Phi_{\alpha}^{0-1} : \Phi^0_{\alpha}(
U_{\alpha})
\longrightarrow \Psi^0_{\gamma}(V_{\gamma})\;\;\;
$$
is smooth by Prop.3.2.2. It follows that the underlying map of a smooth
morphism of graded coalgebras is a smooth map of Fr\'{e}chet manifolds.
\bigskip

\begin{DD}
Let  $X = X_0 \oplus X_1$ be a $\ZZ$-graded Fr\'{e}chet space,
and $\{ U_{\a},\vp_{\a} \}_{\a \in I}$ 
an admissible atlas on a smooth manifold $M$ modelled on the
Fr\'{e}chet space $X_0$.
Let $\{ \Psi_{\a\b} \}_{\a \in I}$ be a collection of maps
such that,
\begin{enumerate}
\item for all $\a \in I, \b\in I(\a)\equiv \{ \b \in I: U_{\a} \cap U_{\b}
\neq \emptyset \}$ 
$$
\Psi_{\a\b}:  {\cal D}_X(\vp_{\a}(U_{\a} \cap U_{\b})) 
\lra
  {\cal D}_X(\vp_{\b}(U_{\a} \cap U_{\b}))
$$
is an isomorphism of smooth open coalgebras such that $\Psi^0_{\a\b} =
\vp_{\b}\circ\vp_{\a}^{-1}$;
\item for all  $\a \in I$, $ \Psi_{\a\a} = {\rm id}_{\R\vp_{\a}(U_{\a})
\otimes S(X)}$;
\item for all $\a,\b,\gamma \in I$ such that $U_{\a} \cap U_{\b}
 \cap U_{\gamma}\neq \emptyset$, 
$$
\Psi_{\b\gamma}\circ\Psi_{\a\b} = \Psi_{\a\gamma}
$$
on ${\cal D}_X(\vp_{\a}(U_{\a} \cap U_{\b} \cap U_{\gamma}))$.
\end{enumerate}
A collection $\{ \Psi_{\a\b} \}_{\a \in I}$ 
with the properties above is called an $X$-cocycle of
transition ${\bf sc}$-morphisms  over  the atlas 
$\{ U_{\a},\vp_{\a} \}_{\a \in I}$ on $M$. 
\end{DD} 

\noindent
One has the following  "reconstruction theorem" which is 
very useful in constructing new smooth coalgebras,
\begin{PP}
Let $\{ \Psi_{\a\b} \}_{\a \in I}$ 
be  an $X$-cocycle of
transition ${\bf sc}$-morphisms on $M$. Then there exists a 
unique smooth coalgebra ${\cal M}$ with the underlying manifold
$M$  and with the admissible $X$-atlas 
$\{ (U_{\a},\Psi_{\a}) \}_{\a \in I}$ such
that
\begin{equation}
\label{coatl}
\Psi_{\a\b}\;=\; \Psi_{\b}\circ 
{\Psi_{\a}^{-1}}_{|  {\cal D}_X( \Psi_{\a}^0(U_{\a} \cap U_{\b}) )}
\;\;\;,
\end{equation}
for all $\a\in I$, $\b\in I(\a)$. 
\end{PP}

\noindent
The smooth coalgebra ${\cal M}$ and
the $X$-atlas 
$\{ ( U_{\a},\Psi_{\a}) \}_{\a \in I}$ of the proposition above 
are said to be {\it generated } by  the $X$-cocycle 
$\{ \Psi_{\a\b} \}_{\a \in I}$.\bigskip
   
\noindent{\bf Proof.}
Let $\{ \Psi_{\a\b} \}_{\a \in I}$ be an $X$-cocycle related to a
smooth atlas $\{ U_{\a},\vp_{\a} \}_{\a \in I}$ on $M$.
 For $p\in M$,
consider the space of pairs $(\a,m)_p$, where 
$\a\in I$ is such that $p\in U_{\a}$ and 
$m \in {\cal D}_{X\vp_{\a}(p)}= \R\{\vp_{\a}(p)\}\otimes S(X)$. 
Two such pairs are equivalent
$$
(\a,m)_p \sim (\a',m')_p
$$
if $m' =\Psi_{\a\a'}( m)$. Using the cocycle properties 
 (Def.4.1.7) one easily shows this is an equivalence 
relation. Let ${\cal M}_p$ denote the space of
 equivalence classes of this relation and let
 $[(\a,m)_p]_{\sim}$ denote the equivalence class of
 $(\a,m)_p$.
Since $\Psi_{\a\a'}$ are $\ZZ$-graded coalgebra morphisms ${\cal M}_p$
acquires the structure of irreducible pointed
$\ZZ$-graded cocomutative coalgebra:
\begin{eqnarray*}
\Delta[(\a,m)_p]_{\sim} &\equiv & \sum\limits_{(m)}\;
[(\a,m_{(1)})_p]_{\sim} \otimes [(\a,m_{(2)})_p]_{\sim}\;\;\;,\\
\varepsilon ( [(\a,m)_p]_{\sim} )& \equiv & 
\varepsilon_{\vp_{\a}(U_{\a})}(m)\;\;\;.
\end{eqnarray*}
Let ${\cal M}$ be the direct sum of irreducible
$\ZZ$-graded coalgebras
$$
{\cal M} \equiv \parbox{30pt}{\scriptsize $
                               \hspace{7pt} \bigoplus \\
                               p\in M $} \!\!
{\cal M}_p\;\;\;.
$$
For  each $\alpha \in I$ we define
$$
{\Psi}_{\alpha} :  {\cal M}(U_{\alpha}) \ni  
 [(\a,m)_u]_{\sim}
\longrightarrow m \in  {\R}\vp_{\a}(U_{\a})\otimes S(X) \;\;\;.
$$
Using the cocycle properties (Def.4.1.7) one verifies that   the collection 
$\left\{ (U_{\alpha}, \Psi_{\alpha}) \right\}_{\alpha \in I} $
 is an $X$-atlas on ${\cal M}$ with the required transition
${\bf sc}$-morphisms. By Prop.4.1.1 it defines a unique
$X$-smooth structure on ${\cal M}$.

Suppose that there exists another smooth coalgebra ${\cal M}'$ over $M$
with an admissible $X$-atlas 
$\left\{ ( U'_{\alpha}, \Psi'_{\alpha}) \right\}_{\alpha \in I} $
satisfying the condition (\ref{coatl}).
Then the map defined for each $p\in M$
by
$$
{\cal M}'_p \ni m \lra  \Psi_{\a}^{-1}\circ\Psi'_{\a} \in {\cal M}_p 
$$
extends by linearity to the diffeomorphism of smooth coalgebras over
the identity map. Hence ${\cal M}'={\cal M}$ by Rem.4.1.2.
\hfill $\Box$ \bigskip

\noindent 
Two cocycles 
$\{\Psi'_{\a'\b'} \}_{\a' \in I}$,
$\{ {\Psi''}_{\a''\b''} \}_{\a'' \in I''}$ 
of transition ${\bf sc}$-morphisms on $M$ are said to be {\it compatible}
if there exists a third one
$\{ \Psi_{\a\b} \}_{\a \in I}$
such that 
\begin{eqnarray*}
\{ \vp'_{\a'} \}_{\a' \in I'}\cup\{ \vp''_{\a''} \}_{\a'' \in I''}
&\subset &\{ \vp_{\a} \}_{\a \in I}\;\;\;,\\
\{\Psi'_{\a'\b'} \}_{\a' \in I'}\cup
\{\Psi''_{\a''\b''} \}_{\a'' \in I''}
&\subset& \{\Psi_{\a\b} \}_{\a \in I}\;\;\;,
\end{eqnarray*}
as sets of maps. Using the construction of  Prop.4.1.3 for
all three cocycles of the definition above
and comparing the resulting smooth coalgebras
one gets 

\begin{PP}
Compatible $X$-cocycles of transition ${\bf sc}$-morphisms on $M$ 
generate
the same $X$-smooth coalgebra ${\cal M}$ and  compatible 
$X$-atlases on ${\cal M}$.
\end{PP}

\begin{DD}
Let ${\cal M}$ be an $X$-smooth coalgebra. A linear functional $f$ on 
${\cal M}$ is called a superfunction if for each $p\in M$
there exists an admissible chart $( U_{\a}, \Phi_{\a})$
such that $p\in U_{\a}$, and the functional $(\Phi^{-1}_{\a})' f$ is
a superfunction on the open coalgebra ${\cal D}_X(\Phi_{\a}^0(U_{\a}))$.
\end{DD}

\begin{PP} Let ${\cal M}'$ be the full algebraic dual of ${\cal M}$
endowed with the  $\ZZ$-graded algebra structure dual to the 
$\ZZ$-graded coalgebra structure on ${\cal M}$.
 
The subspace ${\cal M}^{\wedge} \subset {\cal M}'$ 
consisting of all superfunctions on an  $X$-smooth coalgebra 
${\cal M}$ is a $\ZZ$-graded subalgebra of
${\cal M}'$.
\end{PP}
 Let  $\left\{
( U_{\alpha}, \Phi_{\alpha}) \right\}_{\alpha \in I} $
be an admissible $X$-atlas on a smooth coalgebra ${\cal M}$ 
and let $U$ be an open (with respect
to the induced topology ${\cal T}_M$) subset of $M$. Then  the collection
\begin{equation}
\label{opena}
\left\{ (U \cap  U_{\alpha}, 
\Phi_{\alpha  | {\cal M}(U) \cap {\cal M}(U_{\alpha})}
) \right\}_{\alpha \in I} \;\;\;,
\end{equation}
is an $X$-atlas on the subcoalgebra
$$
{\cal M}(U) = \parbox[t]{35pt}{\scriptsize $
                               \hspace{7pt} \bigoplus \\
                               p \in U $}
\!\!\!{\cal M}_p\;\;\;,
$$
 Compatible $X$-atlases on ${\cal M}$ induce compatible
$X$-atlases
on ${\cal M}(U)$. The subcoalgebra ${\cal M}(U)\subset {\cal M}$ 
with the 
smooth structure defined by  the induced atlas (\ref{opena}) is 
called an {\it open  subcoalgebra} of $ {\cal M}$.

Assigning to each open substet $U\subset M$ the $\ZZ$-graded algebra
${\cal M}(U)^{\wedge}$ of superfunctions on ${\cal M}(U)$
 and introducing the restriction maps as
duals to the inclusions ${\cal M}({U'})\subset {\cal M}(U)$, ($U'\subset U$)
one gets a sheaf of $\ZZ$-graded algebras ${\cal M}_M^{\wedge}=
(M,{\cal M}(\: .\:)^{\wedge})$. ${\cal M}_M^{\wedge}$ is called
the {\it sheaf of superfunctions on} ${\cal M}$.

Let $\Phi:{\cal M}\ra{\cal N}$ be a morphism of smooth coalgebras.
By Rem.3.5.2 for each superfunction $g \in {\cal N}^{\wedge}$,
the functional $\Phi' g$ is a superfunction on ${\cal M}$.
$\Phi^*g =\Phi' g$ is called the {\it pull-back} of $g$. For each open subset 
$V\subset N$ we define
$
\Phi^*_V :{\cal N}(V)^{\wedge} \ra {\cal M}(U)^{\wedge}$ as a map
dual to 
$$
\Phi_V : {\cal M}({\Phi^0}^{-1}(V)) \ni\mu \lra
\Phi(\mu) \in {\cal N}(V)\;\;\;.
$$
The collection of maps $\{\Phi^*_.\}$ 
defines  a morphism of sheaves of $\ZZ$-graded algebras
$\Phi^* =(\Phi^0,\Phi^*_.) : {\cal M}_M^{\wedge}\ra {\cal N}_N^{\wedge}$.
One has the following global version of Prop.3.5.4
\begin{PP}
The correspondence
\begin{eqnarray*}
{\cal M} &\lra & {\cal M}_M^{\wedge}=(M,{\cal M}(\: .\:)^{\wedge})
\;\;\;,\\
\Phi &\lra & \Phi^* = (\Phi^0,\Phi^*_.)\;\;\;,
\end{eqnarray*}
is a covariant functor from the category ${\bf SC}$ of smooth
coalgebras to the category of sheaves of $\ZZ$-graded algebras.
\end{PP}

\subsection{Direct product}

Let ${\cal M}, {\cal N}$ be smooth coalgebras modelled on graded Fr\'{e}chet
spaces $X$, and $Y$, respectively. One can introduce an 
$(X\oplus Y)$-smooth
structure on ${\cal M}\otimes {\cal N}$ as follows. Let 
$\left\{( U_{\alpha}, \Phi_{\alpha}) \right\}_{\alpha \in I} $ 
be an admissible $X$-atlas on ${\cal M}$, and 
$\left\{
( V_{\beta}, \Psi_{\beta}) \right\}_{\beta \in J} $ an  
admissible $Y$-atlas on ${\cal N}$. Consider the collection 
\begin{equation}
\label{prodsc}
\left\{
(U_{\alpha}\times V_{\beta}, 
\Phi_{\alpha}\otimes \Psi_{\beta}) \right\}_{(\alpha, \beta) 
\in I\times J}\;\;\;.
\end{equation}
One obviously has
$$
M\times  N = \parbox{50pt}{\scriptsize $
                               \hspace{22pt} \bigcup \\
                               (\alpha,\beta) \in I\times J $}
\,U_{\alpha}\times V_{\beta}\;\;\;,
$$
and the $\ZZ$-graded coalgebra isomorphisms
\begin{eqnarray*}
{\cal D}_{X\oplus Y}
(\Phi^0_{\a}\times \Psi^0_{\b} ( U_{\alpha}\times  V_{\b}))
&=& {\cal D}_X(\Phi^0_{\a}(U_{\a}))\otimes 
{\cal D}_Y(\Psi^0_{\b}(V_{\b}))\;\;\;,\\
{\cal D}_{X\oplus Y}
(\Phi^0_{\a}\times \Psi^0_{\b} (( U_{\alpha}\times  V_{\b})
\cap( U_{\a'}\times  V_{\b'}))) &=&
{\cal D}_X(\Phi^0_{\a}( U_{\a}\cap  U_{\a'}))
\otimes 
{\cal D}_Y(\Psi^0_{\b}( V_{\b}\cap V_{\b'}))\;\;\;.
\end{eqnarray*}
By Def.3.3.1 the r.h.s of the equations above can be regarded
as tensor products in the model category while the l.h.s
as open subcoalgebras of  ${\cal D}_{X\oplus Y}$.
Then the $\ZZ$-graded coalgebra morphisms
$$
(\Phi_{\alpha}\otimes \Psi_{\beta})\circ(\Phi_{\alpha'}
\otimes \Psi_{\beta'})^{-1}
= (\Phi_{\alpha}\circ\Phi_{\alpha'}^{-1})\otimes
(\Phi_{\b}\circ\Phi_{\b'}^{-1})
$$
 are smooth by Prop.3.3.1. It follows that the collection
(\ref{prodsc}) defines an $(X\oplus Y)$-atlas on ${\cal M}\otimes
{\cal N}$. One can easily verify that compatible atlases on ${\cal M}$ and 
${\cal N}$ lead by the construction above to compatible atlases on 
${\cal M}\otimes {\cal N}$, hence the following definition:

\begin{DD}
The tensor product of two smooth coalgebras 
${\cal M}, {\cal N} \in {\rm O}{\bf SC}$
modelled on the $\ZZ$-graded Fr\'{e}chet spaces $X, Y$, respectively,
is the tensor product of $\ZZ$-graded cocommutative coalgebras 
${\cal M}\otimes {\cal N}$ endowed with the
$(X\oplus Y)$-smooth structure determined
by the atlas {\rm (\ref{prodsc})}.
\end{DD}

\noindent
As a consequence of Prop.3.3.1 
one gets

\begin{PP}
$\makebox[100pt]{ }$
\begin{enumerate}
\item
Let $\Phi : {\cal M} \rightarrow {\cal N}$, 
$\Phi' : {\cal M'} \rightarrow {\cal N'}$ be  morphisms of smooth coalgebras.
Then the tensor product of graded coalgebra morphisms
$$
\Phi\otimes\Phi' : {\cal M}\otimes{\cal M'}\ni m\otimes m' \longrightarrow 
\Phi(m)\otimes\Phi'(m') \in {\cal N}\otimes{\cal N'}
$$
is a morphism of smooth coalgebras.
\item
Let ${\cal M}\otimes {\cal N}$ be the tensor product of smooth coalgebras.
Let $\varepsilon_M, \varepsilon_N$ be counits in the coalgebras 
${\cal M}, {\cal N}$, respectively.
Then the maps
\begin{eqnarray*}
P_M\!\!\!& :&\!\!\! {\cal M}\otimes {\cal N} \ni m\otimes n \longrightarrow
\varepsilon_N(n)\cdot m \in {\cal M}\;\;\;,\\
P_N\!\!\!& :&\!\!\! {\cal M}\otimes {\cal N} \ni m\otimes n \longrightarrow
\varepsilon_M(m)\cdot n \in {\cal N}\;\;\;,
\end{eqnarray*}
are morphisms of smooth coalgebras.
\item
The comultiplication $\Delta : {\cal M} \rightarrow {\cal M}
\otimes{\cal M}$
in a smooth coalgebra ${\cal M}$ is a morphism of smooth coalgebras.
\end{enumerate}
\end{PP}

\noindent {\bf Remark 4.2.1} Let us observe that the $X\oplus Y$-atlas 
(\ref{prodsc}) induces the smooth atlas
$$
\left\{ (U_{\alpha}\times  V_{\beta}, 
\Phi^0_{\alpha}\times \Psi^0_{\beta}) \right\}_{(\alpha, \beta) 
\in I\times J}
$$
on $M\times N$. It follows that the underlying manifold of 
${\cal M}\otimes {\cal N}$ is the cartesian product $M\times N$
 of Fr\'{e}chet manifolds. The underlying parts of smooth coalgebra morphisms
from Prop.4.2.1 are given by
\begin{eqnarray*}
(\Phi\otimes\Phi')^0\!\! \!&: &\!\!\!M\times M'\ni( u,  u')\longrightarrow 
(\Phi^0(u),{\Phi'}^0(u')) \in  N\times N'\;\;\;,\\
P^0_M\!\!\!& :&\!\!\! M\times N \ni (u,v) \longrightarrow
u \in M\;\;\;,\\
P^0_N\!\!\!& :&\!\!\! M\times N \ni (u,v) \longrightarrow
v \in N\;\;\;,\\
\Delta^0\!\!\! &: & \!\!\! M \ni u \longrightarrow (u,u) \in M\times M\;\;\;.
\end{eqnarray*}
\bigskip

\noindent
By Th.3.3.1 and Prop.4.2.1 one gets

\begin{TT}
 $({\cal M}\otimes {\cal N}, P_M,
P_N)$ is the direct product in the category of smooth coalgebras, i.e. for
every smooth coalgebra ${\cal E}$ and smooth coalgebra morphisms
$\Phi_M : {\cal E} \rightarrow {\cal M}$, 
$\Phi_N : {\cal E} \rightarrow {\cal N}$ there exists a unique morphism of 
smooth coalgebras $\Phi : {\cal E} \rightarrow {\cal M}\otimes {\cal N}$ 
making the diagram 
\begin{center}
\begin{picture}(120,100)(0,10)
\put(40,93){\makebox(40,10){$
%
   {\cal M}\otimes {\cal N}
$}}
\put(5,75){\makebox(20,10)[r]{\small $
%
 P_M
$}}
\put(95,75){\makebox(20,10)[l]{$
%
 P_N
$}}
\put(-20,48){\makebox(20,10)[br]{\small$
%
 {\cal M}
$}}
\put(120,48){\makebox(20,10)[bl]{$
%
  {\cal N}
$}}
\put(5,18){\makebox(20,10)[tr]{\small$
%
 \Phi_M
$}}
\put(95,18){\makebox(20,10)[tl]{\small$
%
 \Phi_N
$}}
\put(65,50){\makebox(20,10)[l]{\small$
%
 \Phi
$}}
\put(50,2){\makebox(20,10)[t]{\small$
%
 {\cal E}
$}}
\put(50,90){\vector(-3,-2){48}}
\put(70,90){\vector(3,-2){48}}
\put(50,15){\vector(-3,2){48}}
\put(70,15){\vector(3,2){48}}
\put(60,81){\vector(0,1){8}}
\multiput(60,17)(0,8){8}{\line(0,1){4}}
\end{picture}
\end{center}
commute. 
\end{TT} 

\noindent {\bf Remark 4.2.2} The unique morphism $\Phi$ in the theorem above
is given by the composition 
$$
\Phi = (\Phi_M \otimes \Phi_N) \circ \Delta_E\;\;\;,
$$
and its underlying part by
$$
\Phi^0 : E \in u \longrightarrow (\Phi_M^0(u),\Phi_N^0(u)) \in M\times N\;\;\;.
$$

\subsection{Subcategory of smooth Fr\'{e}chet manifolds}

Let us denote by ${\rm O}{\bf FM}$ and ${\rm M}{\bf FM}$ 
the collections
of objects and  morphisms of the category ${\bf FM}$ of smooth Fr\'{e}chet
manifolds. In this subsection we shall show that 
the category ${\bf FM}$ is equivalent to the category ${\bf SC}_0$
of even smooth coalgebras. ${\bf SC}_0$ is defined as the full subcategory
of  ${\bf SC}$ consisting of all smooth coalgebras modelled on purely
even graded Fr\'{e}chet spaces $X = X \oplus \{o\}$. By definition
for all ${\cal M},{\cal N}\in {\bf SC}_0$
$$
{\rm M}{\bf SC}_0({\cal M},{\cal N}) \equiv 
{\rm M}{\bf SC}({\cal M},{\cal N})
\;\;\;.
$$
Note that by Th.4.2.1 ${\bf SC}_0$ inherits the direct product from 
${\bf SC}$.\bigskip

\noindent
Gathering together Remarks 4.1.1, 4.1.3, 4.2.1, and 4.2.2 one gets

\begin{PP}
The correspondence
\begin{eqnarray*}
{\rm O}{\bf SC} \ni {\cal M} &\longrightarrow & 
M \in {\rm O}{\bf FM}\;\;\;,\\
{\rm M}{\bf SC}({\cal M},{\cal N}) \ni \Phi &
\longrightarrow & \Phi^0 \in {\rm M}{\bf FM}(M,N)\;\;\;
\end{eqnarray*}
is a covariant functor respecting the direct product.
\end{PP}

\noindent
Let $\{ (U_{\a},\vp_{\a}) \}_{\a \in I}$ be an admissible atlas on 
 a Fr\'{e}chet manifold $M$ modelled on a Fr\'{e}chet space $X$.
By Rem.3.4.1 the family 
$\{ \Phi_{\a\b} \}_{\a \in I}$ 
of ${\bf sc}$-morphisms  defined by
\begin{equation}
\label{starcoc}
\Psi_{\a\b}\;=\; {(\vp_{\b}\circ 
{\vp_{\a}^{-1}})_*}_{| \R \Psi_{\a}^0(U_{\a} \cap U_{\b}) \otimes S(X)}
\;\;\;,\;\;\;\b\in I(\a)\;\;\;,
\end{equation}
is an $(X\oplus \{o\})$-cocycle of transition ${\bf sc}$-morphisms 
over
the atlas $\{ (U_{\a},\vp_{\a})\}_{\a\in I}$ on $M$. 
By the same token cocycles constructed from compatible smooth
atlases on $M$ by formula (\ref{starcoc}), are compatible.
Thus,  by Prop.4.1.3 and Prop.4.1.4, the following definition is not 
ambiguous 

\begin{DD}
The $(X\oplus \{o\})$-smooth coalgebra ${\cal D}(M)$ 
generated by the cocycle {\rm (\ref{starcoc})} is
called the {\it smooth coalgebra of the Fr\'{e}chet manifold $M$}.
\end{DD}

\noindent
For each admissible atlas $\{ (U_{\a},\vp_{\a}) \}_{\a \in I}$ on $M$,
 the $(X\oplus \{o\})$-atlas on ${\cal D}(M)$
generated by the cocycle (\ref{starcoc}) will
be  denoted by $\{ ( U_{\a},{\vp}_{\a*}) \}_{\a \in I}$.
Applying this construction to the maximal atlas on
$M$ we define for each admissible chart $(U,\vp)$ on $M$
the corresponding admissible chart $(U,\vp_*)$
on ${\cal D}(M)$. One easily verifies that the definition of
$( U,\vp_*)$ is independent of the choice of
admissible atlas containing $(U,\vp)$.

\begin{PP}
Let ${\cal M} \in {\rm O}{\bf SC}_0$. Then ${\cal M}$ is the smooth 
coalgebra of its underlying manifold $M$, i.e.
${\cal M}={\cal D}(M).$
\end{PP}

\noindent {\bf Proof.}
Let $\left\{( U_{\alpha}, \Phi_{\alpha}) \right\}_{\alpha \in I} $ 
be an admissible 
$(X\oplus \{o\})$-atlas on ${\cal M}$. Then by
Rem.4.1.1 the atlas $\left\{
(U_{\alpha}, \Phi^0_{\alpha}) \right\}_{\alpha \in I} $ 
is an admissible smooth atlas on the underlying manifold $M$. 
For all $\alpha, \beta \in I(\a)$, 
$\Phi_{\alpha}\circ \Phi_{\beta}^{-1}$ are morphisms of 
the model category ${\bf sc}_0$ and by  Rem.3.4.1
$
\Phi_{\alpha}\circ \Phi_{\beta}^{-1} = 
(\Phi^0_{\alpha}\circ {\Phi^0_{\beta}}^{-1})_*
$.
Then by the construction of ${\cal D}(M)$ the map defined for
each $p\in M$ by 
$$
{\cal M}_p \ni \mu \longrightarrow ({\Phi}^0_{\a*})^{-1}
\circ \Phi_{\alpha}(\mu) \in {\cal D}(M)_p
$$
extends by linearity to a diffeomorphism of smooth coalgebras over 
the identity map.
Hence ${\cal M}={\cal D}(M)$ by Rem.4.1.2. 
 \hfill $\;\;\;\Box$ \bigskip

\noindent
Let $\phi : M \rightarrow N$ be a smooth map of Fr\'{e}chet manifolds.
Then for each
$p \in M$ there are admissible charts $(U_{\alpha}, \vp_{\alpha})$ at
 $p \in M$
and $(V_{\beta}, \psi_{\beta})$ at $\phi(p) \in N$
 such that the composition 
$$
\psi_{\beta}\circ\phi\circ \vp_{\alpha}^{-1} : 
\vp_{\alpha}(U_{\alpha}) \lra \psi_{\beta}(V_{\beta})
$$
is a smooth map between open subsets of
Fr\'{e}chet spaces (morphism in ${\bf fm}$).

For all $\mu \in {\cal D}(M)_p$ we define
\begin{equation}
\label{pstar}
\phi_{*p}(\mu) \equiv \psi^{-1}_{\beta*} \circ (
\psi_{\beta}\circ\phi\circ \vp_{\alpha}^{-1})_* \circ 
\vp_{\alpha*}(\mu)
\;\;\;.
\end{equation}
By Rem.3.4.1 the definition above is independent of the choice 
of admissible charts at $p \in M$ and $\phi(p) \in N$.  
Extending the formula (\ref{pstar}) by linearity in $p$  one gets 
the smooth morphism of graded coalgebras
$$
\phi_* : {\cal D}(M) \longrightarrow  {\cal D}(N)\;\;\;,
$$
with underlying part $(\phi_*)^0 = \phi$.

\begin{PP}
Let ${\cal M}, {\cal N} \in {\rm O}{\bf SC}_0$. For all 
smooth morphisms of
graded coalgebras $\Phi : {\cal M}\rightarrow {\cal N}$,  
$
\Phi = (\Phi^0)_*
$.
\end{PP}

\noindent{\bf Proof.}
By Def.4.1.5  for each $p\in M$  there exist admissible
charts $( U_{\alpha}, \Phi_{\alpha})$ on ${\cal M}$,
 and
 $(V_{\gamma}, \Psi_{\gamma})$ 
on  ${\cal N}$, such that $p\in U_{\a}$, 
$\Phi^0(U_{\alpha})
\subset V_{\g}$ and the composition
$
\Psi_{\g} \circ\Phi\circ(\Phi_{\alpha})^{-1}
$
is an ${\bf sc}_0$-morphism. Then  by Rem.3.4.1
$$
\Psi_{\g} \circ\Phi\circ(\Phi_{\alpha})^{-1} =
(\Psi_{\g}^0 \circ\Phi^0\circ(\Phi_{\alpha}^0)^{-1})_*
\;\;\;.
$$
By  Prop.4.3.2 one can assume $\Phi_{\a} =\Phi^0_{\a*}$
and $\Psi_{\g} =\Psi^0_{\g*}$.
Hence for all $p\in M$, $\mu\in {\cal D}(M)_p$
$$
\Phi(\mu) = (\Psi^0_{\g*})^{-1}\circ
(\Psi_{\g}^0 \circ\Phi^0\circ(\Phi_{\alpha}^0)^{-1})_*
\circ \Phi^0_{\a*}(\mu) \;\;\;,
$$
and $\Phi =(\Phi^0)_*$. \hfill$\;\;\;\Box$\bigskip

\noindent Prop.4.3.1,
 Prop.4.3.2,  and Prop.4.3.3 imply the following global counterpart 
of Prop.3.4.1 and Rem.3.4.1.

\begin{TT}
The correspondence
\begin{eqnarray*}
{\rm O}{\bf FM} \ni M &\longrightarrow & 
{\cal D}(M) \in {\rm O}{\bf SC}_0\;\;\;,\\
{\rm O}{\bf FM}(M,N) \ni \phi &
\longrightarrow &  \phi_* 
 \in {\rm M}{\bf SC}_0({\cal D}(M),{\cal D}(N))\;\;\;
\end{eqnarray*}
is  an equivalence of categories ${\bf FM}$ and ${\bf SC}_0$. Moreover
it is the right inverse to the functor of {\rm Prop.4.3.1}.
\end{TT}

\noindent
The result above means that the category of smooth Fr\'{e}chet manifolds
can be regarded
as a full subcategory of the category of smooth coalgebras. This 
partly justifies our
construction of ${\bf SC}$ as a $\ZZ$-graded extension of ${\bf FM}$.
\bigskip

\noindent{\bf Remark.4.3.1}
Composing the functors from Prop.4.3.1 and Th.4.3.1 one gets 
the covariant functor respecting the direct product
\begin{eqnarray*}
{\rm O}{\bf SC} \ni {\cal M} &\longrightarrow & M\;\longrightarrow\;
\widetilde{\cal M}\equiv {\cal D}(M) \in {\rm O}{\bf SC}_0\;\;\;,
\\
{\rm M}{\bf SC}({\cal M},{\cal N}) \ni \Phi &
\longrightarrow &  \Phi^0\;\longrightarrow \; \widetilde{\Phi} 
\equiv (\Phi^0)_* 
 \in {\rm M}{\bf SC}_0(\widetilde{\cal M},\widetilde{\cal N})
\;\;\;.\nonumber
\end{eqnarray*}
The functor above is called {\it the underlying functor}.

Let us note that $\widetilde{\cal M}$ is canonically embedded in ${\cal M}$.
Let 
$
i_0: S(X_0) \longrightarrow S(X_0\otimes X_1)
$
be the canonical embedding defined as the universal extension of the composition
$$
X_0 \;
\begin{picture}(25,15)(0,0)
\put(0,4){\vector(1,0){25}}
\put(7,8){\makebox(10,10)[b]{\small$
\rho_0 
$}}
\end{picture}
\; X_0\oplus X_1 \;
\begin{picture}(25,15)(0,0)
\put(0,4){\vector(1,0){25}}
\put(7,8){\makebox(10,10)[b]{\small$
\theta 
$}}
\end{picture}
\;S(X_0\oplus X_1 )\;\;\;.
$$
For each $p\in M$ we define
$$
\widetilde{\cal M}_p \ni \mu \longrightarrow \Phi^{-1}_{\alpha} \circ
({\rm id}_{\Phi^{0}_{\alpha}U_{\alpha}} \otimes i_0) \circ 
{\Phi}^{0}_{\alpha*}(\mu) \in {\cal M}_p\;\;\;,
$$
where $( U_{\alpha},\Phi_{\alpha})$ is an admissible chart of 
${\cal M}$
at $ p\in M$. One easily verifies that the definition above is independent
 of the choice of an admissible chart at $p$ and extends by linearity 
to the  morphism of smooth coalgebras
$
\widetilde{i} : \widetilde{\cal M} \longrightarrow {\cal M},
$
with $\widetilde{i}^0 = {\rm id}_M$. By Rem.4.1.2 
$\widetilde{\cal M}$ can be regarded as a subcoalgebra of 
${\cal M}$. $\widetilde{\cal M}$ is called
{\it the underlying subcoalgebra of} ${\cal M}$. \bigskip

\noindent{\bf Remark 4.3.2} We shall briefly discuss the geometric
interpretation of the smooth coalgebra of a Fr\'{e}chet manifold $M$.
By definition, ${\cal D}(M)$ is the direct sum of its
irreducible components 
$$
{\cal D}(M) = \parbox{35pt}{\scriptsize $
                               \hspace{7pt} \bigoplus \\
                               p \in M $}
\!\!\!{\cal D}(M)_p\;\;\;.
$$
With respect to the graded coalgebra structure each irreducible
component ${\cal D}(M)_p$ is isomorphic with $S(X\oplus \{o\})$.
For each $p\in M$ let 
$$
{\cal D}(M)_p = \parbox{35pt}{\scriptsize $
                               \hspace{7pt} \bigcup \\
                               k\geq 0 $}
\!\!\!{\cal D}(M)_{p}^{(k)}\;\;\;
$$
be  the coardical filtration of ${\cal D}(M)_p$.

The smooth structure on ${\cal D}(M)$ is related to the smooth 
structures of
the $k$-th order co-jet vector bundles over $M$ in the following way. 
For $k\geq 0$ we define the subset
$$
{\cal T}^{(k)}(M) \equiv \parbox{30pt}{\scriptsize $
                               \hspace{7pt} \bigcup \\
                               p\in M $} \!\!
{\cal D}(M)_p^{(k)}x \;\subset\;{\cal D}(M) \;\;\;,
$$
and the projection
$$
\pi^{(k)} : {\cal T}^{(k)}(M) \supset {\cal D}(M)_p^{(k)} \ni \mu
\longrightarrow p \in M\;\;\;.
$$
Let  $\left\{(U_{\alpha}, \vp_{\alpha}) \right\}_{\alpha \in I} $ 
be an admissible atlas  on $M$ and
$\left\{(U_{\a}, \vp_{\a*}) \right\}_{\alpha \in I} $ 
the corresponding 
$(X\oplus \{o\})$-atlas on ${\cal D}(M)$. 
For each $k\geq 0$, the collection 
$\left\{(U_{\alpha}, \tau^{(k)}_{\alpha}) \right\}_{\alpha \in I}$
where
\begin{equation}
\label{trivi}
\tau^{(k)}_{\alpha} : (\pi^{(k)})^{-1}(U_{\alpha})\ni \mu 
\longrightarrow
{\vp}_{\alpha*}(\mu) \in U_{\alpha} \times S^{(k)}(X)\;\;\;,
\nonumber
\end{equation}
is a trivializing covering of $\pi^{(k)} : {\cal T}^{(k)}(M) \rightarrow
M$ (we are using the terminology of \cite{Lang}).
One easily verifies that compatible atlases on ${\cal D}(M)$ yield compatible
trivializing coverings and
that $\pi^{(k)} : {\cal T}^{(k)}(M) \rightarrow M$
acquires the structure of a smooth vector bundle over $M$ with standard
fibre $S^{(k)}(X)$.

In the case of a finite-dimensional Fr\'{e}chet space $X\approx \Rn$ the bundle
$\pi^{(k)} : {\cal T}^{(k)}(M) \rightarrow M$ constructed above is called 
the $k$-th order co-jet bundle over $M$ and is dual to the bundle of $k$-th
order jets on $M$.

In the case of an infinite-dimensional Fr\'{e}chet space $X$, for $k\geq 2$ the standard
fibre $S^{(k)}(X)$ is not complete with respect to the direct sum and the projective
tensor product topologies on
$$
S^{(k)}(X) = \parbox{30pt}{\scriptsize $ \hspace{8pt} k\\
                               \makebox[2pt]{} \bigoplus \\
                               i=0 $} S^i(X)\;\;\;,
$$
and $S^i(X)$, respectively. Since the transition maps of the trivializing covering
(\ref{trivi}) are continuous with respect to this topology, the bundle
$\pi^{(k)} : {\cal T}^{(k)}(M) \rightarrow M$ admits a unique extension to a bundle
with a complete standard fibre. Let us stress that in the present coalgebraic
approach this completion will not be used.\bigskip

\noindent {\bf Remark 4.3.3}
The construction discussed in the previous remark applies also to the 
subset of all primitive elements of the coalgebra ${\cal D}(M)$. 
Let ${\cal T}_p(M)$ be the space of all primitive, 
with respect to  $p\in M$, elements of ${\cal D}(M)$. 
For each $p\in M$ one has the invariant decomposition 
$$
{\cal D}(M)^{(1)}_p = \R\oplus {\cal T}_p(M)\;\;\;.
$$
Let us introduce the set
$$
{\cal T}(M) \equiv \parbox{30pt}{\scriptsize $
                               \hspace{7pt} \bigcup \\
                               p\in M $} \!\!
{\cal T}_p(M) \; \;\;,
$$
with the projection $\pi : {\cal T}(M) \ra M$ given by 
$\pi({\cal T}_p(M))=p$. 
In this case the smooth structure 
on ${\cal D}(M)$ induces the canonical
smooth structure of the tangent bundle of $M$.\bigskip

\noindent Let $M$
be a smooth Fr\'{e}chet manifold,
$({\cal D}(M),\Delta_M, \varepsilon_M )$ the smooth coalgebra of $M$, 
and
$(C^{\infty}(M),M,u)$ the algebra of smooth functions on $M$. For each
$p \in M$ let us consider the pairing
$$
\langle\;.\; ,\;.\; \rangle_p : {\cal D}(M)_p\times C^{\infty}(M)
\longrightarrow \R\;\;\;,
$$
given by 
\begin{equation}
\label{pap}
\langle \mu_p , f\rangle_p \equiv \langle{\vp}_{\alpha*}(\mu_p)
 , f\circ \vp^{-1}_{\alpha}\rangle_{\vp_{\alpha}(U_{\alpha})}\;\;\;,
\end{equation}
where $(U_{\alpha},\vp_{\alpha})$ is an admissible chart on $M$ at $p$.
One easily verifies that the definition above is independent of the choice
of an admissible chart at $p$. Extending formula (\ref{pap}) by linearity
in the first variable one gets the pairing
\begin{equation}
\label{pa}
\langle\;.\; , \;.\;\rangle_M : {\cal D}(M)\times C^{\infty}(M)
\longrightarrow \R\;\;\;.
\end{equation}
 Using  Rem.3.4.1 and Prop.3.4.2 one gets 

\begin{PP}
Let ${\cal D}(M)$ be the smooth coalgebra of a Fr\'{e}chet manifold
$M$ and $C^{\infty}(M)$ the algebra of smooth functions on $M$.
\begin{enumerate}
\item
For all $f,g \in C^{\infty}(M)$, $\mu \in {\cal D}(M)$ 
\begin{eqnarray*}
\langle \mu, f\cdot g\rangle_M &=& 
\sum\limits_{(\mu)} \langle \mu_{(1)}, f\rangle_M 
\langle \mu_{(2)}, g\rangle_M\;\;\;,\\
\langle \mu, 1\rangle_M & = & \varepsilon_M(\mu )\;\;\;,
\end{eqnarray*}
where $\Delta_M\mu = \sum\limits_{(\mu)} \mu_{(1)}\otimes \mu_{(2)}$.
\item
Let $\phi :M\rightarrow N$ be a smooth map of  Fr\'{e}chet manifolds.
Then
$$
\langle \phi^*\mu, f\rangle_N = \langle \mu, f\circ \phi\rangle_M\;\;\;,
$$
for all $\mu \in {\cal D}(M), f \in C^{\infty}(N)$.
\end{enumerate}
\end{PP}
As in the case of open even coalgebras one can show that the pairing 
(\ref{pa}) is nonsingular. Then 
 the proposition above implies the following
\begin{TT}
The map
$$
{\cal D}(M)\ni \mu \lra \langle\mu,\;.\;\rangle \in 
C^{\infty}(M)^{\circ}\;\;\;
$$
is an injective morphism of $\ZZ$-graded coalgebras.
\end{TT}

\subsection{Subcategory of supermanifolds}

We define the category ${\bf SC}^<$ as a subcategory of ${\bf SC}$
consisting of all smooth coalgebras modelled on finite-dimensional
$\ZZ$-graded Fr\'{e}chet spaces and all ${\bf SC}$-morphisms
between them. By definition ${\bf SC}^<$ is a full subcategory
of ${\bf SC}$ i.e.
for all ${\cal M}, {\cal N} \in {\rm O}{\bf SC}^<$ 
$$
{\rm M}{\bf SC}^<({\cal M}, {\cal N}) =
{\rm M}{\bf SC}({\cal M}, {\cal N})\;\;\;.
$$
By Th.4.2.1 ${\bf SC}^<$  inherits the direct product from ${\bf SC}$.
In this subsection we shall construct an equivalence of
the category of BLK supermanifolds ${\bf SM}$ introduced in
Subsect.2.4 with the category of finite-dimensional smooth coalgebras 
${\bf SC}^<$.\bigskip

\noindent
Let ${\cal A}_M$ be a supermanifold. For each $(m,n)$-atlas 
$\{ ( U_{\a},F_{\a}) \}_{\a \in I}$ on ${\cal A}_M$  
the family of maps  $\{ F_{\a\b} \}_{\a \in I}$
given for all $\a\in I$, $\b\in I(\a)$ by  
\begin{equation}
\label{smcoc}
F_{\a\b}\;=\; F_{\b}\circ 
{F_{\a}^{-1}} :
{\cal S}^{n,m}_{F^0_{\a}(U_{\a} \cap U_{\b})} \lra
{\cal S}^{n,m}_{F^0_{\b}(U_{\a} \cap U_{\b})}
\;\;\;.
\end{equation}
is an 
 $(m,n)$-cocycle of  ${\bf sm}$-transition maps over the
smooth atlas $\{ ( U_{\a},F_{\a}^0) \}_{\a \in I}$ on
the underlying manifold $M$. Applying the functor of Prop.3.4.4
to the cocycle (\ref{smcoc}) one gets $(\Rm\!\oplus\!\Rn)$-cocycle
of ${\bf sc}$-transition maps 
\begin{equation}
\label{sccoc}
F_{\a\b*}\;=\; (F_{\b}\circ 
{F_{\a}^{-1}})_* : 
{\cal D}_{m,n}(F^0_{\a}(U_{\a} \cap U_{\b}))\lra
{\cal D}_{m,n}(F^0_{\b}(U_{\a} \cap U_{\b}))
\;\;\;,
\end{equation}
over the same smooth atlas on $M$. By Prop.4.1.3 the
cocycle (\ref{sccoc}) generates a unique smooth coalgebra 
${\cal D}({\cal A}_M)$ and the admissible $(\Rm\!\oplus\!\Rn)$-atlas
$\{ (U_{\a},F_{\a*}) \}_{\a \in I}$
on ${\cal D}({\cal A}_M)$.   
One easily verifies that different $(m,n)$-atlases on ${\cal A}_M$
lead by the above construction to compatible
$(\Rm\!\oplus\!\Rn)$-atlases on the same smooth
coalgebra ${\cal D}({\cal A}_M)$. Applying the construction
to the maximal $(m,n)$-atlas on ${\cal A}_M$, we define
for each chart $( U,F)$ on ${\cal A}_M$ the
corresponding  admissible chart $(U,F_*)$ on
${\cal D}({\cal A}_M)$.

\begin{DD}
The smooth coalgebra ${\cal D}({\cal A}_M)$ 
generated by the cocycle {\rm  (\ref{sccoc})} is called
the smooth coalgebra of the supermanifold ${\cal A}_M$.
\end{DD}

\noindent 
Reversing the construction above and using the reconstruction theorem
for supermanifolds
(Prop.2.3.3) one can show that the correspondence
$$
{\rm O}{\bf SM} \ni {\cal A}_M \lra {\cal D}({\cal A}_M)\in
{\rm O}{\bf SC}^<
$$
is bijective. Moreover for each admissible chart 
$( U,\Phi)$ on ${\cal D}({\cal A}_M)$ there exists
a chart $( U,F)$ on ${\cal A}_M$ such that $\Phi=F_*$.
\bigskip

\noindent
Let $F =(F^0,F_.):{\cal A}_M \ra {\cal B}_N$ 
be a morphism of supermanifolds. 
Then for each
$p \in M$ there are  charts $(U_{\alpha}, F_{\alpha})$ on
${\cal A}_M$
and $(V_{\g}, G_{\g})$ on ${\cal B}_N$
 such that $p \in U_{\alpha}$,  $F^0(U_{\alpha})\subset V_{\g}$,
and the composition 
$$
G_{\g}\circ F\circ F_{\alpha}^{-1} : 
{\cal S}_{m,n}(F^0_{\alpha}(U_{\alpha})) 
\lra 
{\cal S}_{m,n}(G^0_{\g}(V_{\g}))
$$
is a  morphism in the model category ${\bf sm}$.

Let ${\cal D}({\cal A}_M)_p$ be the irreducible component of
 ${\cal D}({\cal A}_M)$  containing the group-like element $p\in M$.
For all $\mu \in {\cal D}({\cal A}_M)_p$ we define
\begin{equation}
\label{sstar}
F_{*p}(\mu) \equiv G^{-1}_{\g*} \circ (
G_{\g} \circ F \circ F_{\alpha}^{-1})_* \circ 
F_{\alpha*}(\mu)
\;\;\;.
\end{equation}
By Prop.3.4.4 the definition above is independent of the choice 
of  admissible charts at $p \in M$ and $\Phi^0(p) \in N$.  
Extending formula (\ref{sstar}) by linearity in $p$  one gets
the smooth morphism of graded coalgebras
$$
F_* : {\cal D}({\cal A}_M) \longrightarrow  
{\cal D}({\cal B}_N)\;\;\;,
$$
with the underlying part $(F_*)^0 = F^0$.
As a consequence of Prop.3.4.4 one gets

\begin{TT}
The correspondence
\begin{eqnarray}
{\rm O}{\bf SM} \ni {\cal A}_M &\longrightarrow & 
{\cal D}({\cal A}_M) \in {\rm O}{\bf SC}^<\;\;\;,\label{fsmsc}\\
{\rm O}{\bf SM}({\cal A}_M,{\cal B}_N) \ni F &
\longrightarrow &  F_* 
 \in {\rm M}{\bf SC}^<(
{\cal D}({\cal A}_M),{\cal D}({\cal B}_N))\;\;\;\nonumber
\end{eqnarray}
is  an equivalence of categories ${\bf SM}$ and ${\bf SC}^<$. 
\end{TT}

\noindent
It follows that the category of BLK supermanifolds
can be identified with 
the full subcategory of the category of 
finite-dimensional smooth coalgebras. Since by Th.4.3.1
the category of Fr\'{e}chet manifolds can be identified with
subcategory of even smooth coalgebras, ${\bf SC}$ provides
a correct  extension of both categories.
\bigskip

\noindent{\bf Remark 4.4.1} As in the case of category ${\bf FM}$
of Fr\'{e}chet manifolds (Rem.4.3.2, Rem.4.3.3) the smooth coalgebra
of a supermanifold admits the geometrical interpretation in
terms of co-jet vector superbundles \cite{Batchelor}.
Since the constructions of these bundles requires some
techniques of algebraic geometry \cite{Ruiperez} not used
in the present paper, we refer  to the original paper
\cite{Batchelor} for the discussion of this point.
\bigskip

\noindent Let 
$({\cal D}({\cal A}_M),\Delta_M, \varepsilon_M )$ be
the smooth coalgebra of a supermanifold ${\cal A}_M$, and
$({\cal A}(M),M,u)$ the algebra of superfunctions on ${\cal A}_M$. 
For each
$p \in M$ the pairing
$$
\langle\;.\; ,\;.\; \rangle_p : {\cal D}({\cal A}_M)_p\times
 {\cal A}(M)
\longrightarrow \R\;\;\;,
$$
is defined  by 
\begin{equation}
\label{spap}
\langle \mu , f\rangle_p \equiv \langle
F_{\alpha*}(\mu)
 ,  (F^{-1}_{\alpha})_{U_{\a}}(f_{|U_{\a}})
\rangle_{F^0_{\alpha}(U_{\alpha})}\;\;\;,
\end{equation}
where $(U_{\alpha},F_{\alpha})$ is a chart on ${\cal A}_M$ at $p$.
By Prop.3.4.4 the definition above is independent of the choice
of a chart at $p$. Extending formula (\ref{spap}) by linearity
 one gets the pairing
\begin{equation}
\label{spa}
\langle\;.\; , \;.\; \rangle_M : {\cal D}({\cal A}_M)\times {\cal A}(M)
\longrightarrow \R\;\;\;.
\end{equation}
As a consequence of Prop.3.4.4 and Prop.3.4.5  one has

\begin{PP} Let ${\cal D}({\cal A}_M)$ be the smooth coalgebra of a 
supermanifold ${\cal A}_M$.
\begin{enumerate}
\item
For all $f,g \in {\cal A}(M)_0 \cup{\cal A}(M)_1$, 
$\mu \in {\cal D}({\cal A}_M)$ 
\begin{eqnarray*}
\langle \mu, f\cdot g\rangle_M &=& 
\sum\limits_{(\mu)} (-1)^{|f|\cdot |\mu_{(2)}|}
 \langle \mu_{(1)}, f\rangle_M 
\langle \mu_{(2)}, g\rangle_M\;\;\;,\\
\langle \mu, 1\rangle_M & = & \varepsilon_M(\mu )\;\;\;,
\end{eqnarray*}
where $\Delta_M\mu = \sum\limits_{(\mu)} \mu_{(1)}\otimes \mu_{(2)}$.
\item
Let $F =(F^0,F_.):{\cal A}_M \ra {\cal B}_N$ 
be a morphism of supermanifolds. 
Then
$$
\langle F_*\mu, f\rangle_N = \langle \mu, F_N f
\rangle_M\;\;\;,
$$
for all $\mu \in {\cal D}({\cal A}_M), f \in {\cal B}(N)$.
\end{enumerate}
\end{PP}
Using Prop.A.5.6 and the above proposition one obtains
the following global  version of Prop.3.4.6
\cite{Kostant,Batchelor}

\begin{TT}
The map 
$$
{\cal D}({\cal A}_M) \ni \omega \lra 
\langle\omega,\: . \:\rangle_M \in {\cal A}(M)^{\circ}
$$
is an isomorphism of $\ZZ$-graded coalgebras.
\end{TT}

\noindent  Each superfunction
$f$ on a supermanifold ${\cal A}_M$ can be identified via the pairing
(\ref{spa}) with a superfunction $\langle \:.\:, f \rangle_M$ on
${\cal D}({\cal A}_M)$. Assuming  this identification one has 

\begin{PP}
The functor of {\rm Prop.4.1.6} restricted to the subcategory 
${\bf SC}^{<}$ of finite-dimensional smooth coalgebras
\begin{eqnarray*}
{\rm O}{\bf SC}^< \ni {\cal M} &\lra & 
{\cal M}_M^{\wedge}=(M,{\cal M}(\: .\:)^{\wedge})
\in {\rm O}{\bf SM}
\;\;\;,\\
{\rm M}{\bf SC}^< \ni \Phi &\lra & \Phi^* = (\Phi^0,\Phi^*_.)
\in {\rm M}{\bf SM}\;\;\;,
\end{eqnarray*}
is an equivalence of categories  and  the  inverse to the functor
of {\rm Th.4.4.1}.
\end{PP}

\section*{Appendix}
\renewcommand{\thesection}{A}
\setcounter{subsection}{0}

\subsection{Graded spaces}

A {\it $\ZZ$-graded space} is a vector space $V$ with distinguished
subspaces $V_0, V_1$ such that $V$ is the direct sum
$V_0 \oplus V_1$ of $V_0$ and $V_1$.
$V_0, V_1$ are called the {\it even} and the {\it odd part} of $V$,
respectively. Similarly, an  element $v\in V$ is called {\it even} if
$v\in V_0$ and {\it odd} if $v\in V_1$.  Any element $v\in V$
has a unique representation as a sum $v=v_0 + v_1$ of its
{\it even} $v_0\in V_0$ and {\it odd} $v_1\in V_1$ {\it components}.
An element $v\in V_0 \cap V_1$ is called {\it homogeneous}.
If $v\in V_i,v\neq 0$ the parity $|v|$ of a homogeneous element
$v$ is defined by $|v|=i \in \ZZ$.\bigskip

\noindent
Let $V, W$ be $\ZZ$-graded spaces. The space ${\rm Hom}(V,W)$
of all linear maps from $V$ to $W$ gets the natural grading
$$
{\rm Hom}(V,W) = {\rm Hom}(V,W)_0 \oplus {\rm Hom}(V,W)_1\;\;\;,
$$
where $f\in {\rm Hom}(V,W)_i$ if $f(V_j) \subset W_{i+j}$.
A {\it morphism of $\ZZ$-graded spaces} is an even linear map
(i.e. an element of ${\rm Hom}(V,W)_0$).\bigskip

\noindent
A {\it graded subspace} $W\subset V$ of a $\ZZ$-graded space $V$
is a vector subspace of $V$ with the $\ZZ$-grading given by
$W = (W\cap V_0) \oplus (W\cap V_1)$.\bigskip

\noindent
The {\it direct sum $V\oplus W$  of $\ZZ$-graded spaces} $V, W$ is the
direct sum $V\oplus W$ of vector spaces $V, W$ with
the $\ZZ$-grading
$$
(V\oplus W)_0 = V_0 \oplus W_0\;\;\;,\;\;\;
(V\oplus W)_1 = V_1 \oplus W_1\;\;\;.
$$

\noindent
The {\it tensor product $V\otimes W$ of $\ZZ$-graded spaces} $V, W$ is
the tensor product $V\otimes W$ of vector spaces with the $\ZZ$-grading
$$
(V\otimes W)_k =
               \parbox[c]{30pt}{\scriptsize $\makebox(5,13)[b]{ } \bigoplus  \\
                                i+j =k                           $}
 V_i \otimes W_j\;\;\;.
$$

\noindent
Let $V, W$be $\ZZ$-graded spaces. The {\it twisting morphism}
$T: V\otimes W \rightarrow W\otimes V$ is a morphisms of graded spaces
defined by
$$
T( v \otimes w ) = (-1)^{|v||w|} w\otimes v
$$
for all $ v\in V_0 \cap V_1$, $w \in W_0 \cap W_1$.

\subsection{Graded algebras}

\begin{DD}
A triple
 $({\cal A}, \mu, u)$ where ${\cal A}$ is a $\ZZ$-graded space
 and $\mu, u$ are morphisms of $\ZZ$-graded spaces
$$
\begin{array}{rlll}
\mu &:& {\cal A}\otimes {\cal A}
\longrightarrow {\cal A} & {\it (multiplication)}\\
u &:& \R \longrightarrow {\cal A} & {\it (unit)}
\end{array}
$$
is called a {\it $\ZZ$-graded algebra} if the diagrams

\begin{center}
\begin{picture}(220,100)(0,10)
\put(40,93){\makebox(40,10){$
%
   {\cal A}\otimes {\cal A} \otimes {\cal A}
$}}
\put(5,75){\makebox(20,10)[r]{\small $
%
 \mu \otimes {\rm id}
$}}
\put(95,75){\makebox(20,10)[l]{$
%
 {\rm id}\otimes \mu
$}}
\put(-20,48){\makebox(20,10)[br]{\small$
%
 {\cal A}\otimes {\cal A}
$}}
\put(120,48){\makebox(20,10)[bl]{$
%
   {\cal A}\otimes {\cal A}
$}}
\put(5,18){\makebox(20,10)[tr]{\small$
%
 \mu
$}}
\put(95,18){\makebox(20,10)[tl]{\small$
%
 \mu
$}}
\put(50,2){\makebox(20,10)[t]{\small$
%
 {\cal A}
$}}
\put(170,48){\makebox(80,10)[bl]{\it
(associativity)}}
\put(50,90){\vector(-3,-2){48}}
\put(70,90){\vector(3,-2){48}}
\put(3,45){\vector(3,-2){48}}
\put(117,45){\vector(-3,-2){48}}
\end{picture}\bigskip

\begin{picture}(220,100)(0,10)
\put(40,93){\makebox(40,10){$
%
   {\cal A}\otimes {\cal A}
$}}
\put(5,75){\makebox(20,10)[r]{\small $
%
 u\otimes {\rm id}
$}}
\put(95,75){\makebox(20,10)[l]{$
%
 {\rm id}\otimes u
$}}
\put(-20,48){\makebox(20,10)[br]{\small$
%
\R \otimes {\cal A}
$}}
\put(120,48){\makebox(20,10)[bl]{$
%
  {\cal A} \otimes \R
$}}
\put(5,18){\makebox(20,10)[tr]{\small$
%
 \equiv
$}}
\put(95,18){\makebox(20,10)[tl]{\small$
%
 \equiv
$}}
\put(65,50){\makebox(20,10)[l]{\small$
%
 \mu
$}}
\put(50,2){\makebox(20,10)[t]{\small$
%
 {\cal A}
$}}
\put(170,48){\makebox(20,10)[bl]{\it
(unitarity)}}
\put(20,70){\line(-3,-2){15}}
\put(20,70){\vector(3,2){30}}
\put(100,70){\line(3,-2){15}}
\put(100,70){\vector(-3,2){30}}
\put(20,35){\vector(-3,2){15}}
\put(20,35){\vector(3,-2){30}}
\put(100,35){\vector(3,2){15}}
\put(100,35){\vector(-3,-2){30}}
\put(60,40){\line(0,1){50}}
\put(60,40){\vector(0,-1){23}}
\end{picture}
\end{center}

\noindent commute.
\end{DD}

\noindent
A $\ZZ$-graded algebra $A$ is called {\it commutative} if the
following diagram commutes.

\begin{center}
\begin{picture}(120,70)(0,10)
\put(40,63){\makebox(40,10){$
%
   {\cal A}
$}}
\put(10,35){\makebox(20,10)[r]{\small $
%
 \mu
$}}
\put(90,35){\makebox(20,10)[l]{$
%
\; \mu
$}}
\put(3,8){\makebox(20,10)[br]{$
%
{\cal A} \otimes {\cal A}
$}}
\put(98,8){\makebox(20,10)[bl]{$
%
  {\cal A} \otimes {\cal A}
$}}
\put(57,16){\makebox(10,10)[b]{$
%
  T
$}}
\put(30,12){\vector(1,0){60}}
\put(25,30){\line(-1,-1){10}}
\put(25,30){\vector(1,1){27}}
\put(95,30){\line(1,-1){10}}
\put(95,30){\vector(-1,1){27}}
\end{picture}.
\end{center}\bigskip

\noindent
A {\it morphism of $\ZZ$-graded algebras} $F:A \rightarrow B$ is
a morphism of graded spaces such that the diagrams
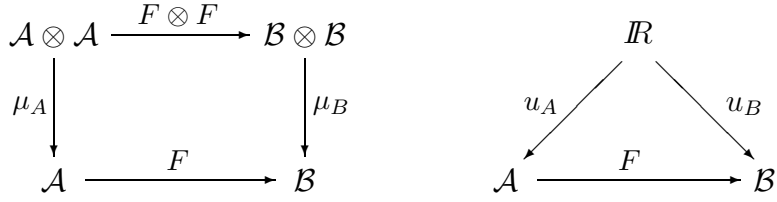
\begin{center}
\begin{picture}(160,70)(0,10)
\put(-17,63){\makebox(40,10)[b]{$
%
   {\cal A} \otimes {\cal A}
$}}
\put(78,63){\makebox(40,10)[b]{$
%
   {\cal B} \otimes {\cal B}
$}}
\put(-19,35){\makebox(20,10)[r]{\small$
%
 \mu_A
$}}
\put(98,35){\makebox(20,10)[l]{\small$
%
\; \mu_B
$}}
\put(-7,8){\makebox(20,10)[b]{$
%
{\cal A}
$}}
\put(88,8){\makebox(20,10)[b]{$
%
  {\cal B}
$}}
\put(45,16){\makebox(10,10)[b]{\small$
%
  F
$}}
\put(45,71){\makebox(10,10)[b]{\small$
%
  F\otimes F
$}}
\put(45,67){\vector(1,0){32}}
\put(45,67){\line(-1,0){20}}
\put(45,12){\vector(1,0){42}}
\put(45,12){\line(-1,0){30}}
\put(3,30){\vector(0,-1){9}}
\put(3,30){\line(0,1){28}}
\put(98,30){\vector(0,-1){9}}
\put(98,30){\line(0,1){28}}
\end{picture}
\begin{picture}(120,70)(0,10)
\put(40,63){\makebox(40,10){$
%
   \R
$}}
\put(10,35){\makebox(20,10)[r]{\small $
%
 u_A
$}}
\put(90,35){\makebox(20,10)[l]{$
%
\; u_B
$}}
\put(0,8){\makebox(20,10)[b]{$
%
{\cal A}
$}}
\put(98,8){\makebox(20,10)[b]{$
%
  {\cal B}
$}}
\put(52,16){\makebox(10,10)[b]{\small$
%
  F
$}}
\put(52,12){\vector(1,0){47}}
\put(52,12){\line(-1,0){30}}
\put(27,30){\vector(-1,-1){10}}
\put(27,30){\line(1,1){27}}
\put(94,30){\vector(1,-1){10}}
\put(94,30){\line(-1,1){27}}
\end{picture}
\end{center}
commute. \bigskip

\noindent
The {\it tensor product of $\ZZ$-graded algebras} $(A,\mu_A, u_A)$,
$(B,\mu_B,u_B)$is the tensor product $A\otimes B$ of $\ZZ$-graded spaces
with the $\ZZ$-graded algebra structure given by
\begin{eqnarray*}
\mu_{A\otimes B}& :& A\otimes B\otimes A\otimes B
{ \begin{picture}(70,20)(0,0) \put(5,3){\vector(1,0){60}}
\put(15,10){$\scriptstyle
{\rm id}_A \otimes T \otimes {\rm id}_B
$ }
\end{picture}}
A\otimes A\otimes B\otimes B
{ \begin{picture}(60,20)(0,0) \put(5,3){\vector(1,0){50}}
\put(15,10){$\scriptstyle
\mu_A \otimes \mu_B
$ }
\end{picture}} A\otimes B  \;\;\;,   \\
u_{A\otimes B}&:& \R \equiv \R\otimes \R
{ \begin{picture}(60,20)(0,0) \put(5,3){\vector(1,0){50}}
\put(15,10){$\scriptstyle
u_A \otimes u_B
$ }
\end{picture}}
A\otimes B                        \;\;\;.
\end{eqnarray*}

\begin{DD}
A bigraded algebra $A$ is a $\ZZ$-graded algebra with a
$\Z_+$-grading
$
A =
 \parbox[c]{30pt}{\scriptsize
$\makebox(4,4)[b]{ } \infty \\
\makebox(1,10)[b]{ } \bigoplus  \\
                                i =1 $}
\!\!\!\!A^i
$
such that:
\begin{enumerate}
\item for every $i\in \Z_+$, $A^i$ is a $\ZZ$-graded subspace of $A$;
\item $u(\R)\subset A^0$;
\item for every $i,j \in \Z_+$, $\mu (A^i\otimes A^j) \subset
A^{i+j}$.
\end{enumerate}
\end{DD}

\subsection{Graded coalgebras}
\begin{DD}
A triple
 $({\cal C}, \Delta, \varepsilon)$ where ${\cal C}$ is a $\ZZ$-graded space
 and $\Delta, \varepsilon$ are morphisms of $\ZZ$-graded spaces
$$
\begin{array}{rlll}
\Delta &:& {\cal C} \longrightarrow {\cal C}\otimes {\cal C} &
{\it (comultiplication)}\\
\varepsilon &:& {\cal C} \longrightarrow \R  & {\it (counit)}
\end{array}
$$
is called a {\it $\ZZ$-graded coalgebra} if the diagrams

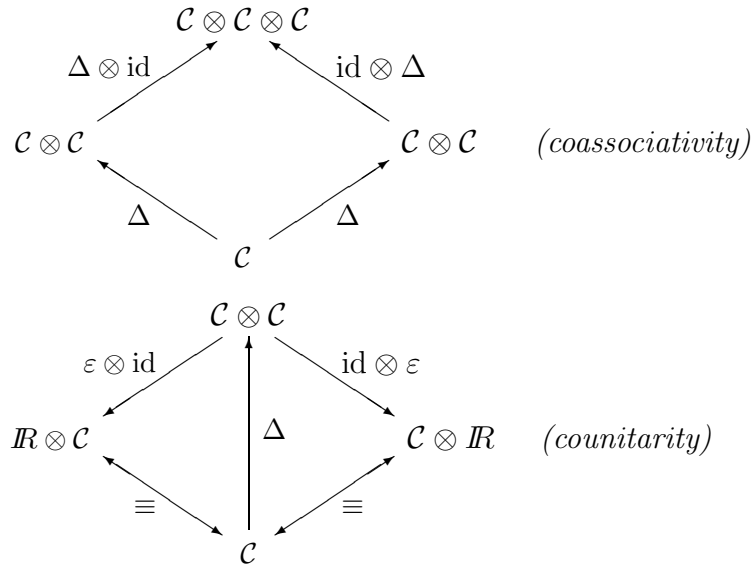
\begin{center}
\begin{picture}(220,100)(0,10)
\put(40,93){\makebox(40,10){$
%
   {\cal C}\otimes {\cal C} \otimes {\cal C}
$}}
\put(5,75){\makebox(20,10)[r]{\small $
%
 \Delta \otimes {\rm id}
$}}
\put(95,75){\makebox(20,10)[l]{$
%
 {\rm id}\otimes \Delta
$}}
\put(-20,48){\makebox(20,10)[br]{\small$
%
{\cal C} \otimes {\cal C}
$}}
\put(120,48){\makebox(20,10)[bl]{$
%
  {\cal C} \otimes {\cal C}
$}}
\put(5,18){\makebox(20,10)[tr]{\small$
%
 \Delta
$}}
\put(95,18){\makebox(20,10)[tl]{\small$
%
 \Delta
$}}
\put(50,2){\makebox(20,10)[t]{\small$
%
 {\cal C}
$}}
\put(170,48){\makebox(20,10)[bl]{\it
(coassociativity)}}
\put(20,70){\line(-3,-2){15}}
\put(20,70){\vector(3,2){30}}
\put(100,70){\line(3,-2){15}}
\put(100,70){\vector(-3,2){30}}
\put(20,35){\vector(-3,2){15}}
\put(20,35){\line(3,-2){30}}
\put(100,35){\vector(3,2){15}}
\put(100,35){\line(-3,-2){30}}
\end{picture} \bigskip

\begin{picture}(220,100)(0,10)
\put(40,93){\makebox(40,10){$
%
   {\cal C}\otimes {\cal C}
$}}
\put(5,75){\makebox(20,10)[r]{\small $
%
 \varepsilon \otimes {\rm id}
$}}
\put(95,75){\makebox(20,10)[l]{$
%
 {\rm id}\otimes \varepsilon
$}}
\put(-20,48){\makebox(20,10)[br]{\small$
%
\R \otimes {\cal C}
$}}
\put(120,48){\makebox(20,10)[bl]{$
%
  {\cal C} \otimes \R
$}}
\put(5,18){\makebox(20,10)[tr]{\small$
%
 \equiv
$}}
\put(95,18){\makebox(20,10)[tl]{\small$
%
 \equiv
$}}
\put(65,50){\makebox(20,10)[l]{\small$
%
 \Delta
$}}
\put(50,2){\makebox(20,10)[t]{\small$
%
 {\cal C}
$}}
\put(170,48){\makebox(20,10)[bl]{\it
(counitarity)}}
\put(20,70){\vector(-3,-2){15}}
\put(20,70){\line(3,2){30}}
\put(100,70){\vector(3,-2){15}}
\put(100,70){\line(-3,2){30}}
\put(20,35){\vector(-3,2){15}}
\put(20,35){\vector(3,-2){30}}
\put(100,35){\vector(3,2){15}}
\put(100,35){\vector(-3,-2){30}}
\put(60,40){\vector(0,1){50}}
\put(60,40){\line(0,-1){23}}
\end{picture}
\end{center}

\noindent commute
\end{DD}

\noindent
A $\ZZ$-graded coalgebra ${\cal C}$ is called {\it cocommutative} if the
following diagram commutes.

\begin{center}
\begin{picture}(120,70)(0,10)
\put(40,63){\makebox(40,10){$
%
   {\cal C}
$}}
\put(10,35){\makebox(20,10)[r]{\small $
%
 \Delta
$}}
\put(90,35){\makebox(20,10)[l]{$
%
\; \Delta
$}}
\put(3,8){\makebox(20,10)[br]{$
%
{\cal C} \otimes {\cal C}
$}}
\put(98,8){\makebox(20,10)[bl]{$
%
  {\cal C} \otimes {\cal C}
$}}
\put(57,16){\makebox(10,10)[b]{$
%
  T
$}}
\put(30,12){\vector(1,0){60}}
\put(25,30){\vector(-1,-1){10}}
\put(25,30){\line(1,1){27}}
\put(95,30){\vector(1,-1){10}}
\put(95,30){\line(-1,1){27}}
\end{picture}.
\end{center}\bigskip

\noindent
A {\it morphism of $\ZZ$-graded coalgebras} $\Phi:{\cal D} \rightarrow
{\cal D}$ is
a morphism of graded spaces such that the diagrams
\begin{center}
\begin{picture}(160,70)(0,10)
\put(-17,63){\makebox(40,10)[b]{$
%
   {\cal C} \otimes {\cal C}
$}}
\put(78,63){\makebox(40,10)[b]{$
%
   {\cal D} \otimes {\cal D}
$}}
\put(-19,35){\makebox(20,10)[r]{\small$
%
 \Delta_C
$}}
\put(98,35){\makebox(20,10)[l]{\small$
%
\; \Delta_D
$}}
\put(-7,8){\makebox(20,10)[b]{$
%
{\cal C}
$}}
\put(88,8){\makebox(20,10)[b]{$
%
  {\cal D}
$}}
\put(45,16){\makebox(10,10)[b]{\small$
%
  \Phi
$}}
\put(45,71){\makebox(10,10)[b]{\small$
%
  \Phi \otimes \Phi
$}}
\put(45,67){\vector(1,0){32}}
\put(45,67){\line(-1,0){20}}
\put(45,12){\vector(1,0){42}}
\put(45,12){\line(-1,0){30}}
\put(3,30){\line(0,-1){9}}
\put(3,30){\vector(0,1){28}}
\put(98,30){\line(0,-1){9}}
\put(98,30){\vector(0,1){28}}
\end{picture}
\begin{picture}(120,70)(0,10)
\put(40,63){\makebox(40,10){$
%
   \R
$}}
\put(10,35){\makebox(20,10)[r]{\small $
%
 \varepsilon_C
$}}
\put(90,35){\makebox(20,10)[l]{$
%
\; \varepsilon_D
$}}
\put(0,8){\makebox(20,10)[b]{$
%
{\cal C}
$}}
\put(98,8){\makebox(20,10)[b]{$
%
  {\cal D}
$}}
\put(52,16){\makebox(10,10)[b]{\small$
%
  f
$}}
\put(52,12){\vector(1,0){47}}
\put(52,12){\line(-1,0){30}}
\put(27,30){\line(-1,-1){10}}
\put(27,30){\vector(1,1){27}}
\put(94,30){\line(1,-1){10}}
\put(94,30){\vector(-1,1){27}}
\end{picture}
\end{center}
commute. \bigskip

\noindent
The {\it tensor product of $\ZZ$-graded coalgebras}
$({\cal C},\Delta_C, \varepsilon_C)$,
$({\cal D},\Delta_D,\varepsilon_D)$
is the tensor product ${\cal C} \otimes {\cal D}$ of $\ZZ$-graded spaces
with the $\ZZ$-graded coalgebra structure given by
\begin{eqnarray*}
\mu_{ C\otimes D}& :& {\cal C}\otimes {\cal D}\otimes
{\cal C}\otimes {\cal D}
{ \begin{picture}(70,20)(0,0) \put(5,3){\vector(1,0){60}}
\put(15,10){$\scriptstyle
{\rm id}_C \otimes T \otimes {\rm id}_D
$ }
\end{picture}}
{\cal C}\otimes {\cal C}\otimes {\cal D}\otimes {\cal D}
{ \begin{picture}(60,20)(0,0) \put(5,3){\vector(1,0){50}}
\put(15,10){$\scriptstyle
\mu_C \otimes \mu_D
$ }
\end{picture}} {\cal C}\otimes {\cal D}  \;\;\;,   \\
u_{A\otimes B}&:& \R \equiv \R\otimes \R
{ \begin{picture}(60,20)(0,0) \put(5,3){\vector(1,0){50}}
\put(15,10){$\scriptstyle
u_C \otimes u_D
$ }
\end{picture}}
{\cal C}\otimes {\cal D}                 \;\;\;.
\end{eqnarray*}

\noindent
Let ${\cal C}$ be a $\ZZ$-graded coalgebra. For any $c\in {\cal C}$,
$\Delta c$ can be written as a (non unique) sum of simple tensors.
It is convenient to use the following so called {\it sigma notation}
$$
\Delta c = \sum\limits_{(c)} c_{(1)}\otimes c_{(2)}
\;\;\;,
$$
where one can assume that all $c_{(1)}, c_{(2)}$ are homogeneous.
Similarly, by the coassociativity one can write
\begin{eqnarray*}
\Delta^k c &=& (\Delta \otimes
\underbrace{{\rm id}\otimes...\otimes {\rm id}}_{k-1} )
\circ ... \circ \Delta \otimes {\rm id} \circ \Delta\\
&=&\sum\limits_{(c)} c_{(1)}\otimes ... \otimes c_{(k+1)}
\;\;\;.
\end{eqnarray*}

\noindent
A $\ZZ$-graded coalgebra ${\cal C}$ is {\it irreducible} if any of two
non-zero $\ZZ$-graded subcoalgebras have non-zero intersection.
A maximal irreducible $\ZZ$-graded subcoalgebra of  ${\cal C}$
is called an {\it irreducible component of ${\cal C}$}. A $\ZZ$-graded
coalgebra is {\it simple} if it has no non-zero proper
$\ZZ$-graded subcoalgebras. A $\ZZ$-graded coalgebra is {\it pointed}
if all simple $\ZZ$-graded subcoalgebras are 1-dimensional.\bigskip

\noindent
The structure theorem for cocomutative coalgebras \cite{sweedler}
is also valid  in the $\ZZ$-graded case
\cite{Kostant,Batchelor}.

\begin{TT}
Any cocomutative $\ZZ$-graded coalgebra is a direct sum of its
irreducible components.
\end{TT}

\noindent
An element $g$ of a $\ZZ$-graded coalgebra is called {\it group-like}
if $\Delta g = g\otimes g$. We denote the set of all group-like
elements of a $\ZZ$-graded coalgebra ${\cal C}$ by
$G({\cal C})$. For all $g\in G({\cal C})$, $I\!\!R g \subset {\cal C}_0$
is a 1-dimensional simple $\ZZ$-graded subcoalgebra of ${\cal C}$.
When a $\ZZ$-graded cocomutative coalgebra is pointed, each irreducible
component ${\cal C}_g$ of ${\cal C}$ is uniquely determined by
a unique group-like element $g$ contained in ${\cal C}_g$.
Then the direct sum decomposition takes the form
$$
{\cal C} = \parbox[c]{35pt}{\scriptsize $
                            \makebox(5,14)[b]{} \bigoplus \\
                            \makebox(2,1){}g\in G({\cal C}) $  }
           {\cal C}_g \;\;\;.
$$       \bigskip

\noindent
In the case of a pointed irreducible  coalgebra
${\cal C}$ some further structure information is encoded
in so called coradical filtration \cite{sweedler}.
We shall briefly describe the $\ZZ$-graded
version of this construction and theorem
\cite{Kostant,Batchelor}.

\begin{DD}
A filtration of a $\ZZ$-graded coalgebra ${\cal C}$ is a
family $\left\{ {\cal C}^{(k)} \right\}^{\infty}_{k=0}$
of $\ZZ$-graded subcoalgebras such that
\begin{enumerate}
\item For any $k\leq k'$, ${\cal C}^{(k)}$
is a $\ZZ$-graded subcoalgebra of ${\cal C}^{(k')}$.
\item $\displaystyle
{\cal C} =      \parbox[c]{35pt}{   \scriptsize $
                                    \makebox(7,12)[b]{} \bigcup \\
                                     \makebox(2,2){}k\geq 0 $  }
{\cal C}^{(k)}\;$.
\item $\displaystyle \Delta {\cal C}^{(k)} =
\sum\limits_{i=0}^k {\cal C}^{(k-i)} \otimes
{\cal C}^{(i)}$, for all $k\geq 0$.
\end{enumerate}
\end{DD}

\noindent
Let ${\cal C}_g$ be a pointed irreducible $\ZZ$-graded
coalgebra and $g$ its unique group-like element. There is
direct sum decomposition
$$
{\cal C}_g = I\!\!R g \oplus {\cal C}^+_g\;\;\;,
$$
where ${\cal C}^+_g = {\rm ker} \varepsilon_{\cal C}$.
Let $\pi^+_g : {\cal C}_g \rightarrow {\cal C}^+_g$
be the projection on the second factor. We define family
$\left\{ {\cal C}^{(k)}_g \right\}_{k=0}^{\infty}$ of
$\ZZ$-graded subcoalgebras of ${\cal C}_g$:
\begin{eqnarray*}
{\cal C}^{(0)}_g &=& I\!\!R g\;\;\;\;\;,\\
{\cal C}^{(k)}_g &=& {\rm ker} \left(
\parbox{30pt}{\scriptsize $\hspace{3pt} k+1 \\
                              \makebox[2pt]{ } \bigotimes\\
                                $                        }
\pi_g^+ \right) \circ \Delta^k\;\;\;,\;\;\;k\geq1\;\;\;.
\end{eqnarray*}
Since both $\pi^+_g$ and $\Delta^k$ are morphisms of
$\ZZ$-graded spaces, ${\cal C}^{(k)}_g$ is a $\ZZ$-graded
subspace of ${\cal C}_g$ for all $k\geq 1$.

\begin{PP}
If $c\in {\cal C}^{(k)+}_g = {\cal C}^{(k)}_g \cap {\cal C}^+_g$,
then
$$
\Delta c = g\otimes c^1 + c^1 \otimes g + y\;\;\;,
$$
where
$$
y\in \sum\limits_{i=1}^{k-1} {\cal C}^{(i)+}_g
\otimes {\cal C}^{(k-i)+}_g\;\;\;.
$$
\end{PP}

\begin{TT}
The family $\left\{ {\cal C}^{(k)}_g \right\}_{k=0}^{\infty}$
is a filtration of the coalgebra ${\cal C}_g$.
\end{TT}

\noindent
The filtration of the theorem above is called the
{\it coradical filtration}.\bigskip

\noindent
Let $g$ be a group-like element of
a $\ZZ$-graded coalgebra  ${\cal C}$.
An element $p\in {\cal C}$   is
called {\it primitive with respect to $g$}
if
$$
\Delta p = p\otimes g + g\otimes p\;\;\;.
$$
We denote the set of all elements
$p\in {\cal C}$ primitive with respect to $g$ by
$P_g({\cal C})$.  Note that $P_g({\cal C}) \subset
{\cal C}_g$, where ${\cal C}_g$ is the irreducible component
containing $g$. By Prop.A.3.1
$$
{\cal C}^{(1)}_g = I\!\!R g \oplus P_g({\cal C}) \;\;\;.
$$
In particular $P_g({\cal C})$ is a $\ZZ$-graded subspace of ${\cal C}$.
In case of a pointed irreducible $\ZZ$-graded coalgebra ${\cal C}$
we denote by $P({\cal C})$
the space of all primitive elements with respect to a unique
group-like element in ${\cal C}$.

\begin{PP}
 Let ${\cal C}, {\cal D}$ be  $\ZZ$-graded cocommutative coalgebras
 and $\Phi, \Psi $ morphisms of $\ZZ$-graded coalgebras
 ${\cal D}\rightarrow{\cal C}$.
 Suppose ${\cal C}$ is pointed irreducible, then $f=g$ if and only if
 ${\rm Im}(\Phi - \Psi)\cap P({\cal C})= \{ o \} $.
 \end{PP}

\begin{DD}
A bigraded coalgebra is a $\ZZ$-graded coalgebra ${\cal C}$ with
a $\Z_+$-grading 

\noindent
$
{\cal C} =  \parbox{35pt}{   \scriptsize $
                                    \hspace{5pt} \bigoplus \\
                                    \;\; k \geq 0   $  }\!\!\!\!\! {\cal C}^k
$
such that:
\begin{enumerate}
\item For every $k\geq0$, ${\cal C}^k$ is a $\ZZ$-graded subspace of ${\cal C}$.
\item $\ve({\cal C}^k) =0$ for all $k\geq 1$.
\item For every $k\geq0$, $\displaystyle 
\Delta({\cal C}^k) \subset
\parbox{30pt}{\scriptsize $\hspace{7pt} k \\
                              \makebox[2pt]{ } \bigoplus\\
                               i = 0 $                        }
\!\!\!\!\! {\cal C}^i \otimes {\cal C}^{k-i}$.
\end{enumerate}
A bigraded coalgebra ${\cal C}$ is called strictly bigraded if
${\cal C}^0 = \R$ and ${\cal C}^1$ coincides with the space $P({\cal C})$ of
all primitive elements of ${\cal C}$.
\end{DD}

\noindent
Note that the condition ${\cal C}^0 = \R$ implies that a strictly
bigraded coalgebra is pointed irreducible. The relation between
the $\Z_+$-grading and the coradical filtration in strictly
bigraded coalgebras is given by the following

\begin{PP}
A bigraded coalgebra ${\cal C}$ with
${\cal C}^0 = \R$ is strictly bigraded if and only if
for every $k\geq 0$
$$
{\cal C}^{(k)} =
\parbox{30pt}{\scriptsize $\hspace{7pt} k \\
                              \makebox[2pt]{ } \bigoplus\\
                               i = 0 $                        }
\!\!\!\!\! {\cal C}^i \;\;\;.
$$
\end{PP}

\subsection{Graded bialgebras}

\begin{DD}
A system $({\cal B}, \mu, u, \Delta, \varepsilon)$ where
${\cal B}$ is a $\ZZ$-graded space and
$\mu, u, \Delta, \varepsilon$ are morphisms of $\ZZ$-graded
spaces
$$
\begin{array}{lll}
\mu: {\cal B}\otimes {\cal B} \longrightarrow {\cal B} &
\;\;\;\;\;\;&
\Delta: {\cal B} \longrightarrow {\cal B}\otimes {\cal B}\\
u: \R \longrightarrow {\cal B}& &
\varepsilon: {\cal B} \longrightarrow \R
\end{array}
$$
is called a $\ZZ$-graded bialgebra if
\begin{enumerate}
\item $({\cal B}, \mu, u)$ is a $\ZZ$-graded algebra.
\item $({\cal B}, \Delta, \varepsilon)$ is a $\ZZ$-graded
coalgebra.
\item $\Delta$ and $\varepsilon$ are morphisms of
$\ZZ$-graded algebras.
\end{enumerate}
\end{DD}

\noindent
Note that condition 3. can be replaced by requirement that
$\mu$ and $u$ are morphisms of $\ZZ$-graded coalgebras.

\begin{DD}
A bigraded bialgebra is a $\ZZ$-graded bialgebra which is
both a bigraded algebra and bigraded coalgebra with
respect to the same $\Z_+$-grading.
\end{DD}

\noindent
A bigraded bialgebra is called cocommutative, pointed,
irreducible, strictly bigraded, if it
is so with respect to its coalgebra
structure.

\begin{DD}
A $\ZZ$-graded bialgebra ${\cal H}$ is called
a $\ZZ$-graded Hopf algebra if there exists a
morphizm of $\ZZ$-graded spaces
$s: {\cal H} \rightarrow {\cal H}$ such that the
diagram

\begin{center}
\begin{picture}(200,80)(0,10)
\put(-17,78){\makebox(40,10)[b]{$
%
   {\cal H} \otimes {\cal H}
$}}
\put(173,78){\makebox(40,10)[b]{$
%
   {\cal H}\otimes {\cal H}
$}}
\put(78,78){\makebox(40,10)[b]{$
%
   {\cal H}
$}}
\put(-17,8){\makebox(40,10)[b]{$
%
{\cal H}\otimes {\cal H}
$}}
\put(88,8){\makebox(20,10)[b]{$
%
  {\cal H}
$}}
\put(173,8){\makebox(40,10)[b]{$
%
   {\cal H}\otimes {\cal H}
$}}
\put(45,82){\line(1,0){40}}
\put(45,82){\vector(-1,0){20}}
\put(50,86){\makebox(10,10)[b]{\small$
  \Delta
$}}
\put(151,82){\vector(1,0){20}}
\put(151,82){\line(-1,0){40}}
\put(141,86){\makebox(10,10)[b]{\small$
  \Delta
$}}
\put(45,12){\line(1,0){40}}
\put(45,12){\vector(-1,0){20}}
\put(50,16){\makebox(10,10)[b]{\small$
  \mu
$}}
\put(151,12){\vector(1,0){20}}
\put(151,12){\line(-1,0){40}}
\put(141,16){\makebox(10,10)[b]{\small$
  \mu
$}}
\put(3,30){\vector(0,-1){9}}
\put(3,30){\line(0,1){43}}
\put(8,43){\makebox(20,10)[l]{\small$
 s\otimes {\rm id}
$}}
\put(193,30){\vector(0,-1){9}}
\put(193,30){\line(0,1){43}}
\put(198,43){\makebox(20,10)[l]{\small$
 {\rm id}\otimes s
$}}
\put(98,75){\vector(0,-1){19}}
\put(105,62){\makebox(20,10)[l]{\small$
\varepsilon
$}}
\put(98,38){\vector(0,-1){18}}
\put(105,25){\makebox(20,10)[l]{\small$
u
$}}
\put(88,43){\makebox(20,10)[b]{$
%
  \R
$}}
\end{picture}
\end{center}

\noindent commutes. The morphism $s$ is called the antipode of ${\cal H}$.
\end{DD}

\noindent
The antipode if exists is unique. One can also show that
$s$ is a $\ZZ$-graded algebra and coalgebra antimorphism
i.e the following diagrams are commutative.

\begin{center}
\begin{picture}(200,80)(0,10)
\put(-17,78){\makebox(40,10)[b]{$
%
   {\cal H} \otimes {\cal H}
$}}
\put(78,78){\makebox(20,10)[b]{$
%
   {\cal H}
$}}
\put(-17,8){\makebox(40,10)[b]{$
%
{\cal H}\otimes {\cal H}
$}}
\put(78,8){\makebox(20,10)[b]{$
%
  {\cal H}
$}}
\put(45,82){\vector(1,0){30}}
\put(45,82){\line(-1,0){20}}
\put(43,86){\makebox(10,10)[b]{\small$
  \mu
$}}
\put(45,12){\vector(1,0){30}}
\put(45,12){\line(-1,0){20}}
\put(43,16){\makebox(10,10)[b]{\small$
  \mu
$}}
\put(88,30){\vector(0,-1){9}}
\put(88,30){\line(0,1){43}}
\put(95,43){\makebox(20,10)[l]{\small$
 s
$}}
\put(3,74){\vector(0,-1){19}}
\put(10,62){\makebox(20,10)[l]{\small$
s\otimes s
$}}
\put(3,38){\vector(0,-1){18}}
\put(10,25){\makebox(20,10)[l]{\small$
T
$}}
\put(-7,43){\makebox(20,10)[b]{$
%
  {\cal H}\otimes {\cal H}
$}}
\put(140,43){\makebox(10,10)[r]{$
%
  \R
$}}
\put(155,53){\vector(1,1){24}}
\put(155,63){\makebox(10,10)[r]{\small$
u
$}}
\put(155,43){\vector(1,-1){24}}
\put(155,23){\makebox(10,10)[r]{\small$
u
$}}
\put(179,78){\makebox(20,10)[b]{$
%
   {\cal H}
$}}
\put(179,8){\makebox(20,10)[b]{$
%
{\cal H}
$}}
\put(189,30){\vector(0,-1){9}}
\put(189,30){\line(0,1){43}}
\put(196,43){\makebox(20,10)[l]{\small$
 s
$}}
\end{picture}

\begin{picture}(200,110)(0,10)
\put(-17,78){\makebox(40,10)[b]{$
%
   {\cal H}
$}}
\put(78,78){\makebox(20,10)[b]{$
%
   {\cal H} \otimes {\cal H}
$}}
\put(-17,8){\makebox(40,10)[b]{$
%
{\cal H}
$}}
\put(78,8){\makebox(20,10)[b]{$
%
  {\cal H} \otimes {\cal H}
$}}
\put(45,82){\vector(1,0){20}}
\put(45,82){\line(-1,0){30}}
\put(37,86){\makebox(10,10)[b]{\small$
  \Delta
$}}
\put(45,12){\vector(1,0){20}}
\put(45,12){\line(-1,0){30}}
\put(37,16){\makebox(10,10)[b]{\small$
  \Delta
$}}
\put(3,30){\vector(0,-1){9}}
\put(3,30){\line(0,1){43}}
\put(10,43){\makebox(20,10)[l]{\small$
 s
$}}
\put(88,74){\vector(0,-1){19}}
\put(95,62){\makebox(20,10)[l]{\small$
s\otimes s
$}}
\put(88,38){\vector(0,-1){18}}
\put(95,25){\makebox(20,10)[l]{\small$
T
$}}
\put(78,43){\makebox(20,10)[b]{$
%
  {\cal H}\otimes {\cal H}
$}}
\put(184,43){\makebox(10,10)[r]{$
%
  \R
$}}
\put(179,53){\vector(-1,1){24}}
\put(169,63){\makebox(10,10)[l]{\small$
u
$}}
\put(179,43){\vector(-1,-1){24}}
\put(169,23){\makebox(10,10)[l]{\small$
u
$}}
\put(135,78){\makebox(20,10)[b]{$
%
   {\cal H}
$}}
\put(135,8){\makebox(20,10)[b]{$
%
{\cal H}
$}}
\put(145,30){\vector(0,-1){9}}
\put(145,30){\line(0,1){43}}
\put(150,43){\makebox(20,10)[l]{\small$
 s
$}}
\end{picture}

\end{center}

\noindent
If ${\cal H}$ is a bigraded bialgebra and the antipode exists
it necessarily respects $\Z_+$-grading.

\subsection{Dual coalgebras}

Let $(A,M,u)$ be an algebra over $\R$ with multiplication
$M: A\otimes A \ra A$
and unit $u:\R \ra A$. We denote by  $A^{\circ}$
 the subspace of the full algebraic dual $A'$
consisting of all elements $\alpha \in A'$ such that
${\rm ker} \alpha$ contains a cofinite ideal of $A$.

\begin{PP}
Let $A, B$ be $\ZZ$-graded algebras and
$F: A \rightarrow B$
a  morphism of  $\ZZ$-graded algebras.
Then:
\begin{enumerate}
\item
$A^{\circ}$ is a linear subspace of $A'$.
\item
Let $F': B' \rightarrow A'$ be dual to $F: A \rightarrow B$.
Then $F'(B^{\circ}) \subset A^{\circ}$.
\item
$A^{\circ}\otimes B^{\circ} = (A\otimes B)^{\circ}$.
\item
Let $M' : A \rightarrow (A\otimes A)'$ be the dual to the multiplication 
in $A$. Then $M'( A^{\circ}) \subset A^{\circ}\otimes A^{\circ}$.
\end{enumerate}
\end{PP}

\begin{PP}
Let $(A,M,u)$  be a $\ZZ$-graded algebra 
over $\R$.  Then the maps
\begin{eqnarray*}
\Delta \equiv M'_{|A^{\circ}} : A^{\circ} &\longrightarrow&
A^{\circ}\otimes A^{\circ}\;\;\;,\\
\ve \equiv u'_{|A^{\circ}} : A^{\circ} &\longrightarrow&\R\;\;\;,
\end{eqnarray*}
define on $A^{\circ}$ a structure of $\ZZ$-graded coalgebra.
If $A$ is $\ZZ$-graded commutative then  $A^{\circ}$
is $\ZZ$-graded cocommutative.
\end{PP}

\begin{DD}
The coalgebra $(A^{\circ},\D,\ve)$
of {\rm Prop.A.5.2} is called the dual coalgebra of $(A,M,u)$.
\end{DD}

\noindent
The following property of $(A^{\circ},\D,\ve)$ may serve as 
independent definition of dual coalgebra.
\begin{PP}
$A^{\circ}$ is the maximal coalgebra in $A'$ i.e. if
for $\a \in A'$, $M'(\a) \in A' \otimes A'$ then $\a \in A^{\circ}$.
\end{PP}

\noindent The dual
coalgebras of  $\ZZ$-graded algebras   of superfunctions
on BLK supermanifolds have been analysed in  \cite{Kostant,Batchelor}.
Here we briefly present structure results used in the main text.

\begin{PP}
Let ${\cal A}(M)$ be the $\ZZ$-graded algebra   of superfunctions
on a supermanifold   ${\cal A}_M$. Then
\begin{enumerate}
\item ${\cal A}(M)^{\circ}$ is a pointed  $\ZZ$-graded cocomutative
coalgebra. 
\item Each group-like element of   ${\cal A}(M)^{\circ}$ is of the
form 
$$
\delta_p=\langle p,\;.\;\rangle : {\cal A}(M) \ni f \lra f^0(p)\in \R\;\;\;,
$$
for some $p \in M$.
\end{enumerate}
\end{PP}

\noindent By the structure theorem for pointed
$\ZZ$-graded cocomutative coalgebras (Th.A.3.1), one has

\begin{PP}
${\cal A}(M)^{\circ}$ is the direct sum of pointed  irreducible 
coalgebras
$$
{\cal A}(M)^{\circ} =  \parbox[c]{35pt}{   \scriptsize $
                                    \makebox(5,13)[b]{} \bigoplus \\
                                     \makebox(2,2){}p\in M $  }
\!\!\!\!{\cal A}(M)^{\circ}_p\;\;\;,
$$
where ${\cal A}(M)^{\circ}_p$ denotes the irreducible component 
containing the group-like element  $\delta_p$.
\end{PP}

\noindent Applying the structure theorem for pointed irreducible
$\ZZ$-graded coalgebras (Th.A.3.2)
one gets for each irreducible component
 ${\cal A}(M)^{\circ}_p$ the coradical filtration
$$
{\cal A}(M)^{\circ}_p =  \parbox[c]{35pt}{   \scriptsize $
                                    \makebox(7,12)[b]{} \bigcup \\
                                     \makebox(2,2){}p\in M $  }
\!\!\!\!{\cal A}(M)^{\circ(k)}_p\;\;\;,
$$
with 
\begin{eqnarray*}
{\cal A}(M)^{\circ(0)}_p &=& \R p \;\;\;,\\
{\cal A}(M)^{\circ(1)}_p &=& \R p \oplus P({\cal A}(M)^{\circ}_p)\;\;\;,
\end{eqnarray*}  
where $P({\cal A}(M)^{\circ}_p)$ is the space of all primitive
with respect to $\delta_p$ elements of ${\cal A}(M)^{\circ}$.
A more detailed description is given by the following
\cite{Kostant,Batchelor}
\begin{PP}
Let $
{\cal A}(M)^{\circ}_p =  \parbox[c]{35pt}{   \scriptsize $
                                    \makebox(7,12)[b]{} \bigcup \\
                                     \makebox(2,2){}k\geq 0 $  }
\!\!\!\!{\cal A}(M)^{\circ(k)}_p
$ be the coradical filtration of the irreducible component
${\cal A}(M)^{\circ}_p$ of the dual coalgebra 
${\cal A}(M)^{\circ}$. Then for each $k\geq 0$
$$
{\cal A}(M)^{\circ(k)}_p = \left\{ \alpha\in {\cal A}(M)^{\circ(k)}:
\langle \alpha, I^{k+1}_p \rangle =0\right\} 
\;=\; (  {{\cal A}(M) /  I^{k+1}_p} )'\;\;\;,
$$
where $I_p$ is a maximal ideal in  ${\cal A}(M)$ consisting of
all superfunctions $f\in {\cal A}(M)$ such that $f^0(p)=0$.
\end{PP}

\end{document}